\documentclass{sig-alternate}

\usepackage[T1]{fontenc}
\usepackage{times}
\usepackage{mathptmx}

\usepackage[latin1]{inputenc}
\usepackage{amssymb, latexsym, amsmath}
\usepackage{enumerate}
\usepackage{epic}
\usepackage{eepic}
\usepackage{pst-tree}
\usepackage{ifthen}
\usepackage{calc}
\usepackage{graphicx}
\usepackage{url}
\usepackage{alltt}

\def\thepage{\arabic{page}}

\newcommand{\nc}{\newcommand}
\nc{\rnc}{\renewcommand}
\nc{\nev}{\newenvironment}

\rnc{\marginpar}[1]{}

\nc{\bs}[1]{\boldsymbol{#1}}

\newcommand*{\Next}{\ensuremath{\textrm{\upshape Next}}}

\nc{\lcm}{\ensuremath{\textrm{\upshape lcm}}}
\rnc{\mod}{\ensuremath{\textrm{\upshape mod}}}

\nc{\bigO}{\ensuremath{{O}}}

\newenvironment{myeqnarray}{\begin{equation}\begin{array}{rcl}\displaystyle}{\end{array}\end{equation}}
\newenvironment{myeqnarray*}{\begin{equation*}\begin{array}{rcl}\displaystyle}{\end{array}\end{equation*}}

\nc{\Randbem}[1]{\marginpar{\footnotesize #1}}

\nc{\ov}[1]{\overline{#1}}
\nc{\vek}{\ov}
\rnc{\vec}{\vek}

\nc{\twodots}{.\,.\,}
\nc{\deff}{:=}
\nc{\parno}{\par\noindent}
\nc{\raus}[1]{}

\nc{\llbrack}{\ensuremath{\lbrack\!\lbrack}}
\nc{\rrbrack}{\ensuremath{\rbrack\!\rbrack}}

\nc{\Blank}{\ensuremath{\Box}}
\nc{\?}{\ensuremath{\circledast}}
\nc{\Nil}{\ensuremath{\circleddash}} 
\nc{\undef}{\ensuremath{\bot}}

\nc{\fertig}{\hfill$\Box$\vspace{\topsep}\par}
\nc{\NN}{\ensuremath{\mathbb{N}}}
\nc{\Al}{\ensuremath{\mathbb{A }}}
\nc{\set}[1]{\ensuremath{\{ #1 \}}}
\nc{\setc}[2]{\set{ #1 : #2}}
\nc{\struc}[1]{\ensuremath{\langle #1 \rangle}}

\nc{\lv}{\ensuremath{\textit{lv}}}
\nc{\cp}{\ensuremath{\textit{cp}}}
\nc{\rcp}{\ensuremath{\textit{rcp}}}

\nc{\ind}{\textit{ind}}
\nc{\skel}{\textit{skel}}

\nc{\rev}{\textup{rev}}

\nc{\config}{\ensuremath{\textit{config}}}
\nc{\tapeconf}{\ensuremath{\textit{tape-config}}}
\nc{\Bin}{\ensuremath{\textit{BIN}}}

\nc{\PartConfT}{\ensuremath{\textit{Conf}_T}}

\nc{\pleft}{\ensuremath{p^{\llbrack}}}
\nc{\pright}{\ensuremath{p^{\rrbrack}}}
\nc{\pact}{\ensuremath{p^{\uparrow}}}

\nc{\phleft}{\ensuremath{\hat{p}^{\llbrack}}}
\nc{\phright}{\ensuremath{\hat{p}^{\rrbrack}}}
\nc{\phact}{\ensuremath{\hat{p}^{\uparrow}}}

\nc{\psleft}{\ensuremath{p'{}^{\llbrack}}}
\nc{\psright}{\ensuremath{p'{}^{\rrbrack}}}
\nc{\psact}{\ensuremath{p'{}^{\uparrow}}}

\nc{\pssleft}{\ensuremath{p''{}^{\llbrack}}}
\nc{\pssright}{\ensuremath{p''{}^{\rrbrack}}}
\nc{\pssact}{\ensuremath{p''{}^{\uparrow}}}

\nc{\Bacc}{\ensuremath{B_{\textit{acc}}}}
\nc{\Brej}{\ensuremath{B_{\textit{rej}}}}

\nc{\Problem}[1]{\ensuremath{\textsc{#1}}}
\nc{\SETEQUALITY}{\Problem{Set-Equality}}
\nc{\MULTISETEQUALITY}{\Problem{Multiset-Equality}}
\nc{\MSEQUALITY}{\Problem{(Multi)Set-Equality}}
\nc{\CHECKSORT}{\Problem{Check-Sort}}
\nc{\CHECK}{\Problem{Check}}
\nc{\SHORT}{\Problem{Short}}

\nc{\uend}{\hfill$\dashv$\par}
\nc{\uendeq}{\eqno\dashv}

\nc{\proofend}{\hfill$\Box$\par}
\nc{\proofendeq}{\eqno\Box}

\rnc{\qed}{\proofend}
\nc{\qedeq}{\proofendeq}

\newcounter{theo}

\nev{defform}[2]{\refstepcounter{theo}\begin{list}{}{%
      \setlength{\labelwidth}{0em}%
      \setlength{\leftmargin}{0em}%
      \setlength{\itemindent}{\labelsep}%
      \setlength{\listparindent}{1.5em}
      \setlength{\topsep}{0.5ex}
      \setlength{\parsep}{0em}}%
      \item[\textbf{#2 \thetheo\ifthenelse{\equal{#1}{}}{}{ (#1)}.}]}%
    {\end{list}
}

\nev{satzform}[2]{\refstepcounter{theo}\begin{list}{}{%
      \setlength{\labelwidth}{0em}%
      \setlength{\leftmargin}{0em}%
      \setlength{\itemindent}{\labelsep}%
      \setlength{\listparindent}{1.5em}
      \setlength{\parsep}{0em}
      \setlength{\topsep}{0.65ex}
    }
  \item[\textbf{#2 \thetheo\if #1\else\ (#1)\fi.}]\itshape}%
  {\end{list}
}

\nev{theorem}[1][]{\begin{satzform}{#1}{Theorem}}{\end{satzform}}
\nev{lemma}[1][]{\begin{satzform}{#1}{Lemma}}{\end{satzform}}
\nev{corollary}[1][]{\begin{satzform}{#1}{Corollary}}{\end{satzform}}
\nev{proposition}[1][]{\begin{satzform}{#1}{Proposition}{}}{\end{satzform}}
\nev{proviso}[1][]{\begin{satzform}{#1}{Proviso}{}}{\end{satzform}}
\nev{fact}[1][]{\begin{satzform}{#1}{Fact}}{\end{satzform}}
\nev{example}[1][]{\begin{defform}{#1}{Example}}{\end{defform}}
\nev{definition}[1][]{\begin{defform}{#1}{Definition}}{\end{defform}}
\nev{remark}[1][]{\begin{defform}{#1}{Remark}}{\end{defform}}

\nev{cenv}[1]{\begin{list}{}{%
      \setlength{\labelwidth}{0em}%
      \setlength{\leftmargin}{0em}%
      \setlength{\itemindent}{\labelsep}%
      \setlength{\listparindent}{1.5em}
      \setlength{\topsep}{0.5ex}
      \setlength{\parsep}{0em}}%
      \item[\textit{#1}]}%
    {\hfill$\dashv$\end{list}}

\newcounter{claimcounter}

\nev{bfenv}[1]{\begin{list}{}{%
      \setlength{\labelwidth}{0em}%
      \setlength{\leftmargin}{0em}%
      \setlength{\itemindent}{\labelsep}%
      \setlength{\listparindent}{1.5em}
      \setlength{\topsep}{0.5ex}
      \setlength{\parsep}{0em}}%
      \item[\textbf{#1}]}%
    {\hfill$\dashv$\end{list}}

\rnc{\proof}[1][]{\noindent\setcounter{claimcounter}{0}\ifthenelse{\equal{#1}{}}{\textit{Proof:
    }}{\textit{Proof of #1: }}}

\nc{\mr}[1]{\marginpar{\footnotesize #1}}
\rnc{\labelenumi}{(\arabic{enumi})}
\rnc{\labelitemi}{--}
\rnc{\phi}{\varphi}
\rnc{\epsilon}{\varepsilon}
\nc{\bigmid}{\;\big|\;}
\nc{\Bigmid}{\;\Big|\;}
\nc{\bin}{\textup{bin}}

\renewcommand{\fboxsep}{2mm}
\newsavebox{\fminibox}
\newlength{\fminilength}
\newenvironment{fminipage}[1][\textwidth]{
  \setlength{\fminilength}{#1-4mm-2\fboxrule}%
  \begin{lrbox}{\fminibox}\begin{minipage}{\fminilength}}{
    \end{minipage}\end{lrbox}\noindent\fbox{\usebox{\fminibox}}
}

\newlength{\pwidth}

\newenvironment{problem}[3]{
  \begin{center}
    \ifthenelse{\equal{#1}{}}%
    {\setlength{\pwidth}{\columnwidth}}%
    {\setlength{\pwidth}{#1cm}}
    \nc{\instance}{\item[Instance]}
    \nc{\parameter}{\item[Parameter]}
    \rnc{\problem}{\item[Problem]}
    \begin{fminipage}[\pwidth]\upshape
      \ifthenelse{\equal{#3}{}}{}{{\scshape #3}}
      \begin{list}{}{
          \ifthenelse{\equal{#2}{}}%
          {\settowidth{\labelwidth}{\textit{Parameter:}}}%
          {\settowidth{\labelwidth}{\textit{#2:}}}
          
          \setlength{\leftmargin}{\labelwidth+\labelsep}
        }
      }{
      \end{list}
    \end{fminipage}
  \end{center}
}

\newenvironment{proofcomment}[1]{\textbf{Proof #1: }}{}

\newtheorem{myclaim}{Claim}{\itshape}{\itshape}

\nc{\ST}{\textup{ST}}
\nc{\NST}{\textup{NST}}
\nc{\RST}{\textup{RST}}
\nc{\FRST}{\textup{\textit{LasVegas}-RST}}

\nc{\PTIME}{\textup{PTIME}}
\nc{\NP}{\textup{NP}}
\nc{\RP}{\textup{RP}}
\nc{\co}{\textup{co-}}

\nc{\TIME}{\textup{TIME}}
\nc{\NTIME}{\textup{NTIME}}
\nc{\RTIME}{\textup{RTIME}}

\title{Randomized Computations on Large Data Sets:\\Tight Lower Bounds\\[2mm]
    {\large\sl -- Full Version --}\vspace{-2mm}}

\numberofauthors{3}

\author{
  \alignauthor Martin Grohe
  \alignauthor Andr\'{e} Hernich
  \alignauthor Nicole Schweikardt
  \end{tabular}
  \begin{tabular}{c}
    \affaddr{Institut f\"ur Informatik, Humboldt-Universit\"at, Berlin, Germany} \\[0.5mm]
    \eaddfnt{\{\,grohe\,|\,hernich\,|\,schweika\,\}@informatik.hu-berlin.de}
}

\begin{document}

\maketitle

\begin{abstract}
  We study the randomized version of a computation model (introduced in
  \cite{grokocschwe05,groschwe05a}) that restricts {random access} to
  external memory and internal memory {space}.
  Essentially, this model can be viewed as a powerful version of a data stream model
  that puts no cost on sequential scans of
  external memory (as other models for data streams) and, in addition, (like
  other external memory models, but unlike streaming models), admits several
  large external memory devices that can be read and written to in parallel.

  We obtain tight lower bounds for the decision problems set equality, multiset
  equality, and checksort. More precisely, we show that any randomized one-sided-error
  bounded Monte Carlo algorithm for these problems must perform $\Omega(\log N)$
  random accesses to external memory devices, provided that the internal memory
  size is at most $O(\sqrt[4]{N} / {\log N})$, where $N$ denotes the size of the
  input data.

  From the lower bound on the set equality problem
  we can infer lower bounds
  on the worst case data complexity of query evaluation for
  the languages XQuery, XPath, and  relational algebra on streaming data.
  More precisely, we show that there exist queries in XQuery, XPath, and relational algebra,
  such that any (randomized) Las Vegas algorithm that evaluates these queries must perform
  $\Omega(\log N)$ random accesses to external memory devices, provided that the internal memory
  size is at most $O(\sqrt[4]{N}/ {\log N})$.
\end{abstract}%


%
%
%

\newpage

\section{Introduction}\label{section:Introduction}

Today's hardware technology provides a hierarchy of storage media from tapes
and disks at the bottom through main memory and (even on-CPU) memory caches at
the top. Storage media from different levels of this memory hierarchy
considerably differ in price, storage size, and access time. Currently, the
most pronounced performance and price (and consequently also size) gap is
between main memory and the next-lower level in the memory hierarchy, usually
magnetic disks which have to rely on comparably slow, mechanical, physically
moving parts.  One often refers to the upper layers above this gap by
\emph{internal memory} and the lower layers of the memory hierarchy by
\emph{external memory}.  The technological reality is such that the time for
accessing a given bit of information in external memory is five to six orders
of magnitude larger than the time required to access a bit in internal memory.
Apart from this, concerning external memory, \emph{random accesses} (which
involve moving the disk head to a particular location) are significantly more
expensive than \emph{sequential scans}.

Modern software and database technology uses clever heuristics to minimize the
number of accesses to external memory and to prefer \emph{streaming} over
\emph{random accesses} to external memory. There has also been a wealth of
research on the design of so-called \emph{external memory algorithms} (cf.,
e.g.\ \cite{mut03,vit01,meysansib03}).  The classes considered in
\emph{computational complexity theory}, however, usually do not take into
account the existence of different storage media.
In \cite{grokocschwe05,groschwe05a}, we introduced a formal model for
such a scenario. The two most significant cost measures in our setting are the
number of random accesses to external memory and the size of the internal
memory. Our model is based on a standard multi-tape Turing machine.  Some of
the tapes of the machine, among them the input tape, represent the external
memory. They are unrestricted in size, but access to these tapes is restricted
by allowing only a certain number $r(N)$ (where $N$ denotes the input size) of
reversals of the head directions. This may be seen as a way of (a) restricting the
number of sequential scans and (b) restricting
random access to these tapes, because each random access can be simulated by
moving the head to the desired position on a tape, which involves at most two
head reversals. The remaining tapes of the Turing machine represent the
internal memory. Access to these internal memory tapes (i.e., the number of
head reversals) is unlimited, but their size is bounded by a parameter $s(N)$.
\marginpar{$\rceil$ \\ !}
We let $\ST(r(N),s(N),O(1))$ denote the class of all problems that can be
solved on such an $\big(r(N),s(N),O(1)\big)$-bounded Turing machine, i.e., a Turing machine
with an arbitrary number of external memory tapes which, on inputs of size $N$, performs less than
$r(N)$ head reversals on the external memory tapes, and uses at most space $s(N)$ on the
internal memory tapes.

The astute reader who wonders if it is realistic to assume that the external
memory tapes can be read in \emph{both} directions (which disks cannot so easily)
and that a sequential scan of an entire external memory tape accounts for only one
head reversal (and thus seems unrealistically cheap) be reminded that this paper's main goal
is not to design efficient external memory algorithms but, instead, to
prove \emph{lower} bounds. Thus, considering a rather powerful computation model
makes our lower bound results only stronger.\marginpar{$\lfloor$}

\marginpar{$\lceil$ \\ ! }
In the present paper, we
prove lower bounds for \emph{randomized} computations (i.e., computations
where in each step a coin may be tossed to determine the next configuration) in a scenario with
several storage media. To this end, we introduce the complexity class $\RST(r(N),s(N),O(1))$, which
consists of all decision problems that can be solved by an
$\big(r(N),s(N),O(1)\big)$-bounded randomized Turing machine with
one-sided bounded error,
where no false positive answers are allowed and the probability of
false negative answers is at most $0.5$ (in the literature, such randomized algorithms are often called
\emph{one-sided-error Monte Carlo algorithms}, cf.\ \cite{Hromkovic}).
To also deal with computation problems where an output (other than just a yes/no answer)
has to be generated, we write $\FRST\big(r(N),s(N),O(1)\big)$ to denote the class
of all functions $f$ for which there exists an $\big(r(N),s(N),O(1)\big)$-bounded
randomized Turing machine that, for every input word $w$, (a) always produces either the
correct output $f(w)$ on one of its external memory tapes or gives the answer \emph{``I don't
know''} and (b) gives the answer \emph{``I don't know''} with probability at most $0.5$
(in the literature, such randomized algorithms are sometimes called
\emph{Las Vegas algorithms}, cf.\ \cite{Hromkovic}).
\marginpar{$\lfloor$}
\bigskip\\
\textbf{Contributions:} \
Our first main result is a lower bound for
three natural decision problems: The \emph{set equality problem} and the
\emph{multiset equality problem} ask whether two given
(multi)sets
of strings are equal, and the \emph{checksort problem} asks, given two
sequences of strings, whether the second is a sorted version of the first.
\\
We show (\mbox{Theorem~\ref{theo:set-equality}}) that neither problem is contained in \linebreak
$\RST(o(\log N),O(\frac{\sqrt[4]{N}}{\log N}),O(1))$.
This lower bound turns out to be \emph{tight} in the following senses:
\begin{enumerate}[$\bullet$]
\item
  If the number of sequential scans (i.e., head reversals) increases from
  $o(\log N)$ to $O(\log N)$, then each of the three
  problems
  can be solved with only constant internal memory and without using
  randomization. In other words (see Corollary~\ref{cor:tight-bounds-for-short-versions}),
  the (multi)set equality problem and the
  checksort problem belong to \linebreak $\ST(O(\log N),O(1),O(1))$.
\item
  When using randomization with the complementary one-sided error model, i.e.,
  machines where no false negative answers are allowed and the probability of
  false positive answers is at most 0.5, then the \emph{multiset equality} problem
  can be solved with just two sequential scans of the input
  (and without ever writing to external memory), and internal memory of size $O(\log N)$.
  In other words (Theorem~\ref{thm:efficient-randomized/ndet-solutions}(a)),
  the multiset equality problem belongs to
  $\co\RST(2,O(\log N),1)$.
\item
  When using nondeterministic machines, then (multi)set equality and checksort can
  be solved with three sequential scans on two external memory tapes and
  internal memory of size $O(\log N)$.
  In other words (Theorem~\ref{thm:efficient-randomized/ndet-solutions}(b)),
  the (multi)set equality problem and the checksort problem belong to
  $\NST(3,O(\log N),2)$.
\end{enumerate}
As a consequence,
we obtain a separation between the deterministic, the randomized,
and the nondeterministic $\ST(\cdots)$ classes
(Cor\-ol\-lary~\ref{cor:DetVsRandVsNdet}).

Our lower bound for the checksort problem, in particular, implies that
the \emph{sorting problem} (i.e., the problem of sorting a sequence
of input strings) does not belong to the complexity class
$\FRST(o(\log N),O(\frac{\sqrt[4]{N}}{\log N}),O(1))$ and thus generalizes the main result
of \cite{groschwe05a} to {randomized} computations.

Our lower bound for the set equality problem leads to the following
lower bounds on the worst case data complexity of database query evaluation problems
in a streaming context:
\begin{enumerate}[$\bullet$]
\item
  There is an \emph{XQuery} query $Q$ such that the problem of evaluating $Q$ on an
  input XML document stream of length $N$ does \emph{not} belong to the class
  $\FRST(o(\log N),O(\frac{\sqrt[4]{N}}{\log N}),O(1))$ (Theorem~\ref{Thm:XQuery}).
\end{enumerate}
Speaking informally, this means that, no matter how many
external memory devices (of arbitrarily large size) are available, as long as the internal memory is of size at most
$O(\frac{\sqrt[4]{N}}{\log N})$, every randomized algorithm that produces the correct
query result with probability at least $0.5$ will
perform $\Omega(\log N)$ random accesses to
external memory.
We obtain analogous results for \emph{relational algebra} queries and for the
node-selecting XML query language \emph{XPath}:
\begin{enumerate}[$\bullet$]
\item
  There is a \emph{relational algebra} query $Q$ such that the problem of evaluating $Q$ on
  a stream consisting of the tuples of the input database relations does
  \emph{not} belong to the complexity class
  $\FRST(o(\log N),O(\frac{\sqrt[4]{N}}{\log N}),O(1))$, where $N$ denotes the total size of the input database
  relations.
  Furthermore, this bound is tight with respect to the number of random accesses to
  external memory, as
  the data complexity of every relational algebra query belongs to
  $\ST(O(\log N),O(1),O(1))$
  (Theorem~\ref{Thm:RelAlgebra}).
\item
  There is an \emph{XPath} query $Q$ such that the problem of filtering an input XML
  document stream with $Q$ (i.e., checking whether at least one node of the document matches
  the query) does \emph{not} belong to the class $\co\RST(o(\log N),O(\frac{\sqrt[4]{N}}{\log N}),O(1))$
  (Theo-\linebreak rem~\ref{Thm:XPath}).
\end{enumerate}
This means that there is an \emph{XPath} query $Q$ such that,
no matter how many
external memory devices (of arbitrarily large size) are available, as long as the internal memory is of size at most
$O(\frac{\sqrt[4]{N}}{\log N})$, every randomized algorithm which
accepts every input document that matches $Q$, and which rejects documents not matching $Q$ with probability $\geq 0.5$,
will perform $\Omega(\log N)$ random accesses to external memory.
\bigskip\\
\textbf{Related Work:} \
Obviously, our model is related to the \emph{bounded reversal Turing
machines}, which have been studied in classical complexity theory (see,
e.g., \cite{wagwec86,cheyap91}). However, in bounded reversal Turing machines, the
number of head reversals is limited on \emph{all} tapes, whereas in our
model there is no such restriction on the internal memory tapes. This makes
our model considerably stronger, considering that in our lower bound results
we allow internal memory size that is polynomially related to the input
size.
\marginpar{$\rceil$ \\ !}
Furthermore, to our best knowledge, all lower bound proofs
previously known for reversal complexity classes on
multi-tape Turing machines go back to the space hierarchy
theorem (cf., e.g., \cite{Papadimitriou,cheyap91}) and thus rely on diagonalization arguments,
and apply only to classes with
$\omega(\log N)$ head reversals. In particular, these lower bounds do not include
the checksort problem and the (multi)set equality problem, as these problems
can be solved with $O(\log N)$ head reversals.

In the classical \emph{parallel disk model} for external memory algorithms
(see, e.g., \cite{vit01,meysansib03,mut03}), the cost measure is simply the
number of bits read from external memory divided by the page size.  Several
refinements of this model have been proposed to include a distinction between
\emph{random access} and \emph{sequential scans} of the external memory, among
them Arge and Bro Miltersen's \emph{external memory Turing machines}
\cite{ArgeBroMilteresn_ExtMemTMs}. We note that their notion of external
memory Turing machines significantly differs from ours, as their machines only
have a single external memory tape and process inputs that consist of a
\emph{constant} number $m$ of input strings. Strong lower bound results (in
particular, for different versions of the \emph{sorting problem}) are known
for the parallel disk model (see \cite{vit01} for an overview) as well as for
Arge and Bro Miltersen's external memory Turing machines
\cite{ArgeBroMilteresn_ExtMemTMs}. However, to the best of our knowledge, all
these lower bound proofs heavily rely on the assumption that the input data
items (e.g., the strings that are to be sorted) are \emph{indivisible} and
that at any point in time, the external memory consists, in some sense, of a
permutation of the input items. We emphasize that the present paper's lower bound
proofs do not rely on such an indivisibility assumption.

Strong lower bounds for a number of problems are known in the context of \emph{data streams}
and for models which permit
a small number of sequential scans of the input data, but no auxiliary
external memory (that is, the version of our model with no extra external
memory tapes apart from the input tape)
\cite{munpat80,alomatsze99,henragraj99,bbdmw02,mut03,barfonjos04,barfonjos05,aggdatrajruh04,grokocschwe05}.
All these lower bounds are obtained by communication complexity.  Note that in
the presence of at least two external memory tapes, communication between
remote parts of memory is possible by simply copying data from one tape to
another and then re-reading both tapes in parallel.  These communication
abilities of our model spoil any attempt to prove lower bounds via
communication complexity, which is the tool of choice both for computation
models permitting few scans but no auxiliary external memory, and for 1-tape
Turing machines.

The deterministic $\ST(\cdots)$-classes were introduced in
\cite{groschwe05a,grokocschwe05}.  In \cite{grokocschwe05} we studied those
classes where only a \emph{single} external memory tape is available and used
methods from communication complexity to obtain lower bounds for these
classes.
The main result of
\cite{groschwe05a} was a lower bound for the \emph{sorting problem} concerning
the deterministic $\ST(\cdots)$-classes with an \emph{arbitrary} number of
external memory tapes.  An important tool for proving this bound was to
introduce deterministic \emph{list machines} as an intermediate machine model.
An overview of the methods used and the results obtained in
\cite{grokocschwe05,groschwe05a} was given in \cite{GKS_FCT05}.  The present
paper builds on \cite{groschwe05a}, as it considers $\ST(\cdots)$-classes with
an \emph{arbitrary} number of external memory tapes and it uses \emph{list
machines} as a key tool for proving lower bound results.  However, the results
presented here go significantly beyond those obtained in
\cite{groschwe05a}. Here we obtain lower bounds for \emph{decision} problems
in the \emph{randomized} versions of the model.  The main result of
\cite{groschwe05a} is that the sorting problem does not belong to $\ST(o(\log
N),\bigO\big(\frac{\sqrt[5]{N}}{\log N}\big),\bigO(1))$, and the proof given
there heavily relies on the fact that the machines are \emph{deterministic}
and the \emph{output} of the sorting problem cannot be generated within the
given resource bounds.  In contrast to the present paper's approach,
the proof method of \cite{groschwe05a} neither works
for decision problems, i.e.\ problems where no output is generated, nor
for randomized computations.  Finally, let us remark that the main result
of \cite{groschwe05a} can be obtained as an immediate corollary of the
present paper's lower bound for the checksort problem.
\bigskip\\
\textbf{Organization:} \
After introducing the deterministic, the nondeterministic, and
the randomized $\ST(\cdots)$ classes in
Section~\ref{section:ClassST}, we formally state our main lower bounds
for decision problems in
Section~\ref{section:MainResults}. In Section~\ref{section:MainResults_QueryEval}
we use these results to derive lower bounds on the data complexity of query evaluation for
the languages
\emph{XQuery}, \emph{XPath}, and \emph{relational algebra}.
The subsequent sections are devoted to the proof of the lower bound on the
decision problems
\emph{(multi)set equality} and \emph{checksort}:
In Section~\ref{section:ListMachines}, \ref{section:SimulationLemma}, and
\ref{section:LowerBoundsForLM} we introduce randomized \emph{list machines},
show that
randomized Turing machines can be simulated by randomized list machines, and
prove that randomized list machines
can neither solve the \emph{(multi)set equality problem} nor the \emph{checksort problem}.
Afterwards, in Section~\ref{section:BoundsForTMs} we transfer these results from
list machines to Turing machines.
We close with
a few concluding remarks and open problems in Section~\ref{section:Conclusion}.

The present paper is the full version of the extended abstract published in the
proceedings of the 25th ACM Sigact-Sigart Symposium on Principles of Database Systems (PODS'06).

%
\section{Complexity Classes}\label{section:ClassST}

We write $\mathbb N$ to denote the set of natural numbers (that is, nonnegative integers).

As our basic model of computation, we use standard multi-tape nondeterministic
Turing machines (NTMs, for short);
cf., e.g., \cite{Papadimitriou}.
The Turing machines we consider will have $t+u$ tapes. We call the first $t$
tapes \emph{external memory tapes} (and think of them as representing $t$
disks). We call the other $u$ tapes \emph{internal memory tapes}. The first
tape is always viewed as the input tape.

Without loss of generality we assume that our Turing machines are
{normalized} in such a way that in each step at most one of its heads
moves to the left or to the right.

Let $T$ be an NTM and $\rho$ a finite run of $T$.
Let $i\ge 1$ be the number of a tape.
We use
\(
  \text{rev}(\rho,i)
\)
to denote the number of times the head on tape $i$ changes
its direction in the run $\rho$.
Furthermore, we let
\(
\textup{space}(\rho,i)
\)
be the number of cells of tape $i$ that are used by $\rho$.

\begin{definition}[$(r,s,t)$-bounded TM]\label{def:boundedTM}
\upshape
  Let $r,s:\mathbb N\to\mathbb N$ and $t\in\mathbb N$.
  A (nondeterministic) Turing machine $T$ is \emph{$(r,s,t)$-bounded}, if every run $\rho$
    of $T$ on an input of length $N$ (for arbitrary $N\in\NN$) satisfies the following conditions:
   (1) \ $\rho$ is finite,
   (2) \ $1+\sum_{i=1}^t\textup{rev}(\rho,i)\le r(N)$, and\footnote{It is
        convenient for technical reasons to add $1$ to the number
        $\sum_{i=1}^t\textup{rev}(\rho,i)$ of changes of the head direction
        here. As defined here, $r(N)$ thus bounds the number of sequential
        scans of the external memory tapes rather than the number of changes
        of head directions.}
   (3) \ $\sum_{i=t+1}^{t+u}\textup{space}(\rho,i)\le s(N)$, where $t+u$ is
      the total number of tapes of $T$. \uend
\end{definition}

\begin{definition}[$\ST(\cdots)$ and $\NST(\cdots)$ classes]\label{def:ST}
\upshape \mbox{} \\
  Let $r,s:\mathbb N\to\mathbb N$ and $t\in\mathbb N$.
  A decision problem belongs to the class $\ST(r,s,t)$ (resp., $\NST(r,s,t)$), if it
  can be decided by a deterministic (resp., nondeterministic) $(r,s,t)$-bounded Turing machine. \uend
\end{definition}

Note that we put no restriction on the running time or the space used on the
first $t$ tapes of an $(r,s,t)$-bounded Turing machine. The following
lemma shows that these parameters cannot get too large.
\begin{lemma}[{\cite{groschwe05a}}]
\label{lem:time}
  Let $r,s:\mathbb N\to\mathbb N$ and $t\in\mathbb N$, and let $T$ be an
  $(r,s,t)$-bounded NTM. Then for every run $\rho=(\rho_1,\ldots,\rho_\ell)$ of $T$ on an input of
  size $N$ we have
  \(
  \ell  \le  N\cdot 2^{O(r(N)\cdot(t+s(N)))}
  \)
  and thus
  $
    \sum_{i=1}^t\textup{space}(\rho,i)  \linebreak \leq   N\cdot  2^{O(r(N)\cdot(t+s(N)))}.
  $
  \uend
\end{lemma}

In \cite{groschwe05a}, the lemma has only been stated and proved for
deterministic Turing machines, but it is obvious that the same proof also applies to
nondeterministic machines (to see this, note that, by definition,
\emph{every} run of an $(r,s,t)$-bounded Turing machine is finite).
\par
In analogy to the definition of \emph{randomized} complexity classes such as the class $\text{RP}$ of
randomized polynomial time (cf., e.g., \cite{Papadimitriou}), we consider the randomized versions $\RST(\cdots)$
and \FRST$(\cdots)$ of the
$\ST(\cdots)$ and $\NST(\cdots)$ classes.
The following definition of randomized Turing machines formalizes the intuition that
in each step, a coin can be tossed to determine which particular successor configuration is chosen in this step.
For a configuration $\gamma$ of an NTM $T$, we write $\Next_T(\gamma)$ to denote the set of all
configurations $\gamma'$ that can be reached from $\gamma$ in a single step. Each such configuration
$\gamma'\in\Next_T(\gamma)$ is chosen with uniform probability, i.e.,
$\Pr(\gamma\to_T\gamma')= 1/|\Next_T(\gamma)|$. For a run $\rho=(\rho_1,\twodots,\rho_\ell)$, the
probability $\Pr(\rho)$ that $T$ performs run $\rho$ is the product of the probabilities
$\Pr(\rho_i\to_T\rho_{i+1})$, for all $i<\ell$.
For an input word $w$, the probability that $T$ accepts $w$ (resp., that $T$ outputs $w'$) is defined as the
sum of $\Pr(\rho)$ for all accepting runs $\rho$ of $T$ on input $w$ (resp., of all runs of $T$ on $w$ that output $w'$).
We say that a decision problem $L$ is solved by
a $(\frac{1}{2},0)$-RTM if, and only if, there is an NTM $T$ such that every run of $T$ has finite length, and
the following is true for all input instances $w$:
If $w\in L$, then $\Pr(T\text{ accepts } w)\geq 1/2$;
if $w\not\in L$, then $\Pr(T\text{ accepts }w) = 0$.
Similarly, we say that a function $f:\Sigma^*\to\Sigma^*$ is computed by a \emph{LasVegas-}RTM if,
and only if, there is an NTM $T$ such that every run of $T$ on every input instance $w$ has finite length and outputs either
$f(w)$ or \emph{``I don't know''}, and
$\Pr(\text{$T$ outputs $f(w)$}) \geq 1/2$.

\begin{definition}[$\RST(\cdots)$ and $\FRST(\cdots)$]\label{def:RST}
\upshape \mbox{}\\
  Let $r,s:\NN\to\NN$ and $t\in\NN$.
 \begin{enumerate}[(a)]
  \item
    A decision problem $L$ belongs to the class $\RST(r,s,t)$, if it can be solved by a
    $(\frac{1}{2},0)$-RTM that is $(r,s,t)$-bounded.
  \item
    A function $f:\Sigma^*\to\Sigma^*$ belongs to $\FRST(r,s,t)$, if it can be solved
    by a \emph{LasVegas}-RTM that is $(r,s,t)$-bounded.
   \uend
 \end{enumerate}
\end{definition}
As a straightforward observation one obtains:
\begin{proposition}\label{prop:DetVsRandVsNdet}
For all $r,s:\NN\to\NN$ and $t\in\NN$, \\
$\ST(r,s,t)  \subseteq  \RST(r,s,t) \subseteq \NST(r,s,t)$.
\uend
\end{proposition}

For classes $R$ and $S$ of functions we let
$\textup{ST}(R,S,t) :=  \bigcup_{r\in R,s\in S}\linebreak\textup{ST}(r,s,t)$ and
$\textup{ST}(R,S,O(1)) := \bigcup_{t\in\NN}\, \textup{ST}(R,S,t)$. Analogous notations are
used for the $\NST(\cdots)$, $\RST(\cdots)$, and $\FRST(\cdots)$ classes, too.

As usual, for every (complexity) class $C$ of decision problems, $\text{co-}C$ denotes the class of all
decision problems whose \emph{complements} belong to $C$.
Note that the $\RST(\cdots)$-classes consist of decision problems that can be solved
by randomized algorithms that allow a moderate number of false negatives, but no false positives.
In contrast to this, the
$\text{co-}\RST(\cdots)$-classes consist of problems that can be solved by randomized
algorithms that
allow a moderate number of false positives, but no false negatives.

From Lemma~\ref{lem:time}, one immediately obtains for all functions $r,s$ with
$r(N)\cdot s(N)\in O(\log N)$ that
$\ST(r,s,O(1))\subseteq \PTIME$, \linebreak $\RST(r,s,O(1))\subseteq \RP$, and
$\NST(r,s,O(1))\subseteq \NP$ (where $\PTIME$, $\RP$, and $\NP$ denote the class of
problems solvable in polynomial time on deterministic, randomized, and nondeterministic
Turing machines, respectively).


\section{Lower Bounds for \\ Decision Problems}\label{section:MainResults}

Our first main result is a lower bound for the \emph{(multi)set equality problem} as well as
for the \emph{checksort problem}.
The \emph{(multi)set equality problem}
asks if two given (multi)sets of strings are the same.
The \emph{checksort problem} asks for two input lists of strings whether the
second list is the lexicographically sorted version of the first list.
We encode inputs as strings over the alphabet $\{0,1,\#\}$.
Formally, the \emph{(multi)set equality} and the \emph{checksort} problem are
defined as follows:
The input instances of each of the three problems are
\begin{description}
\item[\normalfont\textit{\ Instance:}]
  $v_1\#\cdots\# v_m\# v'_1\#\cdots\# v'_m\#$,  \\ where $m{\geq} 0$, and
  $v_i,v'_i\in\{0,1\}^*$ (for all $i\leq m$)
\end{description}
and the task is to decide the following:
\begin{description}
\item[\normalfont\textit{\ $\SETEQUALITY$ problem:}]\mbox{}\\
  Decide if $\{v_1,\ldots,v_m\} = \{v'_1,\ldots,v'_m\}$.
\item[\normalfont\textit{\ $\MULTISETEQUALITY$ problem:}]\mbox{}\\
  Decide if the multisets $\{v_1,\ldots,v_m\}$ and $\{v'_1,\ldots,v'_m\}$ are
  \linebreak equal (i.e., they
  contain the same elements with the same multiplicities).
\item[\normalfont\textit{\ $\CHECKSORT$ problem:}]\mbox{}\\
  Decide if $v_1',\ldots,v_m'$ is the lexicographically sorted (in ascending order)
  version of $v_1,\ldots,v_m$.
\end{description}

\noindent%
For an instance $v_1\#\cdots v_m\# v'_1\#\cdots v'_m\#$ of the above
problems, we usually let $N=2m+\sum_{i=1}^m(|v_i|+|v_i'|)$ denote the size of
the input. Furthermore, in our proofs we will only consider instances where
all the $v_i$ and $v_i'$ have the same length $n$, so that $N=2m\cdot(n+1)$.
\par
The present paper's technically most involved result is the following lower bound:
\begin{theorem}\label{theo:set-equality}
  Let $r,s:\mathbb N\to\mathbb N$ such that
  $r(N)  \in  o(\log N)$ and
  $s(N)  \in  o\left(\sqrt[4]N/r(N)\right)$.
  Then, none of the problems \textup{\CHECKSORT}, \linebreak
  \textup{\SETEQUALITY},  \textup{\MULTISETEQUALITY} belongs to the class \linebreak
  $\RST(r(N),s(N),O(1))$.
 \uend
\end{theorem}

Sections~\ref{section:ListMachines}--\ref{section:BoundsForTMs} are devoted to the proof of
Theorem~\ref{theo:set-equality}. The proof uses an intermediate computation model called
\emph{list machines} and proceeds by (1) showing that randomized Turing machine computations can
be simulated by randomized list machines that have the same acceptance probabilities as the
given Turing machines and (2) proving a lower bound for
$\MSEQUALITY$ and $\CHECKSORT$ on randomized list machines.
\par
By applying the reduction used in \cite[Theorem 9]{groschwe05a}, we obtain that the lower bound
of Theorem~\ref{theo:set-equality} also applies for
the ``$\SHORT$'' versions of $\MSEQUALITY$ and $\CHECKSORT$, i.e., the restrictions of these problems to
inputs of the form $v_1\#\cdots v_m\# \linebreak v'_1\#\cdots v'_m\#$, where each $v_i$ and $v'_i$ is a 0-1-string
of length at most $c\cdot\log m$, and $c$ is an arbitrary constant $\geq 2$.
By using the standard \emph{merge sort} algorithm, one easily obtains that
the ``$\SHORT$'' versions of $\MSEQUALITY$ and $\CHECKSORT$ belong to $\ST(O(\log N),O(\log N),3)$.
Moreover, in \cite[Lemma\;7]{cheyap91} it has been shown that the (general) sorting problem can
be solved by an \linebreak $\big(O(\log N),O(1),2\big)$-bounded deterministic Turing machine. As an
immediate consequence,
we obtain:
\begin{corollary}\label{cor:tight-bounds-for-short-versions}
\SETEQUALITY, \MULTISETEQUALITY, \textsc{Check-\linebreak Sort}, and their
\textsc{\upshape ``$\SHORT$''} versions,
are in $\ST(O(\log N),O(1),2)$, but not in $\RST(o(\log N),O(\sqrt[4]{N}/\log N),O(1))$.
\uend
\end{corollary}

\noindent
A detailed proof can be found in Appendix~\ref{appendix:proofsofcorollaries}.
\\
As a further result, we show that

\begin{theorem}\label{thm:efficient-randomized/ndet-solutions}
\begin{enumerate}[(a)]
  \item\label{thm:efficient-randomized/ndet-solutions:coRST}
    $\MULTISETEQUALITY$ belongs to \\  $\co\RST(2,O(\log N),1)\ \subseteq\ \co\NST(2,O(\log N),1)$.
  \item\label{thm:efficient-randomized/ndet-solutions:NST}
    Each of the problems  $\MULTISETEQUALITY$, $\CHECKSORT$, $\SETEQUALITY$ belongs to \ $\NST(3,O(\log N),2)$.
   \uend
\end{enumerate}
\end{theorem}
\begin{proof}
  \emph{(\ref{thm:efficient-randomized/ndet-solutions:coRST}):}\
  We apply fairly standard \emph{fingerprinting techniques} and show how to
  implement them on a $(2,O(\log N),1)$-bounded randomized Turing machine.
  Consider an instance $v_1\#\ldots\# v_m\#v_1'\#\ldots\#\allowbreak v_m'\#$ of the
  $\MULTISETEQUALITY$ problem.
  For simplicity, let us assume that all the $v_i$ and
  $v_j'$ have the same length $n$. Thus the input size $N$ is $2\cdot
  m\cdot(n+1)$. We view the $v_i$ and $v_i'$ as integers in
  $\{0,\ldots,2^{n}-1\}$ represented in binary.

  We use the following algorithm to decide whether the multisets
  $\{v_1,\ldots,v_m\}$ and $\{v_1',\ldots,v_m'\}$ are equal:
  \begin{enumerate}[(1)]
  \item
    During a first sequential scan of the input, determine the input parameters $n$, $m$, and $N$.
  \item
    Choose a prime $p_1\le k:=m^3\cdot n\cdot\dot\log(m^3\cdot n)$ uniformly at
    random.
  \item
    Choose an arbitrary prime $p_2$ such that $3k<p_2\le 6k$. Such a prime
    exists by Bertrand's postulate.
  \item
    Choose $x\in\{1,\ldots,p_2-1\}$ uniformly at random.
  \item
    For $1\le i\le m$, let $e_i =(v_i\bmod{p_1})$ and $e_i'=(v'_i\bmod{p_1})$.
    If
    \begin{equation}\label{eq:rand1}
    \sum_{i=1}^m x^{e_i}\ \equiv\ \,\sum_{i=1}^m x^{e_i'}\ \bmod{p_2}
    \end{equation}
    then accept, else reject.
  \end{enumerate}
  Let us first argue that the algorithm is correct (for sufficiently large
  $m,n$): Clearly, if the multisets $\{v_1,\ldots,v_m\}$ and
  $\{v_1',\ldots,v_m'\}$ are equal then the algorithm accepts.
  On the other hand, if they are distinct, the probability that the multisets
  $\{e_1,\ldots,e_m\}$ and $\{e_1',\ldots,e_m'\}$ are equal is $O(1/m)$.
  This is due to the following.

  \begin{myclaim}\label{claim:randomprime}
  Let $n,m\in\mathbb N$, $k=m^3\cdot n\cdot\dot\log(m^3\cdot n)$, and
  $0\le v_1,\twodots, v_m,\linebreak v_1',\twodots,v_m'< 2^n$.
  Then for a prime $p\le k$ chosen uniformly at
  random,
  \( \textstyle
  \Pr(\exists i,j\le m\textup{ with }v_i\neq v'_j\text{ and }v_i\equiv
  v_j'\bmod p)\le O\left(1/m\right).
  \)
  \end{myclaim}
  \begin{proof}
  We use the following well-known result (see, for example, Theorem 7.5 of
  \cite{motrag95}):
  Let $n,\ell\in\mathbb N$, $k=\ell\cdot n\cdot\dot\log(\ell\cdot n)$, and
  $0< x< 2^n$. Then for a prime $p\le k$ chosen uniformly at random,
  \[ \textstyle
    \Pr(x\equiv0\bmod p)\ \le\ O\left(\frac{1}{\ell}\right).
  \]
  The claim then follows if we apply this result with $\ell = m^3$
  simultaneously to the at most $m^2$ numbers $x = v_i - v'_j$ with $v_i \neq
  v'_j$.
  \end{proof}
  \mbox{}\\
  To proceed with the proof of Theorem~\ref{thm:efficient-randomized/ndet-solutions}(\emph{\ref{thm:efficient-randomized/ndet-solutions:coRST}}),
  suppose that the two multisets are distinct. Then the polynomial
  \[
  q(X)\ = \ \sum_{i=1}^m X^{e_i}-\sum_{i=1}^m X^{e_i'}
  \]
  is nonzero. Note that all coefficients and the degree of $q(X)$ are at most
  $k< p_2$. We view $q(X)$ as a polynomial over the field $\mathbb
  F_{p_2}$. As a nonzero polynomial of degree at most $p_1$, it has at most
  $p_1$ zeroes. Thus the probability that $q(x)=0$ for the randomly chosen
  $x\in\{1,\ldots,p_2-1\}$ is at most $p_1/(p_2-1)\le1/3$. Therefore, if the multisets
  $\{e_1,\ldots,e_m\}$ and $\{e_1',\ldots,e_m'\}$ are distinct, the algorithm
  accepts with probability at most $1/3$, and the overall acceptance
  probability is at most
  \[ \textstyle
    \frac{1}{3}+O\left(\frac{1}{m}\right)\le\frac{1}{2}
  \]
  for sufficiently large $m$. This proves the correctness of the algorithm.

  Let us now explain how to implement the algorithm on a $(2,\linebreak O(\log
  N),1)$-bounded randomized Turing machine. Note that the binary representations of the primes $p_1$ and
  $p_2$ have length $O(\log N)$. The standard arithmetical operations can be
  carried out in linear space on a Turing machine. Thus with numbers of length
  $O(\log N)$, we can carry out the necessary arithmetic on the internal memory
  tapes of our $(2,O(\log N),1)$-bounded Turing machine.

  To choose a random prime $p_1$ in step (2), we simply choose a random number
  $\le k$ and then test if it is prime, which is easy in linear space. If the
  number is not prime, we repeat the procedure, and if we do this sufficiently
  often, we can find a random prime with high probability. Steps (3) and (4)
  can easily be carried out in internal memory. To compute the number $e_i$ in step
  (5), we proceed as follows: Suppose the binary representation of $v_i$ is
  $v_{i,(n-1)}\ldots v_{i,0}$, where $v_{i,0}$ is the least significant bit.
  Observe that
  \[
    e_i \ = \ \big((\sum_{j=0}^{n-1} 2^j\cdot v_{i,j})\bmod p_1\big).
  \]
  We can evaluate this sum sequentially by taking all terms modulo $p_1$; this
  way we only have to store numbers smaller than $p_1$. This requires one
  sequential scan of $v_i$ and no head reversals.

  To evaluate the
  polynomial $\sum_{i=1}^m x^{e_i}$ modulo $p_2$, we proceed as
  follows: Let $t_i=(x^{e_i}\bmod p_1)$ and $s_i=((\sum_{j=1}^it_i)\bmod
  p_1)$. Again we can compute the sum sequentially by computing $e_i$, $t_i$,
  and $s_i= ((s_{i-1}+t_i)\bmod p_1)$ for $i=1,\ldots, m$. We can evaluate
  $\sum_{i=1}^m x^{e_i'}$ analogously and then test if
  \eqref{eq:rand1} holds.
  This completes the proof of part (\emph{\ref{thm:efficient-randomized/ndet-solutions:coRST}})
  of Theorem~\ref{thm:efficient-randomized/ndet-solutions}.

 \smallskip\par\emph{(b):}\
 Let $w$ be an input of length $N$,
 $w := v_1\# v_2\# \ldots\# v_m\# \linebreak v'_1\# v'_2\# \ldots\# v'_m\#$.
 Note that the multisets $\set{v_1,\ldots,v_m}$ and $\set{v'_1,\ldots,\linebreak
 v'_m}$
 are equal if and only if there is a permutation $\pi$ of $\set{1,\ldots,m}$
 such that for all $i \in \set{1,\ldots,m}$, $v_i = v'_{\pi(i)}$.
 The idea is to ``guess'' such a permutation $\pi$ (suitably encoded as a
 string over $\set{0,1,\#}$), to write sufficiently many copies of the string
 $u := \pi\# w$ onto the first tape, and finally solve the problem by
 comparing $v_i$ and $v'_{\pi(i)}$ bitwise, where in each step we use the next
 copy of $u$.

 A $(3,\bigO(\log N),2)$-bounded nondeterministic Turing machine $M$ can do
 this as follows.
 In a forward scan, it nondeterministically writes a sequence
 $u_1, u_2, \ldots, u_\ell$ of $\ell := m + N \cdot m$ many strings on its
 first and on its second tape, where
 \[
   u_i :=
   \pi_{i,1} \# \ldots \# \pi_{i,m} \#
   v_{i,1}   \# \ldots \# v_{i,m}   \#
   v'_{i,1}  \# \ldots \# v'_{i,m}  \#
 \]
 for binary numbers $\pi_{i,j}$ from $\set{1,\ldots,m}$, and bit strings
 $v_{i,j}$ and $v'_{i,j}$ of length at most $N$.
 While writing the first $N \cdot m$ strings, it ensures that for every
 $i \in \set{1,\ldots,N \cdot m}$, either $v_{i,\lceil i/N \rceil}$ and
 $v'_{i,\pi_{i,\lceil i/N \rceil}}$ coincide on bit $((i-1)\ \mod\ N) + 1$, or
 that both strings have no such bit at all.
 While writing the last $m$ strings, it ensures that for all
 $i \in \set{1,\ldots,m}$ and $j \in \set{i+1,\ldots,m}$,
 $\pi_{N\cdot m+i,i} \neq \pi_{N\cdot m+i,j}$.
 Finally, $M$ checks in a backward scan of both external memory tapes that
 $u_i = u_{i-1}$ for all $i \in \set{2,\ldots,\ell}$, and that $v_{1,j} = v_j$
 and $v'_{1,j} = v'_j$ for all $j \in \set{1,\ldots,m}$.

 The $\SETEQUALITY$ problem can be solved in a similar way.

 Deciding $\CHECKSORT$ is very similar:
 the machine additionally has to check that $v'_i$ is smaller than or equal to
 $v'_j$ for all $j \in \{i+1,\ldots,m\}$.
 This can be done, e.g., by writing $N \cdot \sum_{i=1}^{m-1} i$ additional copies of
 $u$, and by comparing $v'_i$ and $v'_j$ bitwise on these strings for each $i$
 and $j \in \{i+1,\ldots,m\}$.
\end{proof}

\mbox{}\par
Theorems~\ref{theo:set-equality} and \ref{thm:efficient-randomized/ndet-solutions}, in particular,
immediately lead to the following
separations between the deterministic,  randomized, and  nondeterministic $\ST(\cdots)$ classes:

\begin{corollary}\label{cor:DetVsRandVsNdet}
  Let $r,s:\mathbb N\to\mathbb N$ with
  $r(N) \in o(\log N)$ \ and \ $s(N) \in o\left(\sqrt[4]N/r(N)\right)\cap \Omega(\log N)$.
  Then,
  \begin{enumerate}[(a)]
    \item
        $\RST(O(r),O(s),O(1))\;  \neq \; \co\RST(O(r),O(s),O(1))$,
    \item
        $\ST(O(r),O(s),O(1)) \; \varsubsetneq \; \RST(O(r),O(s),O(1)) \; \varsubsetneq $\\
        \mbox{}\hfill $\NST(O(r),O(s),O(1))$.
  \end{enumerate}
\end{corollary}

\noindent
The (straightforward) proof can be found in Appendix~\ref{appendix:proofsofcorollaries}.

Let us note that the lower bound of Theorem~\ref{theo:set-equality} for the problem $\CHECKSORT$ in particular
implies the following generalization of the main result of \cite{groschwe05a} to
randomized computations:

\enlargethispage{\baselineskip}
\enlargethispage{\baselineskip}

\begin{corollary}\label{cor:sort}
The sorting problem (i.e., the problem of sorting a sequence of input strings)
does not belong to the class
\FRST$(o(\log N),O(\sqrt[4]{N}/\log N),O(1))$.
\uend
\end{corollary}

\noindent
The (straightforward) proof can be found in Appendix~\ref{appendix:proofsofcorollaries}.


\section{Lower Bounds for \\ Query Evaluation}\label{section:MainResults_QueryEval}

Our lower bound for the $\SETEQUALITY$ problem (Theorem~\ref{theo:set-equality}) leads to the following
lower bounds on the worst case data complexity of database query evaluation problems
in a streaming context:
\begin{theorem}[Tight Bound for Relational Algebra]\label{Thm:RelAlgebra}
\hspace{4cm}
\begin{enumerate}[(a)]
\item
  For every relational algebra query $Q$, the problem of evaluating $Q$ on
  a stream consisting of the tuples of the input database relations can be solved in
  $\ST(O(\log N),O(1),O(1))$.
\item
  There exists a relational algebra query $Q'$ such that the problem of evaluating $Q'$ on
  a stream of the tuples of the input database relations does not belong to the class
  \FRST$(o(\log N),O(\sqrt[4]{N}/\log N),O(1))$.
 \uend
\end{enumerate}
\end{theorem}
\begin{proof}
  \emph{(a):} \ It is straightforward to see that for every
  relational algebra query $Q$ there exists a number $c_Q$ such that
  $Q$ can be evaluated within $c_Q$
  sequential scans and sorting steps.
  Every sequential scan accounts for a constant number of head reversals and constant internal memory
  space. Each sorting step can be accomplished using the sorting method of \cite[Lemma\;7]{cheyap91} (which is a variant of the
  merge sort algorithm) with $O(\log N)$ head reversals and constant internal memory space.
  Since the number $c_Q$ of necessary sorting steps and scans is constant (i.e., only depends on the query, but not
  on the input size $N$), the query $Q$ can be evaluated by an $(O(\log N),O(1),O(1))$-bounded deterministic Turing machine.
  \smallskip\par
  \emph{(b):} \ Consider the relational algebra query
  \[
     Q' \ \ \deff \ \ (R_1 - R_2) \cup (R_2 - R_1)
  \]
  which computes the symmetric difference of two relations $R_1$ and $R_2$.
  Note that the query result is empty if, and only if, $R_1=R_2$. Therefore, any algorithm that
  evaluates $Q'$ solves, in particular, the $\SETEQUALITY$ problem. Hence, if $Q'$ could be
  evaluated in $\FRST(o(\log N),O(\sqrt[4]{N}/\log N),O(1))$, then \textsc{Set-Equal\-i\-ty} could be
  solved in $\RST(o(\log N),O(\sqrt[4]{N}/\log N),O(1))$, contradicting Theorem~\ref{theo:set-equality}.
\end{proof}
\mbox{}\smallskip

We also obtain lower bounds on the worst case data complexity of evaluating
\emph{XQuery} and \emph{XPath} queries against XML document streams:

\begin{theorem}[Lower Bound for XQuery]\label{Thm:XQuery}
  There is an \emph{XQuery} \linebreak query $Q$ such that the problem of evaluating $Q$ on an
  input XML document stream of length $N$ does not belong to the class
  \FRST$(o(\log N),O(\sqrt[4]{N}/\log N),O(1))$.
 \uend
\end{theorem}
\begin{theorem}[Lower Bound for XPath]\label{Thm:XPath}
  There is an \emph{XPath} query $Q$ such that the problem of filtering an input XML
  document stream with $Q$ (i.e., checking whether at least one node of the document matches
  the query) does not belong to the class $\co\RST(o(\log N),\linebreak O(\sqrt[4]{N}/\log N),O(1))$.
 \uend
\end{theorem}

For proving the Theorems~\ref{Thm:XQuery} and \ref{Thm:XPath}, we represent
an instance
\(
   x_1\#\cdots \# x_m\#y_1\cdots \#y_m\#
\)
of the $\SETEQUALITY$ problem
by an XML document of the form
\begin{quote}\scriptsize
 \begin{alltt}
<instance>
  <set1>
    <item> <string> \(x\sb1\) </string> </item>
      \(\cdots\)
    <item> <string> \(x\sb{m}\) </string> </item>
  </set1>
  <set2>
    <item> <string> \(y\sb1\) </string> </item>
      \(\cdots\)
    <item> <string> \(y\sb{m}\) </string> </item>
  </set2>
</instance>
\end{alltt}
\end{quote}
(For technical reasons, we enclose every string $x_i$ and $y_j$ by a \linebreak \texttt{string}-element \emph{and} an \texttt{item}-element.
For the proof of Theorem~\ref{Thm:XQuery}, one of the two would suffice, but for the proof of Theorem~\ref{Thm:XPath} it is
more convenient if each $x_i$ and $y_j$ is enclosed by two element nodes.)

It should be clear that, given as input $x_1\#\cdots \# x_m\#y_1\cdots \#y_m\#$, the above
XML document can be produced by using a constant number of sequential scans, constant internal memory space,
and two external memory tapes.
\\
\parno
\begin{proofcomment}{of Theorem~\ref{Thm:XQuery}}\mbox{}\\[1mm]
The $\SETEQUALITY$ problem can be expressed by the following XQuery query $Q:=$
{\scriptsize
\begin{alltt}
<result>
  if ( every $x in /instance/set1/item/string satisfies
         some $y in /instance/set2/item/string satisfies
           $x = $y )
     and
     ( every $y in /instance/set2/item/string satisfies
         some $x in /instance/set1/item/string satisfies
           $x = $y )
  then <true/>
  else ()
</result>
\end{alltt}
}%
\noindent%
Note that if $\set{x_1,\twodots,x_m}=\set{y_1,\twodots,y_m}$, then
$Q$ returns the document
{\footnotesize \verb!<result><true/></result>!},
and otherwise $Q$ returns the ``empty'' document
{\footnotesize \verb!<result></result>!}.
Thus, if $Q$ could be evaluated in
\FRST$(o(\log N),O(\sqrt[4]{N}/\log N),O(1))$, then \linebreak the
$\SETEQUALITY$ problem
could
be solved in $\RST(o(\log N),\linebreak O(\sqrt[4]{N}/\log N),O(1))$, contradicting
Theorem~\ref{theo:set-equality}.
\proofend
\end{proofcomment}%
\mbox{}\smallskip\\
\noindent
\begin{proofcomment}{of Theorem~\ref{Thm:XPath}}\mbox{}\\
The \textit{XPath} query $Q$
of Figure~\ref{fig:XPath-query}
\begin{figure*}\small
\begin{fminipage}
\begin{alltt}
        \textrm{descendant::}set1\textrm{\,/\,child::}item\textrm{\,[\,not \,child::}string \textrm{=}
                                  \textrm{\!ancestor::}instance\textrm{\,/\,child::}set2\textrm{\,/\,child::}item\textrm{\,/\,child::}string\textrm{\,]}
\end{alltt}
\end{fminipage}\vspace{-2ex}
{\footnotesize\caption{The \emph{XPath} query $Q$ used in the proof of Theorem~\ref{Thm:XPath}.}\label{fig:XPath-query}}
\end{figure*}
selects all \texttt{item}-nodes below \texttt{set1} whose string content does \emph{not} occur
as the string content of some \texttt{item}-node below \texttt{set2}
(recall the  ``existential'' semantics of \linebreak \textit{XPath}~\cite{XPathSemantics}).
In other words: $Q$ selects all (nodes that represent) elements in $X-Y$, for
$X\deff\set{x_1,\twodots,x_m}$ and $Y\deff\set{y_1,\twodots,y_m}$.

Now assume, \emph{for contradiction}, that the problem of filtering an input XML document stream
with the \textit{XPath} query $Q$ (i.e., checking whether at least one document node is selected by $Q$)
belongs to the class $\co\RST(o(\log N),O(\sqrt[4]{N}/\log N),O(1))$. Then, clearly, there exists an
$\big(o(\log N),O(\sqrt[4]{N}/\log N),O(1)\big)$-bounded randomized Turing machine $T$ which has the
following properties for every
input $x_1\#\cdots\#x_m\#y_1\#\cdots\#y_m\#$ (where
$X\deff\set{x_1,\twodots,x_m}$ and $Y\deff\set{y_1,\twodots,y_m}$):
\begin{enumerate}[(1)]
\item[(1)]
  If $Q$ selects at least one node (i.e., $X-Y\neq\emptyset$, i.e., $X\not\subseteq Y$), then
  $T$ accepts with probability $1$.
\item[(2)]
  If $Q$ does not select any node (i.e., $X-Y=\emptyset$, i.e., $X\subseteq Y$), then
  $T$ rejects with probability $\geq 0.5$.
\end{enumerate}
This machine $T$ can be used to solve the $\SETEQUALITY$ problem by a machine $\tilde{T}$ as follows:
First, $\tilde{T}$ starts $T$ with input $x_1\#\cdots\#x_m\#\linebreak y_1\#\cdots\#y_m\#$.
Afterwards, $\tilde{T}$ starts $T$ with input
$y_1\#\cdots\#y_m\#x_1\#\cdots\#\linebreak x_m\#$.
If both runs reject, then $\tilde{T}$ accepts its entire input.
Otherwise, $\tilde{T}$ rejects.
\\
Let us analyze the acceptance/rejectance probabilities of $\tilde{T}$:
\begin{enumerate}[(--)]
 \item[(i)]
   If $X\neq Y$, then either $X\not\subseteq Y$ or $Y\not\subseteq X$, and thus, due to (1), at least one of the two runs of $T$ has
   to accept. The machine $\tilde{T}$ will therefore reject with probability 1.
 \item[(ii)]
   If $X=Y$, then $X\subseteq Y$ and $Y\subseteq X$.
   Due to (2), we therefore know that each of the two runs of $T$ will accept with probability $\geq 0.5$ and thus, in total,
   $\tilde{T}$ will accept with probability $\geq 0.25$.
\end{enumerate}
To increase the acceptance probability to 0.5, we can start two independent runs of $\tilde{T}$ and accept if at least one
of the two runs accept. In total, this leads to a
$\big(o(\log N),O(\sqrt[4]{N}/\log N),O(1)\big)$-bounded randomized Turing machine which, on every input
$x_1\#\cdots\#x_m\#\linebreak y_1\#\cdots\#y_m\#$,
\begin{enumerate}[--]
\item
  accepts with probability $\geq 0.5$,  if $\set{x_1,\twodots,x_m}=\set{y_1,\twodots,y_m}$,
\item
  rejects with probability 1, otherwise.
\end{enumerate}
In other words: This machine shows that the $\SETEQUALITY$ problem belongs to
$\RST(o(\log N),O(\sqrt[4]{N}/\log N),O(1))$, contradicting Theorem~\ref{theo:set-equality}.
Therefore, the problem of filtering an input XML document stream
with the \textit{XPath} query $Q$ does not belong to the class $\co\RST(o(\log N),O(\sqrt[4]{N}/\log N),O(1))$.
\qed
\end{proofcomment}


\section{List Machines}\label{section:ListMachines}%

This section as well as the subsequent sections are devoted to the proof of Theorem~\ref{theo:set-equality}.
For proving Theorem~\ref{theo:set-equality}
we use \emph{list machines}.
The important
advantage that these list machines have over the original Turing machines is that
they make it fairly easy to track the ``flow of information'' during a
computation.

In \cite{groschwe05a} we introduced the notion of deterministic list machines
with output. In what follows, we propose a nondeterministic version of such
machines without output, i.e., nondeterministic list machines for solving
decision problems.
To introduce \emph{nondeterminism} to the notion of \cite{groschwe05a} requires some
care --- the straightforward approach where, instead of the transition functions used in \cite{groschwe05a},
transition \emph{relations} are allowed, will lead to a machine model that is too weak for adequately
simulating nondeterministic Turing machines. Therefore, instead of using transition \emph{relations},
the following notion of nondeterministic list machines allows explicit nondeterministic choices in transitions.

\begin{definition}[Nondeterministic List Machine]
\label{def:LM}  \upshape \mbox{}\\
  A \emph{nondeterministic list machine (NLM)} is a tuple
  \begin{eqnarray*}
    M & = & (t,m,I,C,A,a_0,\alpha,B,\Bacc)
  \end{eqnarray*}
  consisting of
  \begin{enumerate}[--]
  \item a $t\in\NN$, the \emph{number of lists}.
  \item an $m\in\NN$, the \emph{length of the input}.
  \item a finite set $I$ whose elements are called \emph{input numbers}
    (usually, $I\subseteq\NN$ or $I\subseteq\set{0,1}^*$).
  \item a finite set $C$ whose elements are called \emph{nondeterministic\linebreak choices}.
  \item a finite set $A$ whose elements are called \emph{(abstract) states}.

    We assume that $I$, $C$, and $A$ are pairwise disjoint and do not contain the two special
    symbols `$\langle$' and `$\rangle$'.
    \\
    We call $\Al :=I\cup C\cup A\cup\{\langle,\rangle\}$ \ the \emph{alphabet} of the machine.
  \item an \emph{initial state} $a_0\in A$.
  \item a \emph{transition function}
    \[
    \alpha \ \ : \ \ (A\setminus B) \times \big(\Al^*\big)^t \times C
    \ \to \
    \left( A\times \text{Movement}^t \right)
    \]
    with
    \(
    \begin{array}[t]{rl}
    \text{Movement} \ {\deff}\!\!\!\!  &
    \big\{\ \big(\textit{head-direction},\textit{move}\big) \  \bigmid
    \\
     &  \mbox{\quad}  \textit{head-direction}\in\set{-1,+1},\\
     &  \hspace{0.6cm}\mbox{\qquad}\textit{move}\in\set{\textit{true},\textit{false}} \ \big\}.
    \end{array}
    \)
  \item a set $B\subseteq A$ of \emph{final states}.
  \item a set $\Bacc\subseteq B$ of \emph{accepting states}.
  (We use $\Brej$ $\deff$ \mbox{$B\setminus \Bacc$} to denote the set of \emph{rejecting} states.)
 \uend
  \end{enumerate}
\end{definition}

\begin{figure}[htbp]
 \begin{fminipage}[\columnwidth]
  \includegraphics[scale=0.3]{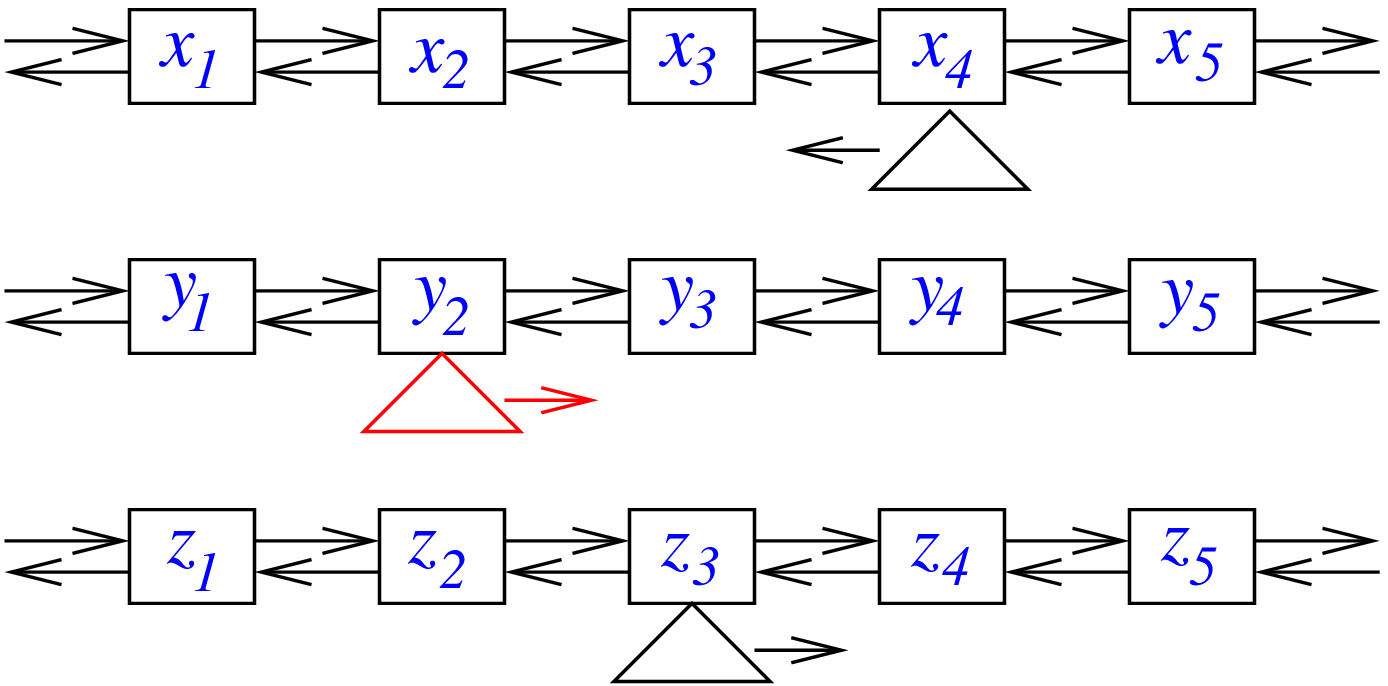}
  \mbox{}\hfill\raisebox{1.1cm}{${\text{\large$\Longrightarrow$}} \atop {\raisebox{-3mm}{\scriptsize $w\ :=\ a\langle
  x_{4}\rangle\langle y_2\rangle\langle z_3\rangle \langle c\rangle$}}$}\hfill\mbox{}
  \\[4ex]
  \mbox{}\hfill\mbox{}
  \includegraphics[scale=0.3]{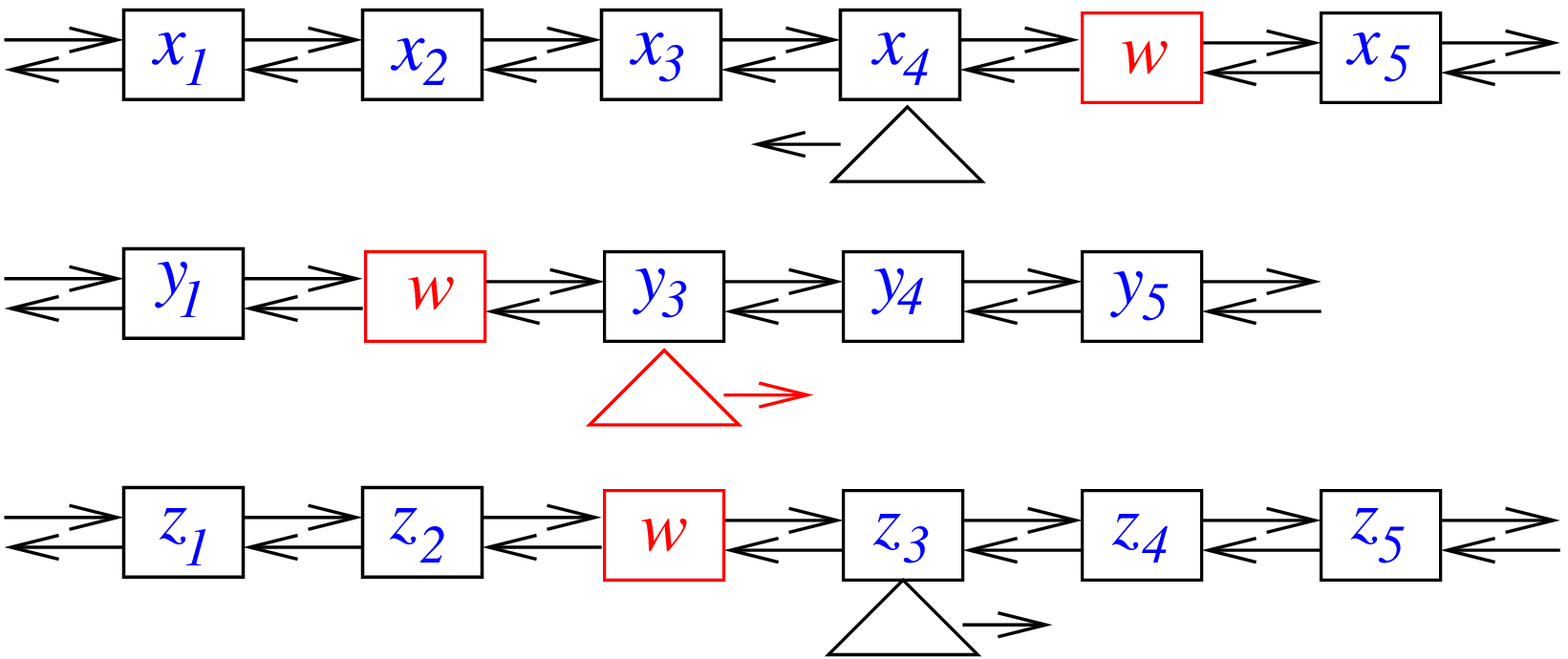}
 \end{fminipage}\vspace{-2ex}
  \caption{A transition of an NLM. The example transition is of the form
  $\big(a,x_4,y_2,z_3,c\big)\to$ $\big(b,(-1,\textit{false}),(1,\textit{true}),(1,\textit{false})\big)$.
  The new string $w$ that is written in the
  tape cells consists of the current state $a$, the content of the list
  cells read before the transition, and the nondeterministic choice $c$.
  }
  \label{fig:uebergang}
\end{figure}

Intuitively, an NLM $M=(t,m,I,C,A,a_0,\alpha,B,\Bacc)$ operates as
follows: The input is a sequence $(v_1,\ldots,v_m)\in I^m$.  Instead of tapes
(as a Turing machine), an NLM operates on $t$ lists. In particular, this means that
a new list cell can be inserted between two existing cells.
As for tapes, there is a read-write head operating on each list. Cells of the lists store strings in
$\Al^*$ (and not just symbols from $\Al$). Initially, the first list, called
the \emph{input list}, contains $(v_1,\ldots,v_m)$, and all other lists are
empty.  The heads are on the left end of the lists.
The transition function only determines the NLM's new state and the head movements, and
\emph{not what is written into the list cells}.
In each step of the
computation, the heads move according to the transition function, by choosing ``nondeterministically''
an arbitrary element in $C$.
In each computation step, the current state,
the content of all current head positions, and the nondeterministic choice $c\in C$
used in the
current transition, are written \emph{behind} each head.
When a final state is reached,
the machine stops.  If this final state belongs to $\Bacc$, the according run
is \emph{accepting}; otherwise it is \emph{rejecting}.
Figure~\ref{fig:uebergang} illustrates a transition of an NLM.
The formal definition of the semantics of nondeterministic list machines can be found in
Appendix~\ref{appendix:LMSemantics}.

An NLM is called \emph{deterministic} if $|C|=1$.
\\
For every run $\rho$ of an NLM $M$
and for each list $\tau$ of $M$,
we define $\rev(\rho,\tau)$ to be the number of
changes of the direction of the $\tau$-th list's head in run $\rho$.
We say that $M$ is \emph{$(r,t)$-bounded}, for some $r,t\in\mathbb N$, if
it has at most $t$ lists, every run
$\rho$ of $M$ is finite, and \linebreak
\ \(
    1+\sum_{\tau=1}^t\rev(\rho,\tau) \ \ \le \ r.
\)

\emph{Randomized} list machines are defined in a similar way as
randomized Turing machines:
For configurations $\gamma$ and $\gamma'$ of an NLM $M$, the probability
$\Pr(\gamma\to_M\gamma')$ that $\gamma$ yields $\gamma'$ in one step, is defined as
$|\setc{c\in C}{\text{$\gamma'$ is the $c$-successor of $\gamma$}}|/|C|$.
For a run $\rho=(\rho_1,\twodots,\rho_\ell)$, the probability $\Pr(\rho)$ that $M$ performs run $\rho$ is
the product of the probabilities $\Pr(\rho_i\to_M\rho_{i+1})$, for all $i<\ell$.
For an input $\vek{v}\in I^m$, the probability that $M$ accepts $\vek{v}$ is defined as
the sum of $\Pr(\rho)$ for all accepting runs $\rho$ of $M$ on input $\vek{v}$.

The following notation will be very convenient:

\begin{definition}[$\rho_M(\vek{v},\vek{c})$]
\upshape
Let $M$ be an NLM and
let $\ell\in\NN$ such that every run of $M$ has length  $\leq\ell$.
For every input $\vek{v}\in I^m$ and every
sequence $\vek{c}=(c_1,\twodots,c_{\ell})\in C^{\ell}$, we use
\(
   \rho_M(\vek{v},\vek{c})
\)
to denote the run $(\rho_1,\twodots,\rho_k)$  obtained by starting $M$ with
input $\vek{v}$ and by making in its $i$-th step the nondeterministic
choice $c_i$ (i.e., $\rho_{i+1}$ is the $c_i$-successor of $\rho_i$).
\uend
\end{definition}


\section{List machines can simulate \\ Turing machines}\label{section:SimulationLemma}

An important property of list machines is that they can simulate
Turing machines in the following sense:

\begin{lemma}[Simulation Lemma] \label{lemma:TM2LM}
  Let $r,s:\mathbb N\to\mathbb N$, $t\in\mathbb N$, and let
  \(
    T   =   (Q,\Sigma,\Delta,q_0,F,F_{\textit{acc}})
  \)
  be an
  $(r,s,t)$-bounded NTM with a total number of \ $t{+}u$ tapes
  and with $\set{\Blank,\#} \subseteq \Sigma$.
  Then for every $m,n\in\mathbb N$ there exists an
  $(r(m{\cdot}(n{+}1)),t)$-bounded NLM
  \ \(
   M_{m,n} =  M  =  (t,m,I,\linebreak C,A,a_0,\alpha,B,\Bacc)
  \) \
  with $I= \big(\Sigma\setminus\set{\Blank,\#}\big)^n$ and
  $|C| \leq 2^{O( \ell(m\cdot (n+1)))}$, where $\ell(N)$ is an upper bound on the length of
  $T$'s runs on input words of length $N$, \ and
  \begin{equation}\label{eq:TM2LM:NoOfStates}
     |A| \ \ \leq \ \ \displaystyle 2^{d\cdot t^2\cdot r(m\cdot(n+1)) \cdot s(m\cdot(n+1)) \; + \; 3t\cdot \log(m\cdot(n+1)) },
  \end{equation}
  for some number
  $d = d(u,|Q|,|\Sigma|)$ that does \emph{not} depend on $r$, $m$, $n$, $t$,
  such that for all $\vek{v}=(v_1,\twodots,v_m)\in I^m$ we have
  \[
    \Pr(M\text{ accepts }\vek{v}) \ \ = \ \ \Pr(T\text{ accepts } v_1\#\cdots v_m\#)\,.
  \]
  Furthermore, if $T$ is deterministic, then $M$ is deterministic, too.
  \uend
\end{lemma}

In Section~\ref{section:BoundsForTMs} we will use the simulation lemma
to transfer the lower bound results for list machines
to lower bound results for Turing machines.

In \cite{groschwe05a}, the simulation lemma has been stated and proved for
deterministic machines.
For nondeterministic machines, the construction is based on the same idea.
However, some further work is
necessary to assure that the according list machine accepts with the same probability as the
given Turing machine. Throughout the remainder of this section, the proof idea
is given;
a detailed proof of Lemma~\ref{lemma:TM2LM} can be found in Appendix~\ref{appendix:SimulationLemma}.
For proving Lemma~\ref{lemma:TM2LM},
the following straightforward characterization of probabilities for
Turing machines is very convenient.

\begin{definition}[$C_T$ and $\bs{\rho_T(w,\vek{c})}$]\label{def:rhoT}
\upshape
    Let $T$ be an NTM for which \linebreak there exists a function $\ell:\NN\to\NN$ such that every run of $T$ on a length $N$
    input word has length at most $\ell(N)$. Let
    $
       b  \deff  \max \setc{|\Next_T(\gamma)|}{\gamma\text{ is a configuration of $T$}}
    $
    be the maximum branching degree of $T$
    (note that $b$ is finite since $T$'s transition relation is finite).
    Let $b'\deff \lcm\set{1,\twodots,b}$ be the least common multiple of the numbers 1,2,\twodots,$b$, and
    let $C_T\deff \set{1,\twodots,b'}$.
    For every $N\in\NN$, every input word $w\in\Sigma^*$ of length $N$, and
    every sequence $\vek{c}=(c_1,\cdots,c_{\ell(N)}) \in (C_T)^{\ell(N)}$,
    we define
    \(
        \rho_T(w,\vek{c})
    \)
    to be the run $(\rho_1,\twodots,\rho_k)$ of $T$ that is obtained by starting $T$ with input $w$ and
    by choosing in its $i$-th computation step the $\big(c_i\ \mod\ |\Next_T(\rho_i)|\big)$-th of
    the $|\Next_T(\rho_i)|$ possible next configurations.
\uend
\end{definition}
\enlargethispage{\baselineskip}
\begin{lemma}\label{lemma:ProbAndTMs}
    Let $T$ be an NTM for which there exists a function $\ell:\NN\to\NN$ such that every run of $T$ on a length $N$
    input word has length at most $\ell(N)$, and let $C_T$ be chosen according to Definition~\ref{def:rhoT}.
    Then we have for every run $\rho$ of $T$ on an input $w$ of length $N$ that
       \ \[ \textstyle
       \begin{array}{c}
         \Pr(\rho) \ = \
         \frac{|\setc{\vek{c}\in (C_T)^{\ell(N)}\ }{\ \, \rho_T(w,\vek{c})=\rho}|}{|(C_T)^{\ell(N)}|}, \quad \text{and, in total,}
       \medskip\\
         \Pr(T\text{ accepts } w)  \ = \
         \frac{|\setc{\vek{c}\in (C_T)^{\ell(N)}\ }{\ \, \rho_T(w,\vek{c}) \text{ accepts}}|}{|(C_T)^{\ell(N)}|}\,.
       \end{array}
       \]
\end{lemma}

\noindent
The (straightforward) proof can be found in Appendix~\ref{appendix:ProofsOmittedInSec:ClassST}.
\smallskip\par
For proving  Lemma~\ref{lemma:TM2LM},
let $T$ be an NTM. We construct an NLM $M$  that
simulates $T$.
The lists of $M$ represent the external memory tapes of $T$. More precisely,
the cells of the lists of $M$ represent segments, or \emph{blocks}, of the
corresponding external memory tapes of $T$ in such a way that the content of a
block at any step of the computation can be reconstructed from the content of
the cell representing it. The blocks evolve dynamically in a way that is
described below.
$M$'s set $C$ of \emph{nondeterministic choices} is defined as
$C:= (C_T)^\ell$, where $C_T$ is chosen according to Definition~\ref{def:rhoT} and
$\ell:=\ell(m\cdot(n+1))$ is an upper bound on $T$'s running time and tape length, obtained from
Lemma~\ref{lem:time}.
Each step of the list machine corresponds to the sequence of Turing machine steps that
are performed by $T$ while none of its external memory tape heads changes its direction or
leaves its current tape block. Of course, the length $\ell'$ of this sequence of $T$'s steps is bounded by
$T$'s entire running time $\ell$. Thus, if
$c=(c_1,\twodots,c_\ell)\in C=(C_T)^\ell$ is the nondeterministic choice used in $M$'s current step,
the prefix of length $\ell'$ of $c$ tells us, which nondeterministic
choices (in the sense of Definition~\ref{def:rhoT}) $T$ makes throughout the corresponding sequence of $\ell'$ steps.
The \emph{states} of $M$ encode:
\begin{enumerate}[--]
\item The current state of the Turing machine $T$.
\item The content and the head positions of the internal memory tapes
  $t+1,\ldots,t+u$ of $T$.
\item The head positions of the external memory tapes $1,\ldots,t$.
\item For each of the external memory tapes $1,\ldots,t$, the boundaries
  of the block in which the head currently is.
\end{enumerate}
Representing $T$'s current state and the content and head positions of the $u$ internal memory tapes
  requires $|Q|\cdot 2^{O(s(m\cdot(n+1)))}\cdot s(m\cdot(n+1))^u$ states.
The $t$ head positions of the external memory tapes increase the number of states by a factor of $\ell^t$.
  The $2t$ block boundaries
  increase the number of states by another factor of $\ell^{2t}$.
So overall, the number of states is bounded by
\(
 |Q|\cdot 2^{O(s(m\cdot(n+1)))}\cdot s(m\cdot(n+1))^u\cdot  \ell^{3t}.
\)
By
Lemma~\ref{lem:time}, this yields the bound \eqref{eq:TM2LM:NoOfStates}.

Initially, for an input word $v_1\#\cdots v_m\#$, the first Turing machine tape is split into $m$ blocks
which contain the input
segments $v_i\#$ (for $1\le i< m$), respectively, $v_m\#\Blank^{\ell-(n+1)}$ (that is,
the $m$-th input segment is padded by as many blank symbols as the Turing machine may enter throughout its computation).
All other tapes just
consist of one block which contains the blank string $\square^\ell$. The heads
in the initial configuration of $M$ are on the first cells of their lists. Now
we start the simulation: For a particular nondeterministic choice
$c_1 = (c_{11},c_{12},\twodots,c_{1\ell})\in C = (C_T)^\ell$, we start $T$'s run
$\rho_T(v_1\#\cdots,c_{11}c_{12}c_{13}\cdots)$.
As long as no head of the external memory tapes
of $T$ changes its direction or crosses the boundaries of its
current block, $M$ does not do anything.
If a head on a tape $i_0\in\{1,\ldots,t\}$ crosses the
boundaries of its block, the head $i_0$ of $M$ moves to the next cell, and the
previous cell is overwritten with sufficient information so that if it is
visited again later, the content of the corresponding block of tape $i_0$ of
$T$ can be reconstructed. The blocks on all other tapes are split behind the
current head position (``behind'' is defined relative to the current direction
in which the head moves). A new cell is inserted into the lists behind the
head, this cell represents the newly created tape block that is behind the
head. The newly created block starting with the current head position is
represented by the (old) cell on which the head still stands. The case that a
head on a tape $i_0\in\{1,\ldots,t\}$ changes its direction is treated
similarly.

The simulation stops as soon as $T$ has reached a final state; and $M$ accepts if, and
only if, $T$ does.
A close look at the possible runs of $T$ and $M$ shows that $M$ has the same
acceptance probabilities as $T$.


\section{Lower Bounds for List Machines}\label{section:LowerBoundsForLM}

This section's main result is that it provides constraints on a list machine's
parameters, which ensure that list machines which comply to these constraints
can neither
solve the \emph{multiset equality problem} nor the
\emph{checksort problem}.
In fact, we can show a slightly stronger result, the precise formulation of which requires
the following observation.

\begin{definition}[sortedness]\label{def:sortedness}
\upshape
Let $m\in\NN$ and let $\pi$ be a permutation of $\set{1,\twodots,m}$.
We define
\ \(
   \textit{sortedness}(\pi)
\) \
to be the length of the longest subsequence\footnote{A sequence
  $(s_1,\twodots,s_\lambda)$ is a \emph{subsequence} of a
  sequence $(s'_1,\twodots,s'_{\lambda'})$, if there exist indices $j_1<\cdots
  < j_{\lambda}$ such that $s_1= s'_{j_1}$, $s_2= s'_{j_2}$, \ldots,
  $s_\lambda = s'_{j_{\lambda}}$.
}
of $\big(\pi(1),\twodots,\pi(m)\big)$ that
is sorted in either ascending or descending order (i.e., that is a subsequence of
$(1,\twodots,m)$ or of $(m,\twodots,1)$).\uend
\end{definition}

\begin{remark}\label{rem:permut}
It is well-known that for every permutation $\pi$ of \linebreak $\set{1,\twodots,m}$,
\(
   \textit{sortedness}(\pi) \ \in \ \Omega(\sqrt{m}),
\)
and that there exists a particular permutation $\varphi:= \varphi_m$ with
\(
   \textit{sortedness}(\varphi) \ \leq \ 2\sqrt{m}-1\,.
\)
In fact, one way of finding such a permutation is
to let $\big(\varphi(1),\twodots,\varphi(m)\big)$ be the numbers $1,\twodots,m$, sorted
lexicographically by their reverse binary representation.
\uend
\end{remark}

\begin{lemma}[Lower Bound for List Machines]\label{lemma:MainLM}
\mbox{}\\
  Let $k,m,n,r,t\in\mathbb N$ such that
  $m$ is a power of\/ $2$ and
  \ $t \geq  2$, \
  $m \geq  24\cdot (t{+}1)^{4r}+1$, \
  $k \geq  2m+3$, \
  $n\ \geq \ 1+ \ (m^2+1)\cdot\log (2 k)$.
  We let $I\deff\set{0,1}^n$, identify $I$ with the set $\set{0,1,\twodots,2^n{-}1}$, and
  divide it into $m$ consecutive intervals $I_1,\twodots,I_m$, each of length ${2^n}/m$.
  Let $\varphi$ be a permutation of $\set{1,\twodots,m}$ with
  $\textit{sortedness}(\varphi)\leq 2\sqrt{m}-1$, and
  let
  \(
   \mathcal{I} \ \deff \ I_{\varphi(1)} \times \cdots \times I_{\varphi(m)} \times
       \ I_{1} \times \cdots \times I_{m}.
  \)
  \\
  Then there is no $(r,t)$-bounded NLM $M= (t,2m,I,C,A,a_0,\alpha,B,\linebreak \Bacc)$
  with $|A|\leq k$ and $I=\set{0,1}^n$,
  such that for all $\vek{v}=(v_1,\twodots,v_m,\linebreak v'_1,\twodots,v'_m)\in\mathcal{I}$ we have:
  \\
  If $(v_1,\twodots,v_m) \ = \ (v'_{\varphi(1)},\twodots,v'_{\varphi(m)})$, then
  $\Pr(M\text{ accepts }\vek{v})\geq \frac{1}{2}$; \linebreak otherwise \ $\Pr(M\text{ accepts }\vek{v})=0$.
  \uend
\end{lemma}

It is straightforward to see that the above lemma, in particular, implies that
neither the \emph{(multi)set equality problem} nor the \emph{checksort problem} can
be solved by list machines with the according parameters.

The proof of Lemma~\ref{lemma:MainLM} is based on the following ideas (the
detailed proof is given in Appendix~\ref{appendix:ProofBoundLM}):
\begin{enumerate}[1.]
\item
  Suppose for contradiction that $M$ is an NLM that meets the lemma's requirements.
\item\label{sketch:vekc}
  Observe that there exists an upper bound $\ell$ on the length of
  $M$'s runs (Lemma~\ref{lemma:shapeofruns}\,(\ref{item:LM:lengthofruns}) in Appendix~\ref{appendix:ProofBoundLM})
  and a particular sequence $\vek{c}=(c_1,\twodots,c_\ell)\in C^\ell$ of
  nondeterministic choices (Lemma~\ref{lemma:Choose_c}),
  such that for at least half of the inputs
  $\vek{v}\deff (v_1,\twodots,v_m,v'_1,\twodots, v'_m)\in\mathcal{I}$ with
  $(v_1,\twodots,v_m)=(v'_{\varphi(1)},\twodots,v'_{\varphi(m)})$, the particular
  run $\rho_M(\vek{v},\vek{c})$ accepts.
  \\
  We let
  $\mathcal{I}_{\textit{acc},\vek{c}}\deff \setc{\vek{v}\in\mathcal{I}}{\rho_M(\vek{v},\vek{c})\text{ accepts}}$
  and, from now on, we only consider runs that are generated by the fixed sequence
  $\vek{c}$ of nondeterministic choices.
\item\label{sketch:i0}
  Show that, throughout its computation, $M$ can ``mix'' the relative order of its input values
  only to a rather limited extent (cf., Lemma~\ref{lemma:Merge}).
  This can be
  used to show that for every run of $M$ on every input
  $(v_1,\twodots,v_m,v'_1,\twodots,v'_m)\in\mathcal{I}$ there must be an
  index $i_0$ such that $v_{i_0}$ and $v'_{\varphi(i_0)}$ are never compared throughout this run.
\item\label{sketch:skeleton}
  Use the notion of the \emph{skeleton} of a run (cf., Definition~\ref{def:skeleton}),
  which, roughly speaking, is obtained from
  a run by
  replacing every input value $v_i$ with its index $i$ and by replacing every nondeterministic
  choice $c\in C$ with the wildcard symbol ``?''. In particular, the skeleton contains
  input \emph{positions} rather than concrete input \emph{values}; but given
  the skeleton together with the concrete input values and the sequence of nondeterministic choices,
  the entire run of $M$ can be reconstructed.
\item\label{sketch:zeta}
  Now choose $\zeta$ to be a skeleton that is generated by the run
  $\rho_M(\vek{v},\vek{c})$ for as many input instances $\vek{v}\in\mathcal{I}_{\textit{acc},\vek{c}}$
  as possible, and use
  $\mathcal{I}_{\textit{acc},\vek{c},\zeta}$ to denote the set of all those input instances.
\item
  Due to \ref{sketch:i0}. there must be an index $i_0$ such that for all
  inputs from $\mathcal{I}_{\textit{acc},\vek{c},\zeta}$, the values $v_{i_0}$ and
  $v'_{\varphi(i_0)}$ (i.e., the values from the input positions $i_0$ and $m+\varphi(i_0)$)
  are never compared throughout the run that
  has skeleton $\zeta$.
  To simplify notation let us henceforth assume without loss of generality that $i_0=1$.
\item
  Now fix $(v_2,\twodots,v_m)$ such that the number of $v_1$ with
  \\
  \(
    V(v_1) \ \ \deff \ \
    (v_1,v_2,\twodots,v_m,v_{\varphi^{-1}(1)},\twodots,v_{\varphi^{-1}(m)})
    \ \ \in \ \ \mathcal{I}_{\textit{acc},\vek{c},\zeta}
  \)\\
  is as large as possible.
\item
  Argue that, for our fixed $(v_2,\twodots,v_m)$, there must be at least two
  distinct $v_1$ and $w_1$ such that $V(v_1)\in \mathcal{I}_{\textit{acc},\vek{c},\zeta}$
  and \linebreak $V(w_1) \in \mathcal{I}_{\textit{acc},\vek{c},\zeta}$.
  This is achieved by observing that the number of skeletons depends on the machine's
  parameters
  $t$, $r$, $m$, $k$, but \emph{not} on $n$ (Lemma~\ref{lemma:NoOfRCPSkels})
  and by
  using the lemma's assumption on the machine's parameters
  $t,r,m,k,n$.
\item
  Now we know that the input values of $V(v_1)$ and $V(w_1)$ coincide on all
  input positions except
  $1$ and $m{+}\varphi(1)$.
  From \ref{sketch:i0}.\ we know that the values from the positions
  $1$ and $m{+}\varphi(1)$ are never compared throughout $M$'s (\emph{accepting})
  runs $\rho_M\big(V(v_1),\vek{c}\big)$ and  $\rho_M\big(V(w_1),\vek{c}\big)$.
  From this we obtain (cf., Lemma~\ref{lemma:CompositionLemma})
  an accepting run $\rho_M(\vek{u},\vek{c})$ of $M$
  on input
  \[
    \begin{array}[t]{rcl}
      \vek{u} &  \deff  &
      (u_1,\twodots,u_m,u'_1,\twodots,u'_m)
     \\
       &  \deff  &
       \big(v_1,v_2,\twodots,v_m,w_{\varphi^{-1}(1)},w_{\varphi^{-1}(2)},
          \twodots,w_{\varphi^{-1}(m)}\big).
    \end{array}
  \]
  In particular, this implies that $\Pr(M\text{ accepts }\vek{u})>0$.
  However, for this particular input $\vek{u}$ we know that $u_1 = v_1 \neq w_1 = u_{\varphi(1)}$,
  and therefore, $(u_1,\twodots,u_m)\neq (u'_{\varphi(1)},\twodots,u'_{\varphi(m)})$.
  This gives us a contradiction to the assumption
  that $\Pr(M\text{ accepts }\vek{v})=0$  for all inputs $\vek{v}=(v_1,\twodots,v_m,v'_1,\twodots,v'_m)$ with
  $(v_1,\twodots,v_m)\neq (v'_{\varphi(1)},\twodots,\linebreak v'_{\varphi(m)})$.
\end{enumerate}


\section{Lower Bounds for \\ Turing Machines}\label{section:BoundsForTMs}

\begin{lemma}\label{lemma:TMBound}
Let $r,s:\NN\to\NN$ such that
$r(N) = o(\log N)$ and $s(N) = o\big(\sqrt[4]{N}/r(N)\big)$.
Then, there is no $(r,s,O(1))$-bounded $(\frac{1}{2},0)$-RTM that
solves the following problem \textrm{\CHECK-$\varphi$}.
\uend
\end{lemma}
\begin{problem}{}{}{\CHECK-$\varphi$}
  \instance $v_1\#\ldots v_m\# v'_1\#\ldots v'_m\#$,
  \\
  where $m\geq 0$ is a power of $2$, and
  \\
  $(v_1,\ldots,v_m,v_1',\ldots,v_m')\in I_{\varphi(1)}\times
   \cdots\times I_{\varphi(m)}\times
   I_1\times\cdots\times I_m$, where $\varphi:=\varphi_m$ is the permutation of
   $\set{1,\twodots,m}$
   obtained from Remark~\ref{rem:permut}, and the sets
   $I_1,\twodots,I_m\subseteq\set{0,1}^{m^3}$ are obtained as the partition of
   the set $\set{0,1}^{m^3}$ into $m$ consecutive subsets, each of size ${2^{(m^3)}}/{m}$.
\problem Decide if $(v_1,\ldots,v_m)=(v'_{\varphi(1)},\ldots,v'_{\varphi(m)})$.
\end{problem}
\begin{proof}
Suppose for contradiction that there is a $t\in\NN$ such that
the problem $\CHECK$-$\varphi$ is solved by an
$(r,s,t)$-bounded $(\frac{1}{2},0)$-RTM $T$.
Without loss of generality we may assume that $t\ge 2$.

Let $d$ be the constant introduced in Lemma~\ref{lemma:TM2LM} (the
simulation lemma). Let $m$ be a sufficiently large power of $2$ such that
\begin{align}
\label{eq:m1}
m & \ \ge \ 24\cdot (t{+}1)^{4\cdot r(2m\cdot(m^3+1))} + 1
  \quad\text{and}\\
\label{eq:m2}
m^3& \ \ge \
\begin{array}[t]{l}
  1 + \ d\cdot t^2\cdot r(2m\cdot(m^3{+}1))\cdot s(2m\cdot(m^3{+}1))  \\
  + \ 3t\cdot \log (2m\cdot(m^3{+}1)).
\end{array}
\end{align}
Such an $m$ exists because $r(N)=o(\log N)$ (for Equation \eqref{eq:m1}) and $r(N)\cdot
s(N)=o\left(\sqrt[4]N\right)$ (for Equation \eqref{eq:m2}). Let
$n:=m^3$ and $I:=\{0,1\}^n$. By Lemma~\ref{lemma:TM2LM}, there is an
$\big(r(2m{\cdot}(n{+}1)),t\big)$-bounded
NLM $M=(t,2m,I,C,A,a_0,\alpha,B,\Bacc)$ with
\[
k \ \ \leq \ \ 2^{d\cdot t^2\cdot r(2m\cdot(n+1)) \cdot s(2m\cdot(n+1)) \; + \; 3t\cdot
  \log(2m\cdot(n+1)) }
\]
states that simulates $T$ on inputs from $I^{2m}$.
In particular, for all instances
\[
  \vek{v} \ = \ (v_1,\twodots,v_m,v'_1,\twodots,v'_m)\ \in \
  I_{\varphi(1)}\times \cdots\times I_{\varphi(m)}\times I_1\times\cdots\times I_m
\]
the following is true:
\begin{center}
$(*)$: \
\parbox[t]{7.6cm}{If $(v_1,\twodots,v_m) \ = \ (v'_{\varphi(1)},\twodots,v'_{\varphi(m)})$, then
$\Pr(M\text{ accepts }\vek{v})\geq \frac{1}{2}$; \\ otherwise $\Pr(M\text{ accepts }\vek{v})=0$.}
\end{center}
We clearly have $k\geq 2 m+3$. By \eqref{eq:m1}, we have
\[
m \ \ \ge \ \  24\cdot (t{+}1)^{4\cdot r(2m\cdot(m^3+1))} +1
  \ \ =   \ \  24\cdot (t{+}1)^{4\cdot r(2m\cdot(n+1))} +1.
\]
By \eqref{eq:m2}, we have
\begin{align*}
n \ = \ \; & \begin{array}[t]{l} m^3 \end{array} \\
\ge \ \;&
    \begin{array}[t]{l}
      1+ (m^2{+}1)\cdot
      \big(
      1 + \ d\cdot t^2\cdot r(2m\cdot(m^3{+}1))\cdot s(2m\cdot(m^3{+}1))  \\
+ \ 3t\cdot\log (2m\cdot(m^3{+}1))
 \big)
 \end{array}
\\
 = \ \;&
     \begin{array}[t]{l}
       1 + (m^2{+}1)\cdot
       \big(
       1 + \ d\cdot t^2\cdot r(2m\cdot(n{+}1))\cdot s(2m\cdot(n{+}1)) \\
       + \ 3t\cdot \log (2m\cdot(n{+}1))
       \big)
     \end{array}
\\
 \geq \ \;&
1+ (m^2{+}1)\cdot
  \big( \log(2) + \log(k)\big)
\\
 =\ \;&
1 + (m^2{+}1)\cdot \log(2 k).
\end{align*}
Thus, $(*)$ is a contradiction to Lemma~\ref{lemma:MainLM}, and
the proof of \linebreak Lemma~\ref{lemma:TMBound} is complete.
\end{proof}
\medskip\par\noindent
%
\begin{proofcomment}{of Theorem~\ref{theo:set-equality}}
For inputs that are instances of the problem \CHECK-$\varphi$, the problems
{\SETEQUALITY},
{\MULTISETEQUALITY}, {\CHECKSORT}, and
\CHECK-$\varphi$ coincide. Thus, Lemma~\ref{lemma:TMBound} immediately implies
Theorem~\ref{theo:set-equality}.
\proofend
\end{proofcomment}


\section{Conclusion}\label{section:Conclusion}%

We have proved tight lower bounds for the natural decision problems
\emph{(multi)set equality} and \emph{checksort}, in our
Turing machine based computation model for processing large data sets.
These lower bounds do not only hold for deterministic, but even for
\emph{randomized} algorithms with one-sided
bounded error probability.
Our results are obtained by carefully analyzing the flow
of information in a Turing machine computation.

As applications of these lower bound results, we obtained lower bounds
on the worst case data complexity of query evaluation for
the languages \emph{XQuery}, \emph{XPath}, and  \emph{relational
algebra} on data \linebreak streams.

We complement our lower bounds for \emph{checksort} and \emph{(multi)set equality}
by proving that these
problems can be solved by nondeterministic machines and by randomized
machines with complementary one-sided error probabilities.
As a consequence,
we obtain a separation between the deterministic, the randomized, the
co-randomized, and the nondeterministic external memory complexity classes.

A specific problem for which we could not prove lower bounds, even though it
looks very similar to the set equality problem, is the
\emph{disjoint sets problem}, which
asks if two given sets of strings are disjoint.
Another important future task is to develop techniques for proving lower bounds
(a) for randomized computations with two-sided bounded error and
(b) for appropriate problems in a setting where $\Omega(\log N)$ head reversals (i.e.,
sequential scans of external memory devices) are available.
%
%

%
\clearpage
{
\addtolength{\textwidth}{-1.5in}
\addtolength{\oddsidemargin}{0.75in}
\addtolength{\evensidemargin}{0.75in}
\onecolumn

\appendix

\medskip


\section{Turing machine basics}\label{appendix:ProofsOmittedInSec:ClassST}

\begin{definition}[Notation concerning Turing machines]\label{def:TuringMachine}
\upshape  \mbox{}\\
Let $T=(Q,\Sigma,\Delta,q_0,F,F_\textit{acc})$ be a nondeterministic Turing machine (NTM, for short) with
$t{+}u$ tapes,
where $Q$ is the state space, $\Sigma$ the
alphabet, $q_0\in Q$ the start state, $F\subseteq Q$ the set of final states,
$F_\textit{acc}\subseteq F$ the set of \emph{accepting} states,
and
\[
\Delta\ \ \subseteq\ \ (Q\setminus F)\times\Sigma^{t+u}\times Q\times
\Sigma^{t+u}\times\{L,N,R\}^{t+u}
\]
the transition relation.
Here $L,N,R$ are special symbols
indicating the head movements.

We assume
that all tapes are one-sided infinite and have cells numbered $1,2,3,$ etc, and
that $\Blank\in\Sigma$ is the ``blank'' symbol which, at the
beginning of the TM's computation, is the inscription of all empty tape cells.

A \emph{configuration} of $T$ is a tuple
\[
  (q,p_1,\ldots,p_{t+u},w_1,\ldots,w_{t+u}) \ \in \ Q\times\mathbb
  N^{t+u}\times(\Sigma^*)^{t+u},
\]
where $q$ is the current state, $p_1,\ldots,p_{t+u}$ are the positions of the
heads on the tapes, and
$w_1,\ldots,w_{t+u}$ are the contents of the tapes.
For a configuration $\gamma$ we write
\(
    \Next_T(\gamma)
\)
for the set of all configurations $\gamma'$ that
can be reached from $\gamma$ in a single computation step.

A configuration is called \emph{final} (resp., \emph{accepting}) if its current state
$q$ is final (resp., accepting), that
is, $q\in F$ (resp., $q\in F_{\textit{acc}}$).
Note that a {final} configuration does not have a
successor configuration.

A \emph{run} of $T$ is a sequence $\rho=(\rho_j)_{j\in J}$ of configurations $\rho_j$
satisfying the obvious requirements. We are only interested in finite runs
here, where the index set $J$ is $\{1,\ldots,\ell\}$ for an $\ell\in\mathbb
N$, and where $\rho_\ell$ is \emph{final}.

When considering \emph{decision problems}, a run $\rho$ is called \emph{accepting}
(resp., \emph{rejecting}) if its final configuration is accepting (resp., rejecting).
When considering, instead, Turing machines that produce an output,
we say that a run $\rho$ \emph{outputs the word $w'$} if $\rho$ end in an \emph{accepting} state
and $w'$ is the inscription of the last (i.e., $t$-th) external memory tape. If $\rho$ ends in
a \emph{rejecting} state, we say that $\rho$ \emph{outputs ``I don't know''}.

Without loss of generality we assume that our Turing machines are
\emph{normalized} in such a way that in each step at most one of its heads
moves to the left or to the right.  \uend
\end{definition}%

\mbox{}
\parno
\begin{proofcomment}{of Lemma~\ref{lemma:ProbAndTMs}}\mbox{}\\
To prove (a), observe that for every run $\rho = (\rho_1,\twodots,\rho_k)$ of $T$ on $w$ we have $k\leq \ell(N)$, and
\[
 \begin{array}{c}
   {
      \frac{|\setc{\vek{c}\in C_T^{\ell(N)}}{\rho_T(w,\vek{c})=\rho}|}{|C_T^{\ell(N)}|} }
   \ = \
   \frac{1}{|C_T^{\ell(N)}|}\cdot
      \left(
        {\displaystyle \prod_{i=1}^{k-1}} \frac{|C_T|}{|\Next_T(\rho_i)|}
      \right)
      \cdot
         |C_T|^{\ell(N)-(k-1)}
   \ = \
    {\displaystyle \prod_{i=1}^{k-1}} \frac{1}{|\Next_T(\rho_i)|}
   \ = \
    \Pr(\rho)\ .
 \end{array}
\]
(b) follows directly from (a), since
\[
  \begin{array}[b]{ccccc}\displaystyle
    \Pr(T \text{ accepts }w) & \displaystyle
    \ \stackrel{\text{def}}{=} \ & \displaystyle
    \sum_{\rho \ : \ \text{$\rho$ is an accepting} \atop \text{run of $T$ on $w$}} \Pr(\rho) &
    \ \stackrel{\text{(a)}}{=} \ & \displaystyle
    \sum_{\rho \ : \ \text{$\rho$ is an accepting} \atop \text{run of $T$ on $w$}}
         \frac{|\setc{\vek{c}\in C_T^{\ell(N)}}{\rho_T(w,\vek{c})=\rho}|}{|C_T^{\ell(N)}|}
   \\[1cm] \displaystyle
     & \displaystyle
    \ = \ & \displaystyle
    \sum_{\vek{c}\in C_T^{\ell(N)} \ : \atop \text{$\rho_T(w,\vek{c})$ accepts}}
         \frac{1}{|C_T^{\ell(N)}|} & \displaystyle
    \ = \ & \displaystyle
    \frac{|\setc{\vek{c}\in C_T^{\ell(N)}}{\rho_T(w,\vek{c}) \text{ accepts}}|}{|C_T^{\ell(N)}|}\ .
  \end{array}
  \qedeq
\]
\end{proofcomment}


\section{Formal Definition of the Semantics of Nondeterministic List Machines}\label{appendix:LMSemantics}

\smallskip\noindent
Formally, the semantics of nondeterministic list machines are defined as follows:
\begin{definition}[Semantics of NLMs]\label{def:NLM-semantics}\mbox{}\\
\upshape
  \begin{enumerate}[(a)]
  \item
A \emph{configuration} of an NLM
\ $M=(t,m,I,C,A,a_0,\alpha,B,\Bacc)$ \
is
a tuple
\ \(
(a,\vec p,\vec d,X)
\) \
with
\[
\vec p=
  \begin{pmatrix}
    p_1\\\vdots\\p_t
  \end{pmatrix}\ \in\ \NN^t\ ,
\quad
\vec d=  \begin{pmatrix}
    d_1\\\vdots\\d_t
  \end{pmatrix}\ \in\ \{-1,+1\}^t\ ,
\quad
X=\begin{pmatrix}
    \vec x_1\\\vdots\\\vec x_t
  \end{pmatrix}\ \in \ \big((\Al^*)^*\big)^t\ ,
\]
where $\NN=\set{1,2,3,\twodots}$ is the set of positive integers,
\begin{enumerate}[--]
\item $a\in A$ is the \emph{current state},
\item $\vec p
$ is the tuple of \emph{head
  positions},
\item $\vec d
$ is the tuple of \emph{head directions},
\item
  $\vec x_i=(x_{i,1},\ldots,x_{i,m_i})\in(\Al^*)^{m_i}$
    for some $m_i\ge 1$, contains the \emph{content of the cells}.
  (The string $x_{i,j}\in \Al^*$ is the content of the $j$th cell of the $i$th list.)
\end{enumerate}
\item
The
\emph{initial configuration} for input $(v_1,\ldots,v_m)\in I^m$ is a tuple
$(a,\vec p,\vec d,X)$, where $a=a_0$, $\vec p=(1,\ldots,1)^\top$, $\vec d=(+1,\ldots,+1)^\top$,
and $X=(\vec x_1,\ldots,\vec x_t)^\top$ with
\[
\vec x_1 \ \ = \ \ \big(\,\langle v_1 \rangle , \ldots, \langle v_m \rangle\,\big) \ \ \in \ \ (\Al^*)^m
\]
and $\vec x_2=\cdots=\vec x_t= \big(\,\langle \rangle\,\big) \in (\Al^*)^1$.

\item\label{item:NLM-semantics:successor}
For a nondeterministic choice $c\in C$, the
\emph{$c$-successor} of a configuration $(a,\vec p,\vec d, X)$ is the configuration
$(a',\vec p',\vec d',X')$ defined as follows:
Suppose that
\[
\alpha\,\big(\,a,\, x_{1,p_1},\twodots,x_{t,p_t},\, c\,\big) \ = \
  \big(\,b,\,e_1,\ldots,e_t\,\big) \,.
\]
We let $a'=b$.
For $1\le i\le t$, let
$m_i$ be the length of the list $\vec x_i$, and let
\begin{align*}
  e'_i & \ := \ (\textit{head-direction}_i,\textit{move}_i)
 \ := \
\begin{cases}
  (-1,\textit{false})&\text{if }p_i=1\text{ and }e_i= (-1,\textit{true}),\\
  (+1,\textit{false})&\text{if }p_i=m_i\text{ and }e_i=(+1,\textit{true}),\\
  e_i&\text{otherwise}.
\end{cases}
\end{align*}
This will prevent the machine from ``falling off'' the left or right end of a list. I.e., if
the head is standing on the rightmost (resp., leftmost) list cell, it will stay
there instead of moving a further step to the right (resp., to the left).

We fix $f_i\in\set{0,1}$ such that $f_i= 1$ iff
$\big(\textit{move}_i= \textit{true}$ or $\textit{head-direction}_i\neq d_i\big)$.

If $f_i=0$ for \emph{all} $i\in\set{1,\twodots,t}$, then we let $\vec p':=\vec p$,
$\vec d':=\vec d$, and $X' \deff X$ (i.e., if none of the machine's head moves, then
the \emph{state} is the only thing that may change in
the machine's current step).
\\
So suppose that there is at least one $i$ such that
$f_i\neq 0$.
In this case, we let
\begin{myeqnarray*}
y & \ \deff \ & a \; \langle x_{1,p_1} \rangle \cdots \langle x_{t,p_t} \rangle \; \langle c \rangle\,.
\end{myeqnarray*}
For all $i\in\set{1,\twodots,t}$, we let
\[
\vec x_i' \ := \
\begin{cases}
  \big(\,x_{i,1},\,\ldots,\, x_{i,p_{i}-1},\,y,\,x_{i,p_i+1},\,\ldots,\, x_{i,m_i}\,\big)
  & \text{if }\textit{move}_i=\textit{true},\\
  \big(\,x_{i,1},\,\ldots,\, x_{i,p_i-1}, \,y,\, x_{i,p_i},\,x_{i,p_i+1},\,\ldots,\, x_{i,m_i}\,\big)
  & \text{if }d_i=+1 \text{ and }\textit{move}_i=\textit{false},\\
  \big(\,x_{i,1},\,\ldots,\,x_{i,p_i-1},\, x_{i,p_i}, \,y,\, x_{i,p_i+1},\,\ldots,\, x_{i,m_i}\,\big)
  & \text{if }d_i=-1 \text{ and }\textit{move}_i=\textit{false},\\
\end{cases}
\]
and, finally,
\[
p_i' \ := \
\begin{cases}
  p_i+1&\text{if }e'_i=(+1,\textit{true}),\\
  p_i-1&\text{if }e'_i=(-1,\textit{true}),\\
  p_i+1&\text{if }e'_i=(+1,\textit{false}),\\
  p_i&\text{if }e'_i=(-1,\textit{false}).
\end{cases}
\]
\item
A configuration $(a,\vec p,\vec d,X)$ is \emph{final} (\emph{accepting}, resp., \emph{rejecting}),
if $a\in B$ ($\Bacc$, resp., $\Brej:= B\setminus\Bacc$).
\\
A (finite)
\emph{run} of the machine is a sequence $(\rho_1,\ldots,\rho_\ell)$ of
configurations, where $\rho_1$ is the initial configuration for some input,
$\rho_\ell$ is final, and for every $i < \ell$ there is a nondeterministic choice $c_i\in C$
such that $\rho_{i+1}$ is the $c_i$-successor of
$\rho_i$.
\\
A run is called \emph{accepting} (resp., \emph{rejecting}) if its final configuration is
accepting (resp., rejecting).
\item
 An input $(v_1,\twodots,v_m)\in I^m$ is \emph{accepted} by machine $M$ if there
 is at least one accepting run of $M$ on input $(v_1,\twodots,v_m)$.
\uend
  \end{enumerate}
\end{definition}
\mbox{}\\
\parno
It is straightforward to see that

\begin{lemma}\label{lemma:probLM}
Let $M=(t,m,I,C,A,a_0,\alpha,B,\Bacc)$ be an NLM, and let $\ell$ be an upper bound
on the length of $M$'s runs.
\begin{enumerate}[(a)]
\item
 For every run $\rho$of $M$ on an input $\vek{v}\in I^m$, we have
 \[
    \Pr(\rho) \ = \
    \frac{|\setc{\vek{c}\in C^\ell}{\rho_M(\vek{v},\vek{c})=\rho}|}{|C^\ell|}\,.
 \]
\item
 \[
   \Pr(M\text{ accepts }\vek{v})\ = \
   \frac{|\setc{\vek{c}\in C^\ell}{\rho_M(\vek{v},\vek{c}) \text{ accepts}}|}{|C^\ell|}\ .
 \]
\end{enumerate}
\end{lemma}
\begin{proof}
To prove (a), observe that for every run $\rho = (\rho_1,\twodots,\rho_k)$ of $M$ on $\vek{v}$
we have $k\leq \ell$, and
\[
 \begin{array}{rcl}
   {
      \frac{|\setc{\vek{c}\in C^{\ell}}{\rho_M(\vek{v},\vek{c})=\rho}|}{|C^{\ell}|} }
   & \ = \ &
   \frac{1}{|C^{\ell}|}\cdot
      \left(
        {\displaystyle \prod_{i=1}^{k-1}} |\setc{c\in C}{\rho_{i+1} \text{ is the $c$-successor of }\rho_i}|
      \right)
      \cdot
         |C|^{\ell -(k-1)}
   \\
   & \ = \
   &  {\displaystyle \prod_{i=1}^{k-1}} \Pr(\rho_i\to_M \rho_{i+1})
   \ \ = \ \
    \Pr(\rho)\ .
 \end{array}
\]
(b) follows directly from (a), since
\[
  \begin{array}[b]{ccccc}\displaystyle
    \Pr(M \text{ accepts }\vek{v}) & \displaystyle
    \ \stackrel{\text{def}}{=} \ & \displaystyle
    \sum_{\rho \ : \ \text{$\rho$ is an accepting} \atop \text{run of $M$ on $\vek{v}$}} \Pr(\rho) &
    \ \stackrel{\text{(a)}}{=} \ & \displaystyle
    \sum_{\rho \ : \ \text{$\rho$ is an accepting} \atop \text{run of $M$ on $\vek{v}$}}
         \frac{|\setc{\vek{c}\in C^{\ell}}{\rho_M(\vek{v},\vek{c})=\rho}|}{|C^{\ell}|}
   \\[1cm] \displaystyle
     & \displaystyle
    \ = \ & \displaystyle
    \sum_{\vek{c}\in C^{\ell} \ : \atop \text{$\rho_M(\vek{v},\vek{c})$ accepts}}
         \frac{1}{|C^{\ell}|} & \displaystyle
    \ = \ & \displaystyle
    \frac{|\setc{\vek{c}\in C^{\ell}}{\rho(\vek{v},\vek{c}) \text{ accepts}}|}{|C^{\ell}|}\ .
  \end{array}
\]
\end{proof}


\section{Proof of the Simulation Lemma}\label{appendix:SimulationLemma}

\smallskip\noindent
Let $T = (Q,\Sigma,\Delta,q_0,F,F_{\textit{acc}})$ be the given
$(r,s,t)$-bounded nondeterministic Turing machine with $t+u$
tapes, where the tapes $1,\ldots,t$ are the external memory tapes and tapes
$t+1,\ldots,t+u$ are the internal memory tapes. Let $m,n\in\mathbb N$ and
$N=m\cdot (n+1)$.
Every tuple $\vek{v} = (v_1,\twodots,v_m)\in I^m$ corresponds to an
input string
$\tilde{v} := v_1\,\#\,v_2\,\# \cdots v_m\,\#$ of length $N$.
Let $r:=r(N)$ and $s:=s(N)$.

By Lemma~\ref{lem:time}, there is a constant $c_1 = c_1(u,|Q|,|\Sigma|)$, which
does \emph{not} depend on $r$, $m$, $n$, $t$, such that
every run of $T$ on every input $\tilde{v}$, for
any $\vek{v}\in I^m$, has length at most
\begin{equation}\label{eq:TM2LM:SizeHatQ}
\ell(N) \ \ := \ \ N\cdot 2^{c_1\cdot r\cdot(t+s)}
\end{equation}
and throughout each such run,
each of $T$'s external memory tapes $1,\ldots,t$ has length $\leq\ell(N)$.
We let $\ell:= \ell(N)$.

\bigskip
\begin{bfenv}{Step 1:}
\textit{Definition of $M$'s set $C$ of nondeterministic choices.}

\medskip\noindent
$M$'s set $C$ of \emph{nondeterministic choices} is chosen as
$C:= (C_T)^\ell$, where $C_T$ is chosen according to Definition~\ref{def:rhoT}.
\\ \mbox{}
\end{bfenv}

\bigskip
\begin{bfenv}{Step 2:}
\textit{Definition of a superset $\tilde{A}$ of $M$'s state set $A$.}

\medskip\noindent
Let $\hat{Q}$ be the set of potential configurations of
tapes $t{+}1,\ldots,t{+}u$, together with the
current state of $T$, that is,
\begin{displaymath}
  \hat{Q}\deff
      \left\{(q,p_{t+1},\ldots,p_{t+u},w_{t+1},\ldots,w_{t+u})
       \ \left| \ \
       \parbox{5cm}{$q\in Q, \
       p_{t+i}\in\{1,\ldots,s\},\\ w_{t+i} \in \Sigma^{\leq s}
       \mbox{ (for all } i\in\set{1,\ldots,u}\mbox{)}\;$}
       \right.
       \right\}.
\end{displaymath}
Then for a suitable constant $c_2=c_2(u,|Q|,|\Sigma|)$ we have
\begin{equation}
  \label{eq:size-qhat}
  |\hat{Q}| \ \ \leq \ \  2^{c_2\cdot s}.
\end{equation}
We let
\begin{align*}
\tilde{A}\deff\;
& \Big\{
  \big(\hat{q},\; \vek{p}_1,  \twodots, \vek{p}_t  \big)
  \Bigmid
  \hat{q}\in \hat{Q},
  \ \ \mbox{and for each $j\in\set{1,\twodots,t}$, }
\\
& \qquad
    \vek{p}_j = (\pleft_j,\pact_j,\pright_j,\textit{head-direction}_j) \mbox{ with}
\\
& \qquad\quad
   \pact_j \in\set{1,\twodots,\ell}, \
   \textit{head-direction}_j\in\set{+1,-1},\mbox{ \ and \ }
\\
& \qquad\quad
    \mbox{either \ \ } \pleft_j = \pright_j = \Nil,
\\
& \qquad\quad
    \mbox{or \ \ }\pleft_j,\pright_j \in\set{1,\twodots, \ell} \mbox{ \
      with \ }
    \pleft_j \leq \pact_j \leq \pright_j \
\Big\}\,.
\end{align*}
Here, $\Nil$ is a symbol for indicating that $\pleft_j$ and $\pright_j$ are ``undefined'', that is,
that they cannot be interpreted as positions on one of the Turing machine's external memory tapes.

Later, at the end of Step~4, we will specify, \emph{which} particular subset of $\tilde{A}$ will
be designated as $M$'s state set $A$.
With \emph{any}  choice of $A$ as a subset of $\tilde{A}$ we will have
\begin{displaymath}
 |A|
 \ \le \
 |\tilde A|
 \ \le \
 |\hat{Q}| \cdot \big(\ell + 1\big)^{3\cdot t} \cdot 2^t
 \ \le \
 2^{c_2\cdot s} \cdot \big( N\cdot 2^{c_1\cdot r\cdot(t+s)} + 1\big)^{3\cdot t}
   \cdot 2^t\;\le\; 2^{d\cdot t^2\cdot r\cdot s}
\end{displaymath}
for a suitable constant $d=d(u,|Q|,|\Sigma|)$. This completes Step~2.
\end{bfenv}

\bigskip
\begin{bfenv}{Step 3:}
  \textit{Definition of $M$'s initial state $a_0$ and $M$'s
  sets $B$ and $B_{\textit{acc}}$ of final states and accepting states, respectively.}

\medskip\noindent
Let
\[
\hat{q}_0 \ \deff \
(q_0, \underbrace{1,\twodots,1}_u, \underbrace{\Blank^{s},\twodots,\Blank^{s}}_u)
\]
be the part of $T$'s \emph{initial configuration} that describes the (start)
state $q_0$ of $T$ and the head positions and initial (i.e., empty) content of the
tapes $t{+}1,\twodots,t{+}u$ (that is, the tapes that represent internal
memory).
\\
Let
\begin{displaymath}
 \vek{p}_1
 \ \deff \
 (\pleft_1,\pact_1,\pright_1,\textit{head-direction}_1)
 \ \deff \
  \left\{
    \begin{array}{ll}
    (1,1,n{+}1,+1) & \text{ if } m > 1 \\
    (1,1,\ell,+1) & \text{ if } m=0
    \end{array}
  \right.
\end{displaymath}
and, for all $i\in\set{2,\twodots,t}$,
\begin{displaymath}
 \vek{p}_i
 \ \deff \
 (\pleft_i,\pact_i,\pright_i,\textit{head-direction}_i)
 \ \deff \
 (1,1,\ell,+1).
\end{displaymath}
As start state of the NLM $M$ we choose
\begin{myeqnarray*}
a_0 & \deff &
(\hat{q}_0, \vek{p}_1, \vek{p}_2, \twodots, \vek{p}_t )
\end{myeqnarray*}%
As $M$'s sets of \emph{final}, resp., \emph{accepting} states we choose
\ $B \deff \tilde{B}\cap A$, resp.,
\ $B_\textit{acc} \deff \tilde{B}_{\textit{acc}}\cap A$ with
\begin{eqnarray*}
 \tilde{B}
& \ \deff \
& \big\{\big(\hat{q},\; \vek{p}_1, \vek{p}_2, \twodots, \vek{p}_t    \big)\in \tilde{A} \ \bigmid \
  \text{$\hat{q}$ is of the form $(q,\vek{p},\vek{y})\in\hat{Q}$}
  \ \text{for some $q\in F\big\}$}
\\
 \tilde{B}_{\textit{acc}}
& \ \deff \
& \big\{\big(\hat{q},\; \vek{p}_1, \vek{p}_2, \twodots, \vek{p}_t    \big)\in \tilde{A} \ \bigmid \
  \text{$\hat{q}$ is of the form $(q,\vek{p},\vek{y})\in\hat{Q}$}
  \ \text{for some $q\in F_{\textit{acc}}\big\}$}.
\end{eqnarray*}
I.e., a state of $M$ is \emph{final} (resp., \emph{accepting}) if, and only if, the associated state
of the Turing machine $T$ is. This completes Step~3.
\end{bfenv}

\bigskip
\begin{bfenv}{Step 4:}
\textit{Definition of $M$'s transition function
 \ $\alpha \ : \ (A\setminus B) \times \big(\Al^*\big)^t \times C
     \ \to \
     \left( A\times \text{Movement}^t \right).
 $}%

\medskip\noindent
We let
\begin{align*}
   \PartConfT
   \ \deff \
   \Big\{
 & \big( q, p_1,\twodots,p_{t+u}, w_1,\twodots,w_{t+u}\big)
   \ \Bigmid \
   q \in Q, \text{ \ and}
 \\
 & \quad \text{for all $j\in\set{1,\twodots,t{+}u}$,} \ \
   p_j\in\NN,
 \\
 & \quad \text{for all $j\in\set{1,\twodots,t}$,} \ \
   w_j\in \set{\?}^*\Sigma^*\set{\?}^* \
          \mbox{ with } w_{j,p_j}\in\Sigma,
 \\
 & \quad\text{for all $j\in\set{1,\twodots,u}$,} \ \
   w_{t+j}\in \Sigma^* \ \Big\},
\end{align*}
where $\?$ is a symbol \emph{not} in $\Sigma$, and
$w_{j,p_j}$ denotes the $p_j$-th letter in the string $w_j$.

Intended meaning: The symbol $\?$ is used as a \emph{wildcard} symbol that may be
interpreted by any symbol in $\Sigma$.
An element in $\PartConfT$ gives (potentially) incomplete
information on a configuration of $T$, where the contents of tapes $1,\twodots,t$ might be
described only in some part (namely, in the part containing no $\?$-symbols).

We let $\tilde{\Al} \deff I \cup C \cup \tilde{A} \cup \set{\struc{,}}$.
By induction on $i$ we fix, for $i\geq 0$,
\begin{itemize}
\item
  a set $A_i\subseteq \tilde{A}$
\item
  a set $K_i \subseteq (\tilde{A}\setminus \tilde{B}) \times (\tilde{\Al}^*)^t$,
\item
  a set $L_i\subseteq \tilde{\Al}^*$, letting
  \begin{myeqnarray}\label{lemma:TM2LM:DefLi}
     L_i & \deff & \big\{ \
                          a\struc{y_1}\cdots\struc{y_t}\struc{c}
                          \ \ : \ \
                          (a,y_1,\twodots,y_t)\in K_i \text{ and } \ c\in C
                   \ \big\}
  \end{myeqnarray}
\item
  a function
  \[
    \config_i :   K_i \ \to \ \PartConfT \cup \set{\undef}
  \]
  Intended meaning: When the NLM $M$ is in a situation $\kappa\in K_i$, then $\config_i(\kappa)$ is the
  Turing machine's configuration
  at the beginning of $M$'s current step. If $\config_i(\kappa) = \undef$, then $\kappa$ does not
  represent a configuration of the Turing machine.
\item
  the transition function $\alpha$ of $M$, restricted to $K_i$, that is,
  \[
     \alpha_{|\, K_i}  : K_i \times C \ \to \ \tilde{A} \times \text{Movement}^t
  \]
\item
 for every tape $j\in\set{1,\twodots,t}$, a function
 \begin{displaymath}
    \tapeconf_{j,i} \ \ : \ \  L_i \ \to \ \left\{
    \ (w,\pleft,\pright)
    \ \ \left| \ \
    \parbox{5.3cm}{$\text{either }1\leq \pleft \leq \pright \leq \ell
                  \mbox{ \ and} \\
                   \mbox{\quad } w \in
                   \set{\?}^{\pleft-1}\Sigma^{\pright-\pleft+1}\set{\?}^{\ell(N)-\pright}\\
                  \text{or } \pleft > \pright \mbox{ and } w = \varepsilon $}
      \ \right\}
      \right.
 \end{displaymath}
  Intended meaning: When the NLM $M$ is in a situation $\kappa = (a,\struc{y_1},\twodots,\struc{y_t}) \in K_i$
  and nondeterministically chooses $c\in C$ for its current transition,
  then
  \[
     \tapeconf_{j,i}\big(a\struc{y_1}\cdots\struc{y_t}\struc{c}\big)
  \]
  gives information on the inscription from tape cell
  $\pleft$ up to tape cell $\pright$ of
  the $j$-th tape of the Turing machine's configuration
  at the end of $M$'s current step.
\smallskip
\end{itemize}
\textit{Induction base ($i=0$): }
We start with $M$'s start state $a_0$ and choose
\[
  A_0 \deff \set{\,a_0\,}.
\]
If $a_0$ is \emph{final}, then we let $K_0\deff \emptyset$ and $A\deff A_0$.
This then gives us an NLM $M$ which accepts its input without performing a
single step. This is fine, since $a_0$ is final if, and only if, the Turing
machine $T$'s start state $q_0$ is final, that is, $T$ accepts its input
without performing a single step.

For the case that $a_0$ is \emph{not} final, we let
\[
K_0\ \deff \ \Big\{\ (a_0, {y_1},\twodots, {y_t})\Bigmid y_1\in\setc{\struc{v}}{v\in I} \mbox{ and }
     y_2=\cdots = y_t = \struc{} \ \Big\}.
\]
The set $L_0$ is defined via equation (\ref{lemma:TM2LM:DefLi}).

The function $\config_0$ is defined as follows:
For every
\[
   \kappa \ =  \ (a_0,y_1,\twodots,y_t) \ \in \ K_0
\]
with $y_1 = \struc{v}$ (for some $v\in I$), let
\begin{displaymath}
  \config_0 (\kappa) \ \deff \
  \big(q_0, \overbrace{1,\twodots,1}^{t+u},\ v\, \#\, \?^{\ell -(n+1)},
    \underbrace{\Blank^{\ell},\twodots,\Blank^{\ell}}_{t-1}, \
    \underbrace{\Blank^{s},\twodots,\Blank^{s}}_{u}\big).
\end{displaymath}

Let
\[
  \big(\hat{q}_0 , \vek{p}_1,\twodots,\vek{p}_t \big)\ := \ a_0
\]
with $\vek{p}_j = (\pleft_j,\pact_j,\pright_j,\textit{head-direction}_j)$,
for all $j\in\set{1,\twodots,t}$.

For $j\in \set{1,\twodots,t}$ we define
\[
 \big(\phleft_j,\ \pact_j,\ \phright_j \big)\deff \big(\pleft_j,\pact_j,\pright_j\big).
\]
Now let $c=(c_{1},c_{2},\twodots,c_{\ell})\in C=C_T^\ell$ be an arbitrary element from
$M$'s set $C$ of nondeterministic choices.
For defining $\alpha_{|K_0}(\kappa,c)$ and
$\tapeconf_{j,0}(a\struc{y_1}\cdots\struc{y_t}\struc{c})$, consider the following:
Let us start the Turing machine $T$ with a configuration $\gamma_1$ that
\emph{fits} to $\config_0(\kappa)$, i.e., that can be obtained from
$\config_0(\kappa)$ by replacing each occurrence of the wildcard symbol $\?$
by an arbitrary symbol in $\Sigma$.  Let $\gamma_1,\gamma_2,\gamma_3,\ldots$
be the successive configurations of $T$ when started in $\gamma_1$ and using the nondeterministic
choices $c_{1},c_{2},c_3,\twodots$ (in the sense of Definition~\ref{def:rhoT}).
I.e., for all $\nu\geq 1$,
$\gamma_{\nu+1}$
is the $\big(c_\nu\ \mod\ |\Next_T(\gamma_\nu)|\big)$-th of the $|\Next_T(\gamma_\nu)|$ possible
next configurations of $\gamma_\nu$.

Using this notation, the definition of $\alpha_{|K_0}(\kappa,c)$ and
\[
\tapeconf_{j,0}(a\struc{y_1}\cdots\struc{y_t}\struc{c})
\]
can be taken verbatim from the definition of
$\alpha_{|K_{i+1}}(\kappa,c)$ and
\[
\tapeconf_{j,i+1}(a\struc{y_1}\cdots\struc{y_t}\struc{c}),
\]
given below.
This completes the induction base ($i=0$).

\bigskip\noindent
\textit{Induction step ($i\to i{+}1$): }
We let
\begin{displaymath}
 A_{i+1}
 \ \ \deff \ \
 \left\{ \
   b \in \tilde{A}
   \ \ \left| \ \
   \parbox{8.8cm}{$
     \mbox{there are } \kappa\in K_i \mbox{ and }c\in C \mbox{ such that }
     \ \alpha_{|K_i}(\kappa,c) = (b,e_1,\twodots,e_t)
     \\
     \mbox{(for suitable } (e_1,\twodots,e_t)\in\text{Movement}^t)
   $}
   \right.
 \right\}
\end{displaymath}
and
\begin{displaymath}
 K_{i+1}
 \ \ \deff\ \
 \left\{ \
   (a,y_1,\twodots,y_t)
   \ \ \left| \ \
   \parbox{6cm}{
      $a \in A_{i+1}\setminus \tilde{B}$,\\
      $y_1 \in \setc{\struc{v}}{v\in I} \cup \bigcup_{i'\leq i} L_{i'},\mbox{ and}$\\
      $y_j \in \set{\struc{}} \cup \bigcup_{i'\leq i} L_{i'}, \ \mbox{ for all $j\in\set{2,\twodots,t}$}$
   }
   \right.
 \right\}
\end{displaymath}
The set $L_{i+1}$ is defined via equation (\ref{lemma:TM2LM:DefLi}).
\\
The function $\config_{i+1}$ is defined as follows:
Let $c\in C$ and let $\kappa = (a,y_1,\twodots,y_t)\in K_{i+1}$.
Let
\[
  \big( \hat{q} , \vek{p}_1 , \twodots, \vek{p}_t \big) \ := \ a
\]
with $\vek{p}_j = (\pleft_j,\pact_j,\pright_j,\textit{head-direction}_j)$,
for all $j\in\set{1,\twodots,t}$, and
\[
  \hat{q} \ \ = \ \ (q,p_{t+1},\twodots,p_{t+u},w_{t+1},\twodots,w_{t+u}).
\]
Let $j\in\set{1,\twodots,t}$.
\par
{If $y_j\in L_{i'}$} for some $i'\leq i$, then let
\[
  (w'_j,\psleft_j,\psright_j) \ \ := \ \ \tapeconf_{j,i'}(y_j).
\]
We choose $w_j := w'_j$. (This is well-defined, because $\tapeconf_{j,i'}$ and
$\tapeconf_{j,i''}$ operate identically on all elements in $L_{i'}\cap L_{i''}$, for all
$i',i''\leq i$).
\\
Furthermore, we let $(\phleft_j,\phright_j)$ be defined as follows:
\[
(\phleft_j,\phright_j)\deff
 \begin{cases}
   (\pact_j,\psright_j) & \text{if }\pleft_j=\pright_j=\Nil\text{ and }\textit{head-direction}_j=+1 \\
   (\psleft_j,\pact_j) & \text{if }\pleft_j=\pright_j=\Nil\text{ and }\textit{head-direction}_j=-1 \\
   (\pleft_j,\pright_j) & \text{otherwise}.
 \end{cases}
\]
\par
If $y_j\not \in \cup_{i'\leq i}L_{i'}$, then we make a case distinction on~$j$:
In case that $j\in\set{2,\twodots,t}$, we have $y_j = \struc{}$ and
$\textit{head-direction}_j = +1$.
We define $(\phleft_j,\phright_j)$ as follows:
\begin{myeqnarray*}
  \big(\phleft_j, \phright_j\big) &  \deff & \big(\pact_j,\ell\big) \ ,
\end{myeqnarray*}%
 and choose
\[
  w_j \ \ := \ \ \?^{{\phleft_j} -1}\Blank^{\ell- ({\phleft_j} -1)}\ .
\]
In case that $j=1$, we know that $y_j$ must be of the form $\struc{v}$, for some $v\in I$,
and that $\textit{head-direction}_j=+1$.
If $v$ is \emph{not} the $m$-th input item, that is, there is some $\mu\in\set{1,\twodots,m{-}1}$ such that
$(\mu{-}1)\cdot(n{+}1) < \pact_1 \leq \mu\cdot(n{+}1)$, then we define
\begin{myeqnarray*}
   \big(\phleft_1, \phright_1\big) & \deff & \big(\pact_1,\mu\cdot(n{+}1)\big) \ ,
\end{myeqnarray*}%
 and choose
\[
  w_1 \ \ := \ \ \?^{(\mu-1)\cdot(n+1)}\;v \,\#\ \?^{\ell-\mu\cdot(n+1)}.
\]
Otherwise, $v$ must be the $m$-th input item, that is,
\[
   \pact_1 \ \ > \ \ (m{-}1)\cdot(n{+}1).
\]
In this case we define
\begin{myeqnarray*}
   \big(\phleft_1, \phright_1\big) & \deff & \big(\pact_1,\ell\big)
\end{myeqnarray*}%
 and choose
\[
  w_1 \ \ := \ \ \?^{(m-1)(n+1)}\;v \,\#\ \Blank^{\ell-m\cdot(n+1)}.
\]
If, for some $j_0\in\set{1,\twodots,t}$, $w_{j_0}=\varepsilon$, then we define
\begin{align*}
 \config_{i+1}(\kappa)& \ \deff\ \undef,\\
 \tapeconf_{j,i+1}\big(a\struc{y_1}\cdots\struc{y_t}\struc{c}\big)& \ \deff\
  (\varepsilon,2,1),
\end{align*}
and $\alpha_{|K_{i+1}}(\kappa,c) \deff \big(a,e''_1,\twodots,e''_t\big)$,
where for all $j\in\set{1,\twodots,t}$,
\[
 e''_j
 \ \ \deff \ \
 \begin{cases}
    \big(\textit{head-direction}_j,\textit{true}\big) & \text{if } w_j = \varepsilon \\
    \big(\textit{head-direction}_j,\textit{false}\big) & \text{otherwise}.
 \end{cases}
\smallskip
\]
In what follows, we consider the case where $w_j\neq\varepsilon$, for all $j\in\set{1,\twodots,t}$.
We define
\begin{displaymath}
  \config_{i+1}(\kappa)
  \ \ \deff \ \
  \big( q,p_1,\twodots,p_t,p_{t+1},\twodots,p_{t+u},
        w_1,\twodots,w_t,w_{t+1},\twodots,w_{t+u}\big),
\end{displaymath}
where
$q$ and
$p_{t+1},\twodots, p_{t+u},w_{t+1},\twodots,w_{t+u}$ are obtained from $\hat{q}$,
\\
$p_1,\twodots,p_t$ are obtained from $a$ via \ $p_j\deff \pact_j$, for all $j\in\set{1,\twodots,t}$,
and \\
$w_1,\twodots,w_t$ are chosen as above.

Altogether, the description of the definition of
$\config_{i+1}(\kappa)$ is complete.
\\
\parno
For defining $\alpha_{|K_{i+1}}(\kappa,c)$ and $\tapeconf_{j,i+1}(a\struc{y_1}\cdots\struc{y_t}\struc{c})$,
consider the following:

Let us start the Turing machine $T$ with a configuration
$\gamma_1$ that \emph{fits} to $\config_{i+1}(\kappa)$, i.e., that can be obtained from
$\config_{i+1}(\kappa)$ by replacing each occurrence of the wildcard symbol $\?$ by
an arbitrary symbol in $\Sigma$. Letting
\[
   c\ = \ (c_1,c_2,\twodots,c_\ell) \ \in \ C \ = \ C_T^\ell,
\]
we let $\gamma_1,\gamma_2,\gamma_3,\ldots$ be the successive configurations of $T$ when
started in $\gamma_1$ and using the nondeterministic
choices $c_{1},c_{2},c_3,\twodots$ (in the sense of Definition~\ref{def:rhoT}).
I.e., for all $\nu\geq 1$,
$\gamma_{\nu+1}$
is the $\big(c_\nu\ \mod\ |\Next_T(\gamma_\nu)|\big)$-th of the $|\Next_T(\gamma_\nu)|$ possible
next configurations of $\gamma_\nu$.

Then, there is a minimal $\nu>1$ for which there exists a $j_0\in\set{1,\twodots,t,\undef}$
such that throughout the run $\gamma_1\cdots\gamma_{\nu-1}$,
\begin{enumerate}
\item none of the heads $1,\twodots,t$ changes its direction, and
\item none of the heads $j\in\set{1,\twodots,t}$ crosses a border $\phleft_{j}$ or $\phright_{j}$,
\end{enumerate}
and one of the following cases applies:
\begin{description}
 \item[Case 1:]
   $j_0\neq\undef$, and in the transition from $\gamma_{\nu-1}$ to $\gamma_{\nu}$, head $j_0$ crosses one of the borders
   $\phleft_{j_0}$ or $\phright_{j_0}$. That is,
   in $\gamma_{\nu}$, the $j_0$-th head is either at position $\phleft_{j_0}-1$ or
   at position $\phright_{j_0}+1$. \\
   (And none of the heads $j\in\set{1,\twodots,t}\setminus\set{j_0}$ crosses a border
    or changes its direction.\footnote{Recall that w.l.o.g.\ we assume that the
   Turing machine is normalized, cf.\ Definition~\ref{def:TuringMachine}.} )
 \item[Case 2:]
   $j_0\neq\undef$, and in the transition from $\gamma_{\nu-1}$ to $\gamma_{\nu}$, head $j_0$ changes its direction, but
   does not cross one of the borders $\phleft_{j_0}$ or $\phright_{j_0}$.\\
   (And none of the heads $j\in\set{1,\twodots,t}\setminus\set{j_0}$ crosses a border
    or changes its direction.)
 \item[Case 3:]
  $\gamma_{\nu}$ is \emph{final} and none of the cases 1 and 2 apply. Then we let $j_0:= \undef$.
\end{description}
In all three cases we let
  \begin{myeqnarray*}
   (q'',p''_1,\twodots,p''_{t+u},w''_1,\twodots,w''_{t+u}) & \ \deff \ & \gamma_{\nu} \ .
  \end{myeqnarray*}%
We choose
  \begin{myeqnarray*}
    \hat{q}'' & \ := \ & (q'',p''_{t+1},\twodots,p''_{t+u},w''_{t+1},\twodots,w''_{t+u})
  \end{myeqnarray*}%
and define
  \begin{myeqnarray*}
   b & \ \deff \ & (\hat{q}'',\vek{p}''_1,\twodots,\vek{p}''_t) \ ,
  \end{myeqnarray*}%
where
\begin{myeqnarray*}
  \vek{p}''_j & \ = \ & (\pssleft_j,\pssact_j,\pssright_j,\textit{head-direction}''_j)
\end{myeqnarray*}%
will be specified below.
\\
Finally, we define
  \begin{eqnarray*}
   \alpha_{|K_{i+1}} (\kappa,c) & \ \deff \ & (b,e''_1,\twodots,e''_t),
  \end{eqnarray*}
where, for every $j\in\set{1,\twodots,t}$,
\begin{myeqnarray*}
  e''_j & \ \deff \ & \big(\textit{head-direction}''_j,\ \textit{move}''_j \big)
\end{myeqnarray*}%
will be specified below.
\medskip\\
Recall that \ $\kappa = \big(a,y_1,\twodots,y_t\big)\in K_{i+1}$.
For every $j\in\set{1,\twodots,t}$ we define
\begin{displaymath}
 \tapeconf_{j,i+1}\big(a\struc{y_1}\cdots\struc{y_t}\struc{c}\big)
 \ \deff \
 \begin{cases}
   \big( \?^{\pleft_j-1} w''_{j,\pleft_j}\, \cdots\, w''_{j,\pright_j}\, \?^{\ell-\pright_j+1}, \ \pleft_j,\ \pright_j\big) & \text{if } \pleft_j\leq\pright_j
  \\
  \big(\varepsilon,\pleft_j,\pright_j\big) & \text{otherwise}
 \end{cases}
\end{displaymath}
where $\pleft_j$ and $\pright_j$ are specified below.
\\
\parno
For all $j\in\set{1,\twodots,t}\setminus\set{j_0}$ we know (by the choice of $\nu$ and $j_0$) that
throughout the Turing machine's computation $\gamma_0,\twodots,\gamma_\nu$, head $j$  neither changes
its direction nor crosses one of the borders $\phleft_j$, $\phright_j$.
Consequently, we choose
\begin{myeqnarray*}
 \textit{head-direction}''_j & \deff & \textit{head-direction}_j
\\[1ex]
 \textit{move}''_j & \deff & \textit{false}
\\[1ex]
 \pssact_j & \deff & p''_j
\\[1ex]
 \pssleft_j & \deff &
 \left\{
  \begin{array}{ll}
    \pssact_j & \text{ if } \textit{head-direction}_j = +1 \\
    \phleft_j & \text{ if } \textit{head-direction}_j = -1
  \end{array}
 \right. \smallskip
\\[1ex]
 \pssright_j & \deff &
 \left\{
  \begin{array}{ll}
    \phright_j & \text{ if } \textit{head-direction}_j = +1 \\
    \pssact_j & \text{ if } \textit{head-direction}_j = -1
  \end{array}
 \right. \smallskip
\\[1ex]
 \pleft_j & \deff &
 \left\{
  \begin{array}{ll}
    \phleft_j & \text{ if } \textit{head-direction}_j = +1 \\
    \pssact_j +1 & \text{ if } \textit{head-direction}_j = -1
  \end{array}
 \right. \smallskip
\\[1ex]
 \pright_j & \deff &
 \left\{
  \begin{array}{ll}
    \pssact_j -1 & \text{ if } \textit{head-direction}_j = +1 \\
    \pssright_j  & \text{ if } \textit{head-direction}_j = -1
  \end{array}
 \right.
\end{myeqnarray*}%
In \textbf{Case 3} we have $j_0=\undef$, and therefore, $\alpha_{|K_{i+1}}(\kappa,c)$ and
$\tapeconf_{j,i+1}(a\struc{y_1}\cdots\struc{y_t}\struc{c})$ is fully specified.
Furthermore, note that in Case 3 we know that $\gamma_\nu$ is \emph{final},
i.e., $q''$ is a final state of the Turing machine $T$.
Therefore, $b$ is a \emph{final} state of the NLM $M$, and $M$'s run accepts if, and only if,
the simulated Turing machine run accepts (recall the definition of $M$'s set of final and
accepting states at the end of Step~3).
\\
\parno
For \textbf{Case 1} and \textbf{Case 2}, we have $j_0\in\set{1,\twodots,t}$, and
for specifying
\[
  \textit{head-direction}''_{j_0}, \ \
  \textit{move}''_{j_0}, \ \
  \pssleft_{j_0}, \ \
  \pssact_{j_0}, \ \
  \pssright_{j_0}, \ \
  \pleft_{j_0}, \ \ and \ \
  \pright_{j_0},
\]
we distinguish between the two cases:
\\
\parno
\textbf{ad Case 1:} \ In this case, $j_0\neq\undef$, and head $j_0$ crosses one of the borders
 $\phleft_{j_0}$ or $\phright_{j_0}$ in the transition from $\gamma_{\nu-1}$ to $\gamma_\nu$
 (that is, $p''_{j_0}$ is either $\phright_{j_0}+1$ or $\phleft_{j_0}-1$).
 We choose
 \begin{myeqnarray*}
   \big(\pssleft_{j_0},\ \pssact_{j_0},\ \pssright_{j_0}\big) & \ \deff \ &
   \big(\Nil,\ p''_{j_0},\ \Nil\big)
 \\[1ex]
   \big(\pleft_{j_0},\ \pright_{j_0} \big) & \ \deff \ &
   \big(\phleft_{j_0},\ \phright_{j_0}\big)
 \\[1ex]
   \textit{move}''_{j_0} & \ \deff \ &
   \textit{true}
 \\[1ex]
   \textit{head-direction}''_{j_0} & \ \deff \ &
   \left\{
     \begin{array}{ll}
       +1 & \text{ if } p''_{j_0} = \phright_{j_0}+1 \\
       -1 & \text{ otherwise}.
     \end{array}
   \right.
 \end{myeqnarray*}%
\mbox{}\\
\parno
\textbf{ad Case 2:} \ In this case, $j_0\neq \undef$, and head $j_0$ changes its direction, but does not
 cross one of the borders $\phleft_{j_0}$ or $\phright_{j_0}$.
 We only consider the case where the direction of head $j_0$ changes from $+1$ to $-1$
 (the other case is symmetric). \\
 We choose
 \begin{myeqnarray*}
   \big(\textit{head-direction}''_{j_0},\ \textit{move}''_{j_0}\big) & \ \deff \ &
   \big(-1, \textit{false}\big)
  \\[1ex]
   \big(\pssleft_{j_0},\ \pssact_{j_0},\ \pssright_{j_0}\big) & \deff &
   \big(\phleft_{j_0}, \ p''_{j_0}, \ p''_{j_0}+1\big)
 \\[1ex]
   \big(\pleft_{j_0},\ \pright_{j_0} \big) & \deff &
   \big(p''_{j_0}+2,\ \phright_{j_0}\big)
 \end{myeqnarray*}%
 Note that here we might have $p''_{j_0}+1 = \phright_{j_0}$. In this case, by the above
 definition, we obtain $\pleft_{j_0} = \pright_{j_0}+1$.
\mbox{}\\
\parno
Altogether, this completes the induction step.
\\
\parno
Finally, we are ready to fix $M$'s state set $A$ and transition function $\alpha$ as
follows:
\begin{eqnarray*}
 A & \deff &  \bigcup_{i\geq 0} A_i
\\
 K & \deff &  \bigcup_{i\geq 0} K_i
\\
 \alpha & \deff &  \bigcup_{i\geq 0} \alpha_{|K_i}
\end{eqnarray*}%
Note that
\begin{enumerate}[1.]
 \item
   $\alpha$ is well-defined, because $\alpha_{|K_i}$ and $\alpha_{|K_{i'}}$ operate
   identical on all elements in $(K_i \cap K_{i'})\times C$ (for all $i,i'\geq 0$).
 \item
   $K$ consists of all situations $(a,y_1,\twodots,y_t) \in (A\setminus B)\times (\Al^*)^t$
   that may occur in runs of $M$.
 \item
   $\alpha$ remains undefined for elements $(a,y_1,\twodots,y_t)$ in
   $(A\setminus B)\times (\Al^*)^t$ that
   do \emph{not} belong to $K$.
   This is fine, because such a situation $(a,y_1,\twodots,y_t)$ can never occur in an
   actual run of $M$.
\end{enumerate}
This completes Step~4.
\end{bfenv}

\bigskip
Note that finally, the NLM $M$ is fully specified. Due to the construction we know that
$M$ is $(r,t)$-bounded, because it has $t$ lists and
the number of head reversals during each run on an input $\vek{v}=(v_1,\twodots,v_m)\in I^m$ is bounded by
the number $r{-}1 = r(m\cdot(n{+}1)){-}1$ of head reversals of the according run of the Turing machine $T$ on input
$v_1\#\cdots v_m\#$.

\bigskip

\begin{bfenv}{Step 5:}
  \textit{For every input $\vek{v}= (v_1,\twodots,v_m)\in I^m$ we have
 \[
    \Pr(M\text{ accepts } \vek{v}\big) \ = \ \Pr\big(T\text{ accepts } v_1\#\cdots\#v_m\big).
 \]}%
\proof
Let $\ell_M\in\NN$ be an upper bound on the length of runs of the NLM $M$ (such a number $\ell_M$ exists,
because $M$ is $(r,t)$-bounded; see Lemma~\ref{lemma:shapeofruns}\,(\ref{item:LM:lengthofruns}) in
Appendix~\ref{appendix:ProofBoundLM}).

For the remainder of this proof we fix an input $\vek{v}=(v_1,\twodots,v_m)\in I^m$ for the NLM $M$ and
we let $\tilde{v}:= v_1\#\cdots v_m\#$ denote the corresponding input for the Turing machine $T$.

From Lemma~\ref{lemma:ProbAndTMs} we know that
\[
  \Pr(T \text{ accepts }\tilde{v}) \ = \
  \frac{|\setc{\vek{c}_T\in C_T^\ell}{\rho_T(\tilde{v},\vek{c}_T) \text{ accepts}}|}{|C_T^\ell|}
  \ = \
  \frac{|\setc{\vek{c}_T\in C_T^\ell}{\rho_T(\tilde{v},\vek{c}_T) \text{ accepts}}|}{|C|}.
\]
Furthermore, we know from Lemma~\ref{lemma:probLM} that
\[
  \Pr(M \text{ accepts }\vek{v}) \ = \
  \frac{|\setc{\vek{c}\in C^{\ell_M}}{\rho_M(\vek{v},\vek{c}) \text{ accepts}}|}{|C|^{\ell_M}}.
\]
For showing that \
$\Pr(M\text{ accepts } \vek{v}\big)  =  \Pr\big(T\text{ accepts } \tilde{v}\big)$ \
it therefore suffices to show that
\[
  |\setc{\vek{c}\in C^{\ell_M}}{\rho_M(\vek{v},\vek{c}) \text{ accepts}}|
  \ \ = \ \
  |C|^{\ell_M -1} \cdot
  |\setc{\vek{c}_T\in C_T^\ell}{\rho_T(\tilde{v},\vek{c}_T) \text{ accepts}}|.
\]
Consequently, it suffices to show that there is a function
\[
   f\ : \ C^{\ell_M} \ \to \ C_T^\ell
\]
such that
\begin{itemize}
\item
for every $\vek{c}\in C^{\ell_M}$, the list machine run $\rho_M(\vek{v},\vek{c})$ simulates the
Turing machine run $\rho_T(\tilde{v},f(\vek{c}))$, \ and
\item
for every $\vek{c}_T \in C_T^\ell$,
\begin{equation}\label{eqn:simulationlemmacontruns}
   |\setc{\vek{c}\in C^{\ell_M}}{f(\vek{c})= \vek{c}_T}| \ = \ |C|^{\ell_M -1}.
\end{equation}
\end{itemize}
We can define such a function $f$ as follows:

For every sequence
\[
   \vek{c} \ = \ \big( c^{(1)},\ldots,c^{(\ell_M)} \big) \ \in \ C^{\ell_M},
\]
following the construction of the NLM $M$ in Steps~1--4, we obtain for
each $i\in\set{1,\twodots,\ell_M}$ that there is a uniquely defined prefix $\tilde{c}^{(i)}$ of
$M$'s nondeterministic choice
\[
   c^{(i)} \ = \ \big( c^{(i)}_1,\twodots,c^{(i)}_\ell \big) \ \in \ C \ = \ C_T^{\ell},
\]
such that the following is true for
\[
     \tilde{c} \ \deff \ \tilde{c}^{(1)} \tilde{c}^{(2)} \cdots \tilde{c}^{(\ell_M)},
\]
viewed as a sequence of elements from $C_T$:
\begin{enumerate}[(1)]
\item
  The list machine run $\rho_M(\vek{v},\vek{c})$ simulates the Turing machine run
  $\rho_T(\tilde{v},\tilde{c})$, where $M$ uses in its $i$-th step exactly the
  $\tilde{c}^{(i)}$-portion of $c^{(i)}$ for simulating the according Turing machine steps.
\item
  If $\tilde{\ell} \leq \ell$ denotes the length of the run
  $\rho_T(\tilde{v},\tilde{c})=(\rho_1,\twodots,\rho_{\tilde{\ell}})$, then
  $\tilde{c}$ has exactly the length $\tilde{\ell}{-}1$.
  \medskip
\end{enumerate}
Now let $i_0$ denote the maximum element from $\set{1,\twodots,\ell_M}$ such that
$|\tilde{c}^{(i_0)}|\neq 0$ (in particular, this implies that
$\tilde{c} = \tilde{c}^{(1)}\cdots\tilde{c}^{(i_0)}$).
We let $\tilde{\tilde{c}}^{(i_0)}$ be the prefix of $c^{(i_0)}$ of length
$\ell-(\tilde{\ell}-1-|\tilde{c}^{(i_0)}|)$ and define
\[
   \tilde{\tilde{c}} \ \ \deff \ \ \tilde{c}^{(1)}\cdots \tilde{c}^{(i_0-1)}\tilde{\tilde{c}}^{i_0}.
\]
Note that, viewed as a sequence of elements from $C_T$, $\tilde{\tilde{c}}$ has length exactly $\ell$, and
therefore, we can well define
\[
    f(\vek{c}) \ \deff \ \tilde{\tilde{c}}.
\]
Furthermore, to see that (\ref{eqn:simulationlemmacontruns}) is satisfied, note that
$f$ is surjective, i.e., for every $\tilde{\tilde{c}}\in C_T^\ell$ there exists a
$\vek{c}$ with $f(\vek{c})=\tilde{\tilde{c}}$, and
\[
  |\setc{\vek{c}\in C^{\ell_M}}{f(\vek{c})=\tilde{\tilde{c}}}|
  \ = \
  |C_T|^{\ell\cdot \ell_M - \ell}
  \ = \
  |C_T|^{\ell\cdot(\ell_M -1)}
  \ = \
  |C|^{\ell_M -1}.
\]
(For the first equation, note that through $\tilde{\tilde{c}}$,
exactly $\ell$ of the possible $\ell\cdot \ell_M$ $C_T$-components
of $\vek{c}$ are fixed, whereas each of the remaining $\ell\cdot \ell_M - \ell$ components may carry an
arbitrary element from $C_T$.)
\\
This completes Step~5.
\end{bfenv}

\noindent
Altogether, the proof of Lemma~\ref{lemma:TM2LM} is complete.
\proofend


\section{Detailed Proof of Lemma~21}\label{appendix:ProofBoundLM}

\smallskip\noindent
This section is devoted to the proof of Lemma~\ref{lemma:MainLM}.

After pointing out an easy observation concerning randomized list machines in
subsection~\ref{subsec:ProbabilisticLMs},
we formally fix the notion of the \emph{skeleton} of a list machine's run in
subsection~\ref{subsec:Skeletons}.
Then, in subsection~\ref{subsec:BasicPropertiesLMs} we state and prove some basic properties
of list machines concerning the size and shape of runs and the possibility of composing different
runs.
Afterwards, in subsection~\ref{subsec:InformationFlow}, we take a closer
look at the information flow that can occur during
a list machine's computation, and we show that only a small number of
input positions can be compared during an NLM's run.
Finally, in subsection~\ref{subsec:BoundsForLMs},
we prove Lemma~\ref{lemma:MainLM}.

\subsection{An Easy Observation Concerning Randomized List Machines}\label{subsec:ProbabilisticLMs}

\begin{lemma}\label{lemma:Choose_c}
Let $M=(t,m,I,C,A,a_0,\alpha,B,\Bacc)$ be an NLM, let $\ell$ be an upper bound
on the length of $M$'s runs, and let $\mathcal{J}\subseteq I^m$ such that
$\Pr(M\text{ accepts }\vek{v})\geq \frac{1}{2}$, for all
inputs $\vek{v}\in\mathcal{J}$.
Then there is a sequence $\vek{c}=(c_1,\twodots,c_{\ell})\in C^\ell$ such that
 the set
\[
   \mathcal{J}_{\textit{acc}, \vek{c}} \ \deff \
   \setc{\vek{v}\in \mathcal{J}}{\rho_M(\vek{v},\vek{c})\text{ accepts}}
\]
has size \ $|\mathcal{J}_{\textit{acc},\vek{c}}|\geq \frac{1}{2}\cdot |\mathcal{J}|$.
\end{lemma}
\begin{proof}
By assumption we know that
\begin{equation*}
 \sum_{\vek{v}\in \mathcal{J}} \Pr(M\text{ accepts }\vek{v}) \ \geq \ |\mathcal{J}|\cdot \frac{1}{2}.
\end{equation*}
From Lemma~\ref{lemma:probLM} we obtain
\begin{equation*}
 \sum_{\vek{v}\in \mathcal{J}} \Pr(M\text{ accepts }\vek{v}) \ = \
 \sum_{\vek{v}\in \mathcal{J}} \frac{|\setc{\vek{c}\in C^\ell}{\rho_M(\vek{v},\vek{c}) \text{ accepts}}|}{|C^\ell|}\,.
\end{equation*}
Therefore,
\begin{equation*}
 \sum_{\vek{v}\in \mathcal{J}} |\setc{\vek{c}\in C^\ell}{\rho_M(\vek{v},\vek{c}) \text{ accepts}}|
 \ \geq \
 |C^\ell|\cdot \frac{|\mathcal{J}|}{2}\,.
\end{equation*}
On the other hand,
\begin{equation*}
  \sum_{\vek{v}\in \mathcal{J}} |\setc{\vek{c}\in C^\ell}{\rho_M(\vek{v},\vek{c})\text{ accepts}}|
  \ = \
  \sum_{\vek{c}\in C^\ell} |\setc{\vek{v}\in \mathcal{J}}{\rho_M(\vek{v},\vek{c})\text{ accepts}}|\,.
\end{equation*}
Consequently,
\begin{equation*}
  \sum_{\vek{c}\in C^\ell} |\setc{\vek{v}\in \mathcal{J}}{\rho_M(\vek{v},\vek{c})\text{ accepts}}|
  \ \geq \
  |C^\ell|\cdot \frac{|\mathcal{J}|}{2}\,.
\end{equation*}
Therefore, there must exist at least one $\vek{c}\in C^\ell$ with
\[
   |\setc{\vek{v}\in \mathcal{J}}{\rho_M(\vek{v},\vek{c})\text{ accepts}}|
   \ \geq \
   \frac{|\mathcal{J}|}{2}\,,
\]
and the proof of Lemma~\ref{lemma:Choose_c} is complete.
\end{proof}

\pagebreak

\subsection{Skeletons of runs}\label{subsec:Skeletons}

\begin{definition}[$\text{local\_views}(\rho)$, $\text{ndet\_choices}(\rho)$, $\text{moves}(\rho)$]
\upshape
Let $M$ be an NLM.
\begin{enumerate}[(a)]
\item
 The \emph{local view}, $\lv(\gamma)$, of a configuration $\gamma = (a,p,d,X)$ of
 $M$ is defined via
 \[
   \lv(\gamma) \ \ \deff \ \ (a,d,y) \quad\text{with}\quad
   y \ \ \deff \ \ \begin{pmatrix} x_{1,p_1}\\\vdots\\ x_{t,p_t} \end{pmatrix}.
 \]
 I.e., $\lv(\gamma)$ carries the information on $M$'s current state, head directions, and
 contents of the list cells currently being seen.
\item
Let
$\rho = (\rho_1,\twodots,\rho_\ell)$ be a run
of $M$.
We define
\begin{enumerate}[(i)]
\item
\[
   \text{local\_views}(\rho) \ \deff \
    \big(\,\lv(\rho_1),\ldots,\lv(\rho_\ell)\,\big).
\]
\item
  $\text{ndet\_choices}(\rho) \subseteq C^{\ell-1}$ to be the set of all sequences
  $\vek{c}=(c_1,\twodots,c_{\ell-1})$ such that, for all $i<\ell$, $\rho_{i+1}$ is the
  $c_i$-successor of $\rho_i$.
  \smallskip\\
  Note that $\Pr(\rho)= \frac{|\text{ndet\_choices}(\rho)|}{|C|^{\ell-1}}$.
\item
\[
   \text{moves}(\rho) \ \deff \
    \big(\,\text{move}_1,\twodots,\text{move}_{\ell-1} \,\big) \ \in \ \big(\set{0,1,-1}^t\big)^{\ell-1}\,,
\]
where, for every $i<\ell$,
$\text{move}_i = (\text{move}_{i,1},\twodots,\text{move}_{i,t})^\top \allowbreak \in\set{0,1,-1}^t$
such that, for each $\tau\in\set{1,\twodots,t}$, $\text{move}_{i,\tau} = 0$ (resp., 1, resp.,
$-1$) if, and only if, in the transition from configuration $\rho_i$ to configuration
$\rho_{i+1}$, the head on the $\tau$-th list stayed on the same list cell (resp.,
moved to the next cell to the right, resp., to the left).
\uend
\end{enumerate}
\end{enumerate}
\end{definition}
To prove lower bound results for list machines, we use the notion of a
\emph{skeleton} of a run. Basically,
a skeleton describes the information \emph{flow} during a run, in the
sense that it does \emph{not} describe the exchanged data items (i.e., input
values), but instead, it describes which input \emph{positions} the data
items originally came from.
The input positions of an NLM $M=(t,m,I,C,A,a_0,\alpha,B,\Bacc)$ are
simply the indices $i\in\{1,\ldots,m\}$.

\begin{definition}[Index Strings and Skeletons]\label{def:skeleton}
\upshape
Let $M$ be an NLM, let $\vek{v} = (v_1,\twodots,v_m)\in I^m$ be an input for $M$, let
$\rho$ be a run of $M$ for input $\vek v$, and
let $\gamma = (a,p,d,X)$ be one of the configurations in $\rho$.
\begin{enumerate}[(a)]
\item For every cell content $x_{\tau,j}$ in $X$ (for each list $\tau\in\set{1,\twodots,t}$),
 we write
 \[
    \ind(x_{\tau,j})
 \]
 to denote the \emph{index string}, i.e., the string obtained from
 $x_{\tau,j}$ by replacing each occurrence of input number $v_{i}$ by its index
 (i.e., input position) $i\in \set{1,\twodots,m}$, and by replacing each occurrence of a
 nondeterministic choice $c\in C$ by the wildcard symbol ``?''.
\item
 For $y = \big(x_{1,p_1},\twodots,x_{t,p_t}\big){}^\top$ we let
 \[
    \ind(y)\ := \ \big(\ind(x_{1,p_1}),\twodots,\ind(x_{t,p_t})\big){}^\top.
 \]
\item
 The \emph{skeleton} of a configuration $\gamma$'s
 local view $\lv(\gamma)= (a,d,y)$ is defined via
 \[
    \skel(\lv(\gamma)) \ \deff \ \big(a,d,\ind(y)\big).
 \]
\item The \emph{skeleton of a run} $\rho=(\rho_1,\twodots,\rho_\ell)$ of $M$ is defined via
  \[
    \skel(\rho) \ \deff \ \big(s, \text{moves}(\rho)\big),
  \]
  where $s=(s_1,\twodots,s_\ell)$ with $s_1:=\skel(\lv(\rho_1))$, and
  for all $i<\ell$, 
  if $\text{moves}(\rho)=(\text{move}_1,\twodots,\text{move}_{\ell-1})^\top$,
  \[
     s_{i+1} \ \deff \ \left\{
       \begin{array}{ll}
         \skel(\lv(\rho_{i+1})) \ \ & \text{if } \text{move}_{i}\neq (0,0,\twodots,0)^\top
        \\[1ex]
          \text{``?''} & \text{otherwise.}
       \end{array}
     \right.
  \]
\end{enumerate}
\end{definition}

\begin{remark}
Note that, given
an input instance $\vek{v}$ for an NLM $M$, the skeleton $\zeta:= \skel(\rho)$ of
a run $\rho$ of $M$ on input $\vek{v}$, and a sequence $\vek{c}\in \text{ndet\_choices}(\rho)$,
the entire run $\rho$ can be reconstructed.
\uend
\end{remark}

\subsection{Basic Properties of List Machines}\label{subsec:BasicPropertiesLMs}

\smallskip\noindent
In this section we provide some basic properties of list machines
concerning the size and shape of runs, the number of skeletons of runs, and the
possibility of composing different runs.

\pagebreak

\begin{lemma}[List length and cell size]\label{lemma:CellSizeEtc}
\mbox{}\\
Let $M=(t,m,I,C,A,a_0,\alpha,B,\Bacc)$ be an $(r,t)$-bounded NLM.
\begin{enumerate}[(a)]
\item\label{lemma:CellSizeEtc:listlength}
 {\upshape The \emph{total list length} of a configuration of $M$ is defined as the sum of the
 lengths (i.e., number of cells) of all lists in that configuration.\footnote{Note that
 the total list length never decreases during a computation.}}
 \\
  For every $i\in \set{1,\twodots,r}$, the total list length of each configuration that occurs
  before the $i$-th change of a head direction is \ $\leq (t+1)^i\cdot m$.
  \\
  In particular,
  the total list length of each configuration in each
  run of $M$ is \ $\leq (t+1)^{r}\cdot m$.
\item\label{lemma:CellSizeEtc:cellsize}
  {\upshape The \emph{cell size} of a configuration of $M$ is defined as the maximum length of the
  entries of the cells occurring in the configuration (remember that the cell
  entries are strings over $\Al= I\cup C\cup A\cup\set{\struc{,}}$).}
  \\
  The cell size of each configuration in
  each run of $M$ is \ $\leq  11\cdot (\max\set{t,2})^{r}$.
\end{enumerate}
\end{lemma}
\begin{proof}
For \emph{deterministic} list machines, (a) and (b) were proved in \cite{groschwe05a} (cf., Claims~1 and 2 in
the proof of
\cite[Lemma~15]{groschwe05a}). For \emph{nondeterministic} list machines, the
proofs are virtually identical; only the cell size increases, as now
the list entries also contain the nondeterministic choices.
\par
In fact, the proof of (a) is
identical to the proof of \cite[Claim~2 in the proof of Lemma 15]{groschwe05a}:
Let $\gamma$ be a configuration of total list length $\ell$. Then the total list length of a
successor configuration $\gamma'$ of $\gamma$ is at most $\ell+t$, if a head moves or changes
its direction in the transition from $\gamma$ to $\gamma'$, and it remains $\ell$ otherwise.
\\
Now suppose $\gamma'$ is a configuration that can be reached from $\gamma$ without changing
the direction of any head. Then $\gamma'$ is reached from $\gamma$ with at most $\ell-t$ head
movements, because a head can move into the same direction for at most
$\lambda{-}1$ times on a list of length $\lambda$. Thus the total list length of $\gamma'$ is
at most
\begin{equation}\label{eqn:totallistlength}
  \ell + t\cdot (\ell-t)\,.
\end{equation}
The total list length of the initial configuration is $m+t-1$.
A simple induction based on (\ref{eqn:totallistlength}) shows that the
total list length of a configuration that occurs before the $i$-th change of a head direction
is at most
\[
  (t+1)^i\cdot m.
\]
This proves (a).
\par
For the proof of (b), let $\gamma$ be a configuration of cell size $s$. Then the
cell size of all configurations that can be reached from $\gamma$ without changing the
direction of any head is at most
\[
   1+ t\cdot (2+s) + 3 \ \ = \ \ 4 + t\cdot (2+s).
\]
The cell size of the initial configuration is 3. A simple induction shows that the
total cell size of any configuration that occurs before the
$i$-th change of a head direction is at most
\[
  4 +  \sum_{j=1}^{i-1} 6 t^j + 5 t^i \ \ \leq \ \ 11\cdot (\max\set{t,2})^i.
\]
\end{proof}

\begin{lemma}[The shape of runs of an NLM]\label{lemma:shapeofruns}
Let $M=(t,m,I,C,A,a_0,\alpha,B,\Bacc)$ be an $(r,t)$-bounded NLM, and let $k\deff |A|$.
The following is true for every run $\rho=(\rho_1,\twodots,\rho_\ell)$ of $M$ and the
corresponding sequence $\text{moves}(\rho) = (\text{move}_1,\twodots,\text{move}_{\ell-1})$.
\begin{enumerate}[(a)]
\item\label{item:LM:lengthofruns}
  $\ell \ \leq k+ \,k\cdot (t+1)^{r+1}\cdot m$.
\item
  There is a number $\mu\leq (t+1)^{r+1}\cdot m$ and there are
  indices $1\leq j_1<j_2<\cdots < j_\mu < \ell$ such that:
  \begin{enumerate}[(i)]
    \item
       For every $i\in\set{1,\twodots,\ell{-}1}$,
       \[
          \text{move}_i\neq (0,0,\twodots,0)^\top
          \ \iff \
          i\in\set{j_1,\twodots,j_\mu}\,.
       \]
     \item
       If $\mu=0$, then $\ell\leq k$. \\
       Otherwise,
       $j_1\leq k$; \
       $j_{\nu+1}-j_\nu \leq k$, \ for every $\nu\in\set{1,\twodots,\mu{-}1}$; \ and
       $\ell - j_\mu \leq k$.
  \end{enumerate}
\end{enumerate}
\end{lemma}
\begin{proof}
For indices $i<\ell$ with $\text{move}_i=(0,0,\twodots,0)^\top$ we know from
Definition~\ref{def:NLM-semantics}\,(\ref{item:NLM-semantics:successor}) that the
\emph{state} is the only thing in which $\rho_i$ and
$\rho_{i+1}$ my differ.
As $(r,t)$-bounded NLMs are \emph{not} allowed to have an \emph{infinite} run, we obtain
that without moving any of its heads, $M$ can make at most $k$ consecutive steps.
\par
On the other hand, for every $i\in\set{1,\twodots,r}$ we know from
Lemma~\ref{lemma:CellSizeEtc}\,(\ref{lemma:CellSizeEtc:listlength}) that the total list length
of a configuration that occurs before the $i$-th change of a head direction is
\begin{equation}\label{eqn:steps_without_move}
   \leq\  (t+1)^i \cdot m.
\end{equation}
Thus, between the $(i{-}1)$-st and the $i$-th change of a head direction, the number
of steps in which at least one head moves is
\[
   \leq (t+1)^i\cdot m.
\]
Altogether, for every run $\rho$ of $M$, the total number of steps in which at least one head moves
is
\begin{equation}\label{eqn:steps_with_move}
  \leq \ \
  \sum_{i=1}^{r} (t+1)^i\cdot m
  \ \ \leq \ \
  (t+1)^{r+1}\cdot m.
\end{equation}
Hence, we obtain that the total length of
each run of $M$ is
\[
  \leq \ \ k + \,k\cdot (t+1)^{r+1}\cdot m
\]
(namely, $M$ can pass through at most $k$ configurations before moving a head for the first time, it
can move a head for at most $(t+1)^{r+1}\cdot m$ times, and between any two head movements,
it can pass through at most $k$ configurations).
\\
Altogether, the proof of Lemma~\ref{lemma:shapeofruns} is complete.
\end{proof}
\smallskip

\begin{lemma}[Number of Skeletons]\label{lemma:NoOfRCPSkels}
\mbox{}\\
Let $M=(t,m,I,C,A,a_0,\alpha,B,\Bacc)$ be an $(r,t)$-bounded NLM
with $t\geq 2$ and $k\deff |A|\geq 2$.
\\
The number
\[
  |\setc{\skel(\rho)}{\rho\text{ is a run of $M$}}|
\]
of skeletons of runs of $M$ is
\[
  \leq \quad
  \big(m+k+3\big)^{12\cdot m\cdot (t+1)^{2r+2} + 24\cdot (t+1)^r}.
\]
\end{lemma}
\begin{proof}
We first count the number of skeletons of local views of
 \emph{configurations} $\gamma$ of $M$.
\smallskip\\
Let $\gamma$ be a configuration of $M$, and let $\lv(\gamma)$ be of
the form $(a,d,y)$. Then,
\[
  \skel(\lv(\gamma)) \ = \ (a,d,\ind(y)),
\]
where $a\in A$, $d\in \set{-1,1}^t$, and $\ind(y)$ is a string over the
alphabet
\[
   \set{1,\twodots,m}\cup \set{\text{``?''}}\cup A\cup\set{\langle,\rangle}.
\]
Due to Lemma~\ref{lemma:CellSizeEtc}\;(\ref{lemma:CellSizeEtc:cellsize}), the string
$\ind(y)$ has length $\leq 11\cdot t^r$.
Therefore,
\begin{equation}\label{eqn:no_of_skels_of_configs}
   |\setc{\skel(\lv(\gamma))}{\gamma \text{ is a configuration of $M$}}|
    \ \ \leq \ \
    k\cdot 2^t\cdot \big( m+k+3\big)^{11\cdot t^r}.
\end{equation}
\par
From Lemma~\ref{lemma:shapeofruns} we know that for every run
$\rho=(\rho_1,\twodots,\rho_\ell)$ of $M$ there
is a number $\mu\leq (t+1)^{r+1}\cdot m$ and
indices $1\leq j_1<j_2<\cdots < j_\mu < \ell$ such that
for $\text{moves}(\rho) = (\text{move}_1,\twodots,\text{move}_{\ell-1})$
we have:
\begin{enumerate}[(i)]
    \item
       For every $i\in\set{1,\twodots,\ell{-}1}$,
       \ \(
          \text{move}_i\neq (0,0,\twodots,0)^\top
          \ \iff \
          i\in\set{j_1,\twodots,j_\mu}.
       \)
     \item
       If $\mu=0$, then $\ell\leq k$. \\
       Otherwise,
       $j_1\leq k$; \
       $j_{\nu+1}-j_\nu \leq k$, \ for every $\nu\in\set{1,\twodots,\mu{-}1}$; \ and
       $\ell - j_\mu \leq k$.
\end{enumerate}
The total number of possibilities of choosing such $\mu$, $\ell$, $j_1,\twodots,j_\mu$ is
\begin{equation}\label{eqn:numberofj-mu}
  \leq \ \
  \sum_{\mu=0}^{(t+1)^{r+1}\cdot m} k^{\mu+1}
  \ \ \leq \ \
  k^{2+(t+1)^{r+1}\cdot m}.
\end{equation}
For each fixed $\rho$ with parameters $\mu,\ell,j_1,\twodots,j_\mu$,
$\skel(\rho)=(s,\text{moves}(\rho))$ is of the following form:
For every $i\leq \ell$ with $i\not\in\set{j_1,\twodots,j_\mu}$,
$\text{move}_i=(0,0,\twodots,0)^\top$ and $s_{i+1}=\text{``?''}$.
For the remaining indices $j_1,\twodots,j_\mu$, there
are
\begin{equation}\label{eqn:numberofmoves}
 \leq \ \
 3^{t\cdot\mu}
 \ \ \leq \ \
 3^{(t+1)^{r+2}\cdot m}
\end{equation}
possibilities of choosing $(\text{move}_{j_1},\twodots,\text{move}_{j_\mu})\in \big(\set{0,1,-1}^t\big)^\mu$,
and there are
\begin{equation}\label{eqn:numberofconfs}
 \leq \ \
 |\setc{\skel(\lv(\gamma))}{\gamma \text{ is a configuration of $M$}}|^{\mu}
 \ \ \leq \ \
  \big(k\cdot 2^t\cdot \big( m+k+3\big)^{11\cdot t^r}\big)^{(t+1)^{r+1}\cdot m}
\end{equation}
possibilities of choosing
$(s_{j_1+1},\twodots,s_{j_\mu+1}) = \big(\skel(\lv(\rho_{j_1+1})),\twodots,\skel(\lv(\rho_{j_\mu+1}))\big)$.
\par
In total, by computing the product of the terms in (\ref{eqn:numberofj-mu}), (\ref{eqn:numberofmoves}),
and (\ref{eqn:numberofconfs}), we
obtain that the number $|\setc{\skel(\rho)}{\rho\text{ is a run of $M$}}|$ of skeletons of
runs of $M$ is at most
\begin{equation}\label{eqn:numberofskels1}
\begin{array}[b]{ll}
 &
  \Big( k^{2+(t+1)^{r+1}\cdot m} \Big) \cdot
  \Big( 3^{(t+1)^{r+2}\cdot m} \Big) \cdot
  \Big( \big(k\cdot 2^t\cdot \big( m+k+3\big)^{11\cdot t^r}\big)^{(t+1)^{r+1}\cdot m} \Big)
 \\
 \leq \ \
 &
  \Big(
    k\cdot 3 \cdot k\cdot 2^t\cdot \big( m+k+3\big)^{11\cdot t^r}
  \Big)^{2+(t+1)^{r+2}\cdot m}
 \\
 \leq \ \
 &
  \Big(
     k^2\cdot  2^{t+\log 3} \cdot (m+k+3)^{11\cdot t^r}
  \Big)^{2+(t+1)^{r+2}\cdot m}.
\end{array}
\end{equation}
Obviously,
\[
   k^2 \ \leq \ (k+m+3)^2.
\]
Since $(k+m+3)\geq 2^2$, we have
\[
   2^{t+\log 3} \ \leq \ (m+k+3)^{t+1}.
\]
Inserting this into (\ref{eqn:numberofskels1}), we obtain that the number of skeletons of
runs of $M$ is
\[
 \begin{array}{ll}
  \leq \ \
 &
  \big(m+k+3\big)^{(11t^r+t+3)\cdot (2+(t+1)^{r+2}\cdot m)}
 \\
  \leq \ \
 &
  \big(m+k+3\big)^{(12\cdot(t+1)^r)\cdot (2+(t+1)^{r+2}\cdot m)}
 \\
  \leq \ \
 & \big(m+k+3\big)^{ 24\cdot (t+1)^r + 12\cdot (t+1)^{2r+2}\cdot m}.
 \end{array}
\]
This completes the proof of Lemma~\ref{lemma:NoOfRCPSkels}.
\end{proof}

\begin{definition}\label{def:compare}
\upshape
  Let $M=(t,m,I,C,A,a_0,\alpha,B,\Bacc)$ be an NLM and
  let
  \[
    \zeta \ = \ \big( (s_1,\twodots,s_\ell),\,(\text{move}_1,\twodots,\text{move}_{\ell-1})\big)
  \]
  be the skeleton of a run $\rho$ of $M$.
  We say that two input positions $i,i'\in\set{1,\twodots,m}$, are \emph{compared} in $\zeta$
  (respectively, in $\rho$) iff
  there is a $j\leq \ell$ such that $s_j$ is of the form
  \[
     \skel(\lv(\gamma)) \ = \ (a,d,\ind(y)), \text{ \ \ for some configuration $\gamma$}\,,
  \]
  and
  both $i$ and $i'$ occur in $\ind(y)$.\uend
\end{definition}

\begin{lemma}[Composition Lemma]\label{lemma:CompositionLemma}
  Let   $M=(t,m,I,C,A,a_0,\alpha,B,\Bacc)$ be an NLM and let $\ell\in\NN$ be an upper bound
  on the length of $M$'s runs.
  Let $\zeta$ be the skeleton of a run of $M$, and
  let $i,i'$ be input positions of $M$ that are not
  compared in $\zeta$.
  Let $\vek{v}=(v_1,\twodots,v_m)$ and $\vek{w}=(w_1,\twodots,w_m)$ be two
  different inputs for $M$ with
  \[
    w_j=v_j, \ \ \text{for all} \  j\in\set{1,\twodots,m}\setminus\set{i,i'}
  \]
  (i.e., $\vek{v}$ and $\vek{w}$ only differ at the input positions $i$ and $i'$).
  Furthermore, suppose
  there exists a sequence $\vek{c}=(c_1,\twodots,c_\ell)\in C^\ell$
  such that
  \[
     \skel\big(\rho_M(\vek{v},\vek{c})\big) \ \ = \ \
     \skel\big(\rho_M(\vek{w},\vek{c})\big) \ \ = \ \ \zeta,
  \]
  and $\rho_M(\vek{v},\vek{c})$ and $\rho_M(\vek{w},\vek{c})$ either both accept or both reject.
  Then, for the inputs
  $\vek{u}\deff (v_1,\twodots,v_i,\twodots,w_{i'},\twodots,v_m)$ and
  $\vek{u}'\deff (v_1,\twodots,w_i,\twodots,v_{i'},\twodots,v_m)$
  we have
  \begin{center}
    $ \zeta \ \ = \ \
     \skel\big(\rho_M(\vek{u},\vek{c})\big) \ \ = \ \
     \skel\big(\rho_M(\vek{u}',\vek{c})\big)
    $
    \smallskip\\
    and
    \smallskip\\
    $
    \rho_M(\vek{u},\vek{c}) \text{ accepts }
    \iff
    \rho_M(\vek{u}',\vek{c}) \text{ accepts }
    \iff
    \rho_M(\vek{v},\vek{c}) \text{ accepts }
    \iff
    \rho_M(\vek{w},\vek{c}) \text{ accepts.}
    $
  \end{center}
\end{lemma}
\begin{proof}
Let $\zeta = ((s_1,\ldots,s_{\ell'}),
(\textit{move}_1,\ldots,\textit{move}_{\ell'-1}))$ be the skeleton as in the
hypothesis of the lemma.
We show that $\skel(\rho_M(\vek{u},\vek{c})) = \zeta$, and that
$\rho_M(\vek{u},\vek{c})$ accepts if and only if $\rho_M(\vek{v},\vek{c})$ and
$\rho_M(\vek{w},\vek{c})$ accept.
The proof for $\vek{u}'$ instead of $\vek{u}$ is the same.

Let $\skel(\rho_M(\vek{u},\vek{c})) = ((s'_1,\ldots,s'_{\ell''}),
(\textit{move}'_1,\ldots,\textit{move}'_{\ell''-1}))$.
Let $j$ be the maximum index such that
\begin{enumerate}[(i)]
\item
  $(s'_1,\ldots,s'_j) = (s_1,\ldots,s_j)$, and
\item
  $(\textit{move}'_1,\ldots,\textit{move}'_{j-1}) =
  (\textit{move}_1,\ldots,\textit{move}_{j-1})$.
\end{enumerate}
Let $j'$ be the maximum index such that $j' \leq j$ and
$s_{j'} = s'_{j'} \neq \text{``?''}$.
By the hypothesis of the lemma we know that $i$ and $i'$ do not occur both in
$s_{j'}$.
Thus for some $\vek{x} \in \set{\vek{v},\vek{w}}$, $s_{j'}$ contains only input
positions where $\vek{u}$ and $\vek{x}$ coincide.
Let $\rho_M(\vek{x},\vek{c}) = (\rho_1,\ldots,\rho_{\ell'})$, and
let $\rho_M(\vek{u},\vek{c}) = (\rho'_1,\ldots,\rho'_{\ell''})$.
Since $s_{j'}$ contains only input positions where $\vek{u}$ and $\vek{x}$
coincide, we have $\lv(\rho_{j'}) = \lv(\rho'_{j'})$.
Since $\textit{move}_{j''} = \textit{move}'_{j''} = (0,\ldots,0)^\top$ for all
$j'' \in \set{j',\ldots,j-1}$, we therefore have $\lv(\rho_j) = \lv(\rho'_j)$.
This implies that the behavior in the $j$-th step of both runs,
$\rho_M(\vek{x},\vek{c})$ and $\rho_M(\vek{u},\vek{c})$, is the same.

\medskip\noindent\textit{Case 1 ($j = \ell'$): }
In this case there is no further step in the run, from which we conclude that
$\ell' = \ell''$.
Hence both skeletons, $\zeta$ and $\skel(\rho_M(\vek{u},\vek{c}))$, are
equal.
Moreover, $\lv(\rho_j) = \lv(\rho'_j)$ implies that both runs either accept or
reject.

\medskip\noindent\textit{Case 2 ($j < \ell'$): }
In this case we know that $\ell'' \geq j+1$, and that
$\textit{move}_j = \textit{move}'_j$.
By the choice of $j$ we also have $s_{j+1} \neq s'_{j+1}$, which together with
$move_j = \textit{move}'_j$ implies $s_{j+1} \neq \text{``?''}$ and
$s'_{j+1} \neq \text{``?''}$.
Let $s_{j+1} = (a,d,\ind)$ and $s'_{j+1} = (a',d',\ind')$.
Since $\lv(\rho_j) = \lv(\rho'_j)$, and the behavior in the $j$-th step of
both runs is the same, we have $a = a'$ and $d = d'$.
So, $\ind$ and $\ind'$ must differ on some component
$\tau \in \set{1,\ldots,t}$.
Let $\ind_\tau$ be the $\tau$-th component of $\ind$, and let $\ind'_\tau$ be
the $\tau$-th component of $\ind'$.

Since $(s'_1,\ldots,s'_j) = (s_1,\ldots,s_j)$ and
$(\textit{move}'_1,\ldots,\textit{move}'_j) = (\textit{move}_1,\ldots,\textit{move}_j)$,
the list cells visited directly after step
$j'' \in \set{0,\ldots,j}$ of all three runs, $\rho_M(\vek{v},\vek{c})$,
$\rho_M(\vek{w},\vek{c})$ and $\rho_M(\vek{u},\vek{c})$, are the same.
This in particular implies that $\ind_\tau$ and $\ind'_\tau$ describe the same
list cells, though in different runs.
So, if $\ind_\tau = \langle p \rangle$ for some input position $p$, or
$\ind_\tau = \langle \rangle$, then the cell described by $\ind_\tau$
has not been visited during the first $j$ steps of all three runs, and
therefore, $\ind'_\tau = \ind_\tau$.
Now we may assume that $\ind_\tau \neq \langle p \rangle$ for all input
positions $p$, and $\ind_\tau \neq \langle \rangle$.
Then,
$\ind_\tau = a\langle y_1\rangle\ldots\langle y_t\rangle\langle c\rangle$,
where $(a,d,y_1,\ldots,y_t) = s_{j''}$ for some $j'' \in \set{1,\ldots,j}$,
and $c$ is the $j''$-th nondeterministic choice of $\vek{c}$.
Also,
$\ind'_\tau = a'\langle y'_1\rangle\ldots\langle y'_t\rangle\langle c\rangle$,
where $(a',d',y'_1,\ldots,y'_t) = s'_{j''}$.
But $s_{j''} = s'_{j''}$, which contradicts $\ind_\tau \neq \ind'_\tau$.

To conclude, only Case 1 can occur, which gives the desired result of the
lemma.
\end{proof}

\subsection{The information flow during a list machine's run}\label{subsec:InformationFlow}

\smallskip\noindent
In this subsection we take a closer look at the information flow that can occur during
a list machine's computation and, using this,  we show that only a small number of
input positions can be compared during an NLM's run.
\begin{definition}[subsequence]\label{subsequence}
  \upshape
  A sequence $(s_1,\twodots,s_\lambda)$ is a \emph{subsequence} of a
  sequence $(s'_1,\twodots,s'_{\lambda'})$, if there exist indices $j_1<\cdots
  < j_{\lambda}$ such that $s_1= s'_{j_1}$, $s_2= s'_{j_2}$, \ldots,
  $s_\lambda = s'_{j_{\lambda}}$.\\ \mbox{}\uend
\end{definition}
\begin{definition}\label{def:occurs}
\upshape
Let $M=(t,m,I,C,A,a_0,\alpha,B,\Bacc)$ be an NLM.
Let $\gamma = (a,p,d,X)$ be a configuration of $M$ with
$X=(x_1,\twodots,x_t)^\top$ and $x_\tau = (x_{\tau,1},\twodots,x_{\tau,m_\tau})$,
for each $\tau\in\set{1,\twodots,t}$.
Furthermore, let $(i_1,\twodots,i_\lambda)\in \set{1,\twodots,m}^\lambda$, for
some $\lambda\in\NN$, be a sequence of input positions.

We say that the  sequence $(i_1,\twodots,i_\lambda)$ \emph{occurs} in
configuration $\gamma$, if the following is true:
There exists a $\tau\in\set{1,\twodots,t}$ and
list positions $1\leq j_1 \leq \cdots \leq j_{\lambda} \leq m_\tau$ such that, for all
$\mu\in\set{1,\twodots,\lambda}$, the
input position $i_\mu$ occurs in  $\ind(x_{\tau,j_\mu})$.
\uend
\end{definition}
The following lemma gives a closer understanding of the information flow that
can occur during an NLM's run.
\begin{lemma}[Merge Lemma]\label{lemma:Merge}
Let $M=(t,m,I,C,A,a_0,\alpha,B,\Bacc)$ be an $(r,t)$-bounded NLM, let $\rho$ be
a run of $M$, let $\gamma$ be a configuration in $\rho$, and let, for some $\lambda\in\NN$,
$(i_1,\twodots,i_\lambda)\in\set{1,\twodots,m}^\lambda$ be a sequence of input positions that occurs in
$\gamma$.

Then, there exist $t^r$ subsequences ${s_1}$,\twodots,${s_{t^r}}$ of
$(i_1,\twodots,i_\lambda)$
 such that
the following is true, where we let ${s_\mu} = (s_{\mu,1},\twodots,s_{\mu,\lambda_\mu})$,
for every $\mu\in\set{1,\twodots,t^r}$:
\begin{enumerate}[--]
 \item
   $\displaystyle\set{i_1,\twodots,i_\lambda}
    \ \ = \ \
    \bigcup_{\mu=1}^{t^r}\set{s_{\mu,1},\twodots,s_{\mu,\lambda_{\mu}}}$,\quad and
 \item
   for every $\mu\in\set{1,\twodots,t^r}$, ${s_\mu}$ is a
   subsequence either of $(1,\twodots,m)$ or of $(m,\twodots,1)$.
  \uend
\end{enumerate}
\end{lemma}
\begin{proof}
By induction on $r'\in\set{0,\twodots,r}$ we show that for each configuration that occurs
during the $r'$-th \emph{scan} (i.e., between the $(r'{-}1)$-st and the $r'$-th change of a head
direction), the above statement is true for $t^{r'}$ rather\linebreak than $t^r$.

For the induction start $r'=0$ we only have to consider $M$'s start configuration. Obviously,
every sequence $(i_1,\twodots,i_\lambda)$ that occurs in the start configuration, is
 a subsequence of $(1,\twodots,m)$.

For the induction step we note that all that $M$ can do during the $r'$-th scan is
\emph{merge} entries from $t$ different lists produced during the $(r'{-}1)$-st scan.
Therefore, $\set{i_1,\twodots,i_\lambda}$ is the union of $t$ sequences, each of which
is  a subsequence of either
$(i_1,\twodots,i_\lambda)$ or $(i_\lambda,\twodots,i_1)$ (corresponding to a
forward scan or a backward scan, respectively), and each of these $t$ subsequences has been
produced during the $(r'{-}1)$-st scan. By induction hypothesis, each of these
subsequences is the union of $t^{r'-1}$ subsequences of $(1,\twodots,m)$ or $(m,\twodots,1)$.
Consequently, $(i_1,\twodots,i_\lambda)$ must be the union of $t\cdot t^{r'-1}$ such
subsequences.
\end{proof}
\smallskip
We are now ready to show that only a small number of input positions can be compared
during a list machine's run.
%
%
\begin{lemma}[Only few input positions can be compared by an NLM]\label{lemma:CompareFew}
\mbox{}\\
Let $M=(t,2m,I,C,A,a_0,\alpha,B,\Bacc)$ be an NLM with $2m$ input positions.\\
Let $\vek{v}\deff (v_1,\twodots,v_m,v'_1,\twodots,v'_m)\in I^{2m}$ be an input for $M$,
let $\rho$ be a run of $M$ on input $\vek{v}$, and let $\zeta\deff\skel(\rho)$.
Then, for every permutation $\varphi$ of $\set{1,\twodots,m}$, there are at most
\[
   t^{2r}\cdot\textit{sortedness}(\varphi)
\]
different $i\in\set{1,\twodots,m}$
such that the input positions $i$ and $m+\varphi(i)$ are compared in $\zeta$
(i.e., the input values
$v_i$ and $v'_{\varphi(i)}$ are compared in $\rho$).
\end{lemma}
\begin{proof}
For some $\lambda\in\NN$ let $i_1,\twodots,i_\lambda$ be distinct elements from
$\set{1,\twodots,m}$ such that, for all $\mu\in\set{1,\twodots,\lambda}$,
the input positions $i_\mu$ and $m+\varphi(i_\mu)$ are compared in $\zeta$.
From Definition~\ref{def:compare} and~\ref{def:occurs} it then follows that,
for an appropriate permutation
$\pi:\set{1,\twodots,\lambda}\to\set{1,\twodots,\lambda}$, the sequence
\[
  {\iota}\ \ \deff \ \
  \big( \ \ i_{\pi(1)}\ , \ \ m{+}\varphi(i_{\pi(1)}) \ ,  \
        \ i_{\pi(2)}\ , \ \ m{+}\varphi(i_{\pi(2)}) \ , \
        \ \ldots \ ,
        \ i_{\pi(\lambda)} \ , \ \ m{+}\varphi(i_{\pi(\lambda)}) \ \  \big)
\]
\emph{occurs} in some configuration in run $\rho$.
From Lemma~\ref{lemma:Merge} we then obtain that there exist $t^r$ subsequences
${s_1}$,\twodots,${s_{t^r}}$ of
${\iota}$
 such that
the following is true, where we let ${s_\mu} = (s_{\mu,1},\twodots,s_{\mu,\lambda_\mu})$,
for every $\mu\in\set{1,\twodots,t^r}$:
\begin{enumerate}[--]
 \item
   $\displaystyle\set{\ i_1 \ ,\twodots, \ i_\lambda, \ m+\varphi(i_1) \ ,\twodots,
   \ m+\varphi(i_\lambda) \ }
    \ \ = \ \
    \bigcup_{\mu=1}^{t^r}\set{s_{\mu,1},\twodots,s_{\mu,\lambda_{\mu}}}$,\quad and
 \item
   for every $\mu\in\set{1,\twodots,t^r}$, ${s_\mu}$ is a
   subsequence either of $(1,\twodots,2m)$ or of $(2m,\twodots,1)$.
\end{enumerate}
In particular, at least one of the sequences $s_1,\twodots,s_{t^r}$ must contain
at least $\lambda'\deff \lceil \frac{\lambda}{t^r}\rceil$ elements from
$\set{i_1,\twodots,i_\lambda}$.
W.l.o.g.\ we may assume that $s_1$ is such a sequence, containing the elements
$\set{i_1,\twodots,i_{\lambda'}}$.
\\
Considering now the set
$\set{m{+}\varphi(i_1) \ ,\twodots,\ m{+}\varphi(i_{\lambda'})}$,
we obtain by the
same reasoning that one of the sequences $s_1,\twodots,s_{t^r}$ must contain at least
$\lambda''\deff \lceil\frac{\lambda'}{t^r} \rceil\geq \frac{\lambda}{t^{2r}}$ elements from
$\set{m{+}\varphi(i_1) \ ,\twodots,\ m{+}\varphi(i_{\lambda'})}$.
We may assume w.l.o.g.\ that $s_2$ is such a sequence, containing the elements
$m{+}\varphi(i_1) \ ,\twodots,\ m{+}\varphi(i_{\lambda''})$.
\par
Let us now arrange the elements \ $i_1,\twodots,i_{\lambda''},
m{+}\varphi(i_1),\twodots,m{+}\varphi(i_{\lambda''})$ \
in the same order as they appear in the
sequence ${\iota}$. I.e.,
let $\pi':\set{1,\twodots,\lambda''}\to\set{1,\twodots,\lambda''}$ be a permutation
such that
\[
  {\iota'}\ \ \deff \ \
  \big( \ i_{\pi'(1)} \ , \ \ m{+}\varphi(i_{\pi'(1)}) \ ,
        \ \ldots \ ,
        \ i_{\pi'(\lambda'')} \ , \ \
        m{+}\varphi(i_{\pi'(\lambda'')}) \  \big)
\]
is a subsequence of $\iota$.

Since $s_1$ is a subsequence of $\iota$ and a subsequence of either $(1,\twodots,2m)$ or
$(2m,\twodots,1)$, we obtain that
\[
\mbox{either} \quad i_{\pi'(1)} < i_{\pi'(2)} < \cdots < i_{\pi'(\lambda'')}
\quad \mbox{or} \quad i_{\pi'(1)} > i_{\pi'(2)} > \cdots > i_{\pi'(\lambda'')}.
\]
Similarly, since
$s_2$ is a subsequence of $\iota$ and a subsequence of either $(1,\twodots,2m)$ or
$(2m,\twodots,1)$, we obtain that
\[
\mbox{either} \quad m{+}\varphi(i_{\pi'(1)}) < \cdots < m{+}\varphi(i_{\pi'(\lambda'')})
\quad \mbox{or} \quad m{+}\varphi(i_{\pi'(1)}) > \cdots > m{+}\varphi(i_{\pi'(\lambda'')}),
\]
and therefore,
\[
\mbox{either} \quad \varphi(i_{\pi'(1)}) < \cdots < \varphi(i_{\pi'(\lambda'')})
\quad \mbox{or} \quad \varphi(i_{\pi'(1)}) > \cdots > \varphi(i_{\pi'(\lambda'')}).
\]
In other words,
$\big(\varphi(i_{\pi'(1)}),\twodots,\varphi(i_{\pi'(\lambda'')})\big)$ is a subsequence of
$\big(\varphi(1),\twodots,\varphi(m)\big)$ that is
sorted in either ascending or descending order.
According to Definition~\ref{def:sortedness} we therefore have
\[
   \lambda'' \leq \textit{sortedness}(\varphi)\,.
\]
Since $\lambda'' \geq \frac{\lambda}{t^{2r}}$, we hence
obtain that
\[
  \lambda \leq t^{2r}\cdot \textit{sortedness}(\varphi)\,,
\]
and the proof of Lemma~\ref{lemma:CompareFew} is complete.
\end{proof}

\subsection{Proof of Lemma~\ref{lemma:MainLM}}\label{subsec:BoundsForLMs}%

\smallskip\noindent
Finally, we are ready for the proof of Lemma~\ref{lemma:MainLM}.
\medskip\\
\textbf{Lemma~\ref{lemma:MainLM} (Lower Bound for List Machines) --- restated}.
{ \itshape
\mbox{}\\
  Let $k,m,n,r,t\in\mathbb N$ such that
 $m$ is a power of\/ $2$ and
  \[
  t\ \geq \ 2,\quad
  m\ \geq \ 24\cdot (t{+}1)^{4r}+1,\quad
  k\ \geq \ 2m+3,\quad
  n\ \geq \ 1+ \ (m^2+1)\cdot\log (2 k).
  \]
  We let $I\deff\set{0,1}^n$, identify $I$ with the set $\set{0,1,\twodots,2^n{-}1}$, and
  divide it into $m$ consecutive intervals $I_1,\twodots,I_m$ each of length ${2^n}/m$.
  \\
  Let $\varphi$ be a permutation of $\set{1,\twodots,m}$ with
  $\textit{sortedness}(\varphi)\leq 2\sqrt{m}-1$, and
  let
  \[
   \mathcal{I} \quad \deff \quad I_{\varphi(1)} \times \cdots \times I_{\varphi(m)} \times
       \ I_{1} \times \cdots \times I_{m}.
  \]
  Then there is no $(r,t)$-bounded NLM $M= (t,2m,I,C,A,a_0,\alpha,B,\Bacc)$
  with $|A|\leq k$ and $I=\set{0,1}^n$,
  such that for all $\vek{v}=(v_1,\twodots,v_m,v'_1,\twodots,v'_m)\in\mathcal{I}$ we have:
  \\
  If $(v_1,\twodots,v_m) \ = \ (v'_{\varphi(1)},\twodots,v'_{\varphi(m)})$, then
  $\Pr(M\text{ accepts }\vek{v})\geq \frac{1}{2}$; otherwise $\Pr(M\text{ accepts }\vek{v})=0$.
}%
\begin{proof}
Suppose for contradiction that $M$ is a list machine which meets the
requirements of Lemma~\ref{lemma:MainLM}.
\noindent
We let
\[
 \mathcal{I}_{\textit{eq}} \ \ \deff \ \
  \setc{\ (v_1,\twodots,v_m,v'_1,\twodots,v'_m)\in\mathcal{I}\ \ }{\ \
   (v_1,\twodots,v_m)=(v'_{\varphi(1)},\twodots,v'_{\varphi(m)})\ }.
\]
Note that
\[
  \textstyle
  |\mathcal{I}_\textit{eq}| \ \ = \ \ \left(\frac{2^n}{m}\right)^m.
\]
From the lemma's assumption we know that
\[
   \Pr(M\text{ accepts }\vek{v}) \ \geq \ \frac{1}{2},
\]
for every input $\vek{v}\in\mathcal{I}_\textit{eq}$.
Our goal is to show that there is some input
$\vek{u}\in\mathcal{I}\setminus \mathcal{I}_{\textit{eq}}$, for which there exists
an accepting run, i.e., for which $\Pr(M\text{ accepts }\vek{u})> 0$. It should be clear that
once having shown this, the proof of Lemma~\ref{lemma:MainLM} is complete.
\medskip
\\
Since $M$ is $(r,t)$-bounded, we know from Lemma~\ref{lemma:shapeofruns} that there exists a
number $\ell\in\NN$ that is an upper bound on the length of $M$'s runs.
From Lemma~\ref{lemma:Choose_c} we obtain a sequence $\vek{c} = (c_1,\twodots,c_\ell)\in C^\ell$
such that the set
\[
   \mathcal{I}_{\textit{acc},\vek{c}}
   \ \ \deff \ \
   \setc{\vek{v}\in\mathcal{I}_\textit{eq} \ }{ \ \rho_M(\vek{v},\vek{c}) \text{ accepts}}
\]
has size
\[
   |\mathcal{I}_{\textit{acc},\vek{c}}|
   \ \ \geq \ \
   \frac{|\mathcal{I}_\textit{eq}|}{2}
   \ \ \geq \ \
   \frac{1}{2}\cdot\left(\frac{2^n}{m}\right)^m.
\]
Now choose $\zeta$ to be the skeleton of a run of $M$
such that the set
\[
\mathcal{I}_{\textit{acc},\vek{c},\zeta} \ \ \deff \ \
  \{\
    \vek{v}\in\mathcal{I}_{\textit{acc},\vek{c}} \ \ : \ \
    \zeta = \skel(\rho_M(\vek{v},\vek{c}))\
  \}
\]
is as large as possible.

\begin{myclaim}\label{claim:sizeofIaccczeta}
\quad \(\displaystyle
 |\mathcal{I}_{\textit{acc},\vek{c},\zeta}| \ \ \geq \ \
  \frac{|\mathcal{I}_{\textit{acc},\vek{c}}|}{(2 k)^{m^2}}
   \ \ \geq \ \
  \frac{1}{2\cdot (2k)^{m^2}}\cdot \left( \frac{2^n}{m}\right)^m.
\)
\end{myclaim}
\begin{proof}
Let $\eta$ denote the number of skeletons of runs of $M$.
From Lemma~\ref{lemma:NoOfRCPSkels} we know that
\[
  \eta \ \ \leq \ \ \big(2m+k+3\big)^{24\cdot m\cdot (t+1)^{2r+2} + 24\cdot (t+1)^r}.
\]
From the assumption we know that $k\geq 2m+3$, and therefore
\begin{equation}\label{eqn:eta1}
  \eta \ \ \leq \ \
  \big( 2 k\big)^{24\cdot m\cdot (t+1)^{2r+2} + 24\cdot (t+1)^r}.
\end{equation}
From the assumption $m\geq 24\cdot(t{+}1)^{4r}+1$ we obtain that
\begin{equation}\label{eqn:eta2}
  24\cdot m\cdot (t+1)^{2r+2} + 24\cdot (t+1)^r
  \ \ \leq \ \
  24 \cdot m\cdot (t+1)^{2r+2} + m
  \ \ \leq \ \
  m^2\,.
\end{equation}
Altogether, we obtain from (\ref{eqn:eta1}) and (\ref{eqn:eta2}) that
\[
  \eta
 \quad\leq\quad
 (2 k)^{m^2}.
\]
Since the particular skeleton $\zeta$ was chosen in such a way that
$|\mathcal{I}_{\textit{acc},\vek{c},\zeta}|$ is as large as possible, and since the total
number of skeletons is at most $(2 k)^{m^2}$, we conclude that
\[
  |\mathcal{I}_{\textit{acc},\vek{c},\zeta}|
  \ \ \geq \ \
  \frac{|\mathcal{I}_{\textit{acc},\vek{c}}|}{(2k)^{m^2}}
  \ \ \geq \ \
  \frac{1}{2\cdot (2k)^{m^2}}\cdot \left( \frac{2^n}{m}\right)^m.
\]
Hence, the proof of Claim~\ref{claim:sizeofIaccczeta} is complete.
\end{proof}

\begin{myclaim}\label{claim:non-compare-i0}
There is an $i_0\in\set{1,\twodots,m}$ such that the input positions $i_0$ and
$m+\varphi(i_0)$ are not compared in $\zeta$.
\end{myclaim}
\begin{proof}
According to the particular choice of the permutation $\varphi$ we know that
\[
 \textit{sortedness}(\varphi) \ \ \leq \ \ 2\cdot \sqrt{m}-1.
\]
Due to Lemma~\ref{lemma:CompareFew} it therefore suffices to show that
\ \ $m \ > \ t^{2r}\cdot (2\sqrt{m}-1)$.
\\
From the assumption that \ \ $m \ \geq \ 24\cdot(t+1)^{4r}+1$ \ \ we know that, in particular,
$m>4\cdot t^{4r}$, i.e., $\sqrt{m}> 2\cdot t^{2r}$.
Hence, \ \ $t^{2r}\cdot(2\sqrt{m}-1) \ < \ \frac{1}{2}\cdot\sqrt{m}\cdot(2\sqrt{m}-1) \ \ \leq \ m$,
and the proof of Claim~\ref{claim:non-compare-i0} is complete.
\end{proof}%
\smallskip%
Without loss of generality let us henceforth assume that $i_0=1$ (for other $i_0$,
the proof is analogous but involves uglier notation).
\\
Now choose $v_2\in I_{\varphi(2)}$, \ldots, $v_{m}\in I_{\varphi(m)}$ such that
\[
\left| \
\setc{\ v_1\in I_{\varphi(1)}\ }{\
 (v_1,v_2\twodots,v_m,
  v_{\varphi^{-1}(1)},v_{\varphi^{-1}(2)},\twodots,v_{\varphi^{-1}(m)})\in
   \mathcal{I}_{\textit{acc},\vek{c},\zeta}\ }
 \
\right|
\]
is as large as possible.
Then, the number of $v_1$ such that
\[
  (v_1,v_2\twodots,v_m,
   v_{\varphi^{-1}(1)},v_{\varphi^{-1}(2)},\twodots,v_{\varphi^{-1}(m)})
   \ \ \in \ \  \mathcal{I}_{\textit{acc},\vek{c},\zeta}
\]
is at least
\[
\frac{|\mathcal{I}_{\textit{acc},\vek{c},\zeta}|}{\left(\frac{2^n}{m}\right)^{m-1}}
\quad\stackrel{\text{Claim~\ref{claim:sizeofIaccczeta}}}{\geq}\quad
\frac{\left(\frac{2^n}{m}\right)^{m}}{2\cdot (2 k)^{m^2}\cdot\left(\frac{2^n}{m}\right)^{m-1}}
\quad\geq\quad
\frac{2^n}{2m\cdot(2k)^{m^2}}\,.
\]
From the assumption we know that \ $n\geq 1+(m^2+1)\cdot \log(2 k)$.
Therefore,
\[
2^n
\quad\geq\quad
2\cdot (2 k)^{m^2+1}
\quad\geq\quad
2 \cdot (2k)\cdot (2 k)^{m^2}
\quad\stackrel{k\geq m}{\geq}\quad
2\cdot 2m\cdot (2 k)^{m^2}.
\]
Consequently,
\[
\frac{2^n}{2m\cdot(2 k)^{m^2}}
\quad\geq\quad
2.
\]
Thus, there are two different elements $v_1\neq w_1$ such that for
\ $(w_2,\twodots,w_m)\deff (v_2,\twodots,v_m)$ \ we have
\ $\vek{v}\deff(v_1,\twodots,v_m,v_{\varphi^{-1}(1)},\twodots,v_{\varphi^{-1}(m)})\in
     \mathcal{I}_{\textit{acc},\vek{c},\zeta}$ \
and
\ $\vek{w}\deff(w_1,\twodots,w_m,w_{\varphi^{-1}(1)},\twodots,w_{\varphi^{-1}(m)})\in
    \mathcal{I}_{\textit{acc},\vek{c},\zeta}$.
\\
Since the run $\rho_M(\vek{v},\vek{c})$ \emph{accepts},
we obtain from Lemma~\ref{lemma:CompositionLemma} that for the input
\[
 \vek{u}\  \deff \
 (v_1,\twodots,v_m,w_{\varphi^{-1}(1)},\twodots,w_{\varphi^{-1}(m)}) \ \ \in \ \
 \mathcal{I}\setminus\mathcal{I}_{\textit{eq}},
\]
the run $\rho_M(\vek{u},\vek{c})$ has to accept.
Therefore, we have found an input $\vek{u}\in\mathcal{I}\setminus\mathcal{I}_{\textit{eq}}$ with
\[
  \Pr(M\text{ accepts }\vek{u}) \ > \ 0.
\]
This finally completes the proof of Lemma~\ref{lemma:MainLM}.
\end{proof}




\section{Proofs of Lower Bounds for Turing machines}\label{appendix:proofsofcorollaries}

%
%
\smallskip\noindent
\begin{proofcomment}{of Corollary~\ref{cor:tight-bounds-for-short-versions}}\mbox{}\\
The upper bound is easily obtained when using a result of Chen and Yap \cite[Lemma\;7]{cheyap91} which
states that the sorting problem (i.e., the problem of sorting a given sequence of strings) can
be solved with two external memory tapes, $O(\log N)$ head reversals, and only constant internal memory space.

The lower bound for the problems $\CHECKSORT$, $\SETEQUALITY$, and $\MULTISETEQUALITY$ is stated in
Theorem~\ref{theo:set-equality}. To obtain the according lower bound for the ``$\SHORT$'' versions of
these problems,
we reduce the problem $\CHECK$-$\varphi$
(cf., Lemma~\ref{lemma:TMBound}) to the problems $\textsc{Short-}\CHECKSORT$,
$\textsc{Short-}\SETEQUALITY$, and $\textsc{Short-}\MULTISETEQUALITY$ (that is, the restriction
of these problems to inputs of the form
$v_1\#\cdots v_m\linebreak \#v'_1\#\cdots v'_m\#$, where each $v_i$ and each $v'_i$ is a 0-1-string of length at most
$c\cdot\log m$ for some constant $c \geq 2$)
in such a way that the reduction can be carried out in
$\ST(O(1),O(\log N),2)$.
More precisely, we construct a reduction (i.e., a function) $f$ that maps every
instance
\[
   \vek{v} \ := \ v_1\#\cdots v_m\#v'_1\#\cdots v'_m\#
\]
of $\CHECK$-$\varphi$ to an instance
\[
   f(\vek{v})
\]
of $\textsc{Short-}\CHECKSORT$ (respectively, of
$\textsc{Short-}\SETEQUALITY$ or $\textsc{Short-}\MULTISETEQUALITY$), such that
\begin{enumerate}[(1)]
\item the string $f(\vek{v})$ is of length $\Theta(|\vek{v}|)$,
\item $f(\vek{v})$ is a ``yes''-instance of $\textsc{Short-}\CHECKSORT$ (respectively, a
  ``yes''-instance of \linebreak
  $\textsc{Short-}\MSEQUALITY$) if, and only if,
  $\vek{v}$ is a ``yes''-instance of $\CHECK$-$\varphi$, and
\item there is an $(O(1),O(\log N), 2)$-bounded deterministic Turing machine that, when given
  an instance $\vek{v}$ of $\CHECK$-$\varphi$, computes $f(\vek{v})$.
\end{enumerate}
It should be clear that the existence of such a mapping $f$ shows that if
$\textsc{Short-}\CHECKSORT$ (respectively, $\textsc{Short-}\MSEQUALITY$) belongs to the
class $\RST(O(r),O(s),O(1))$, for some $s\in\Omega(\log N)$,
then also $\CHECK$-$\varphi$ belongs to $\RST(O(r),O(s),O(1))$. If $r$ and $s$ are chosen according
to the assumption of Corollary~\ref{cor:tight-bounds-for-short-versions}, this would cause a
contradiction to Lemma~\ref{lemma:TMBound}. Therefore,
$\textsc{Short-}\MSEQUALITY$ and $\textsc{Short-}\CHECKSORT$
do not belong to the class
$\RST(o(\log N), O(\frac{\sqrt[4]{N}}{\log N}), O(1))$.
\medskip\\
Now let us concentrate on the construction of the reduction $f$.
\\
For $i\in\set{1,\twodots,m}$, we subdivide the 0-1-string $v_i\in\set{0,1}^{m^3}$ into
$\mu:=\lceil\frac{m^3}{\log m}\rceil$ consecutive blocks $v_{i,1},\ldots,v_{i,\mu}$,
each of which has length $\log m$ (to ensure that also the last sub block has length $\log m$,
we may pad it with leading $0$s). In the same way, we subdivide the string $v'_i$ into sub blocks
$v'_{i,1},\ldots,v'_{i,\mu}$.
For a number $i\in\set{1,\twodots,m}$ we use $\text{BIN}(i)$ to denote the
binary representation of $i{-}1$ of length $\log m$; and
for a number $j\in\set{1,\twodots,\mu}$ we use $\text{BIN}'(j)$ to denote the
binary representation of $j{-}1$ of length $3\cdot \log m$.
\\
For every $i\in\set{1,\twodots,m}$ and $j\in\set{1,\twodots,\mu}$ we let
\[
  \begin{array}{rcrcc}
    w_{i,j} & \ \ \deff \ \ & \text{BIN}(\varphi(i))\ & \text{BIN}'(j)\ & v_{i,j},
\\
    w'_{i,j} & \ \ \deff \ \ & \text{BIN}(i)\ & \text{BIN}'(j)\ & v'_{i,j},
  \end{array}
\]
for every $i\in\set{1,\twodots,m}$ we let
\[
  \begin{array}{rcc}
     u_i & \ \ \deff \ \ &  w_{i,1}\# w_{i,2} \# \cdots w_{i,\mu} \#,
  \\
     u'_i & \ \ \deff \ \ &  w'_{i,1}\# w'_{i,2} \# \cdots w'_{i,\mu} \#,
  \end{array}
\]
and finally, we define
\[
   f(\vek{v}) \ \ \deff \ \ u_1  \cdots u_m \ u'_1  \cdots u'_m.
\]
Clearly, $f(\vek{v})$ can be viewed as an instance for $\textsc{Short-}\CHECKSORT$ or
$\textsc{Short-}\MSEQUALITY$, where
$m':= \mu\cdot m = \lceil\frac{m^4}{\log m}\rceil$ pairs $w_{i,j}$ and $w'_{i,j}$ of 0-1-strings of length
$5\cdot \log m \leq 2 \cdot \log m'$
are given.
Let us now check that the function $f$ has the properties (1)--(3).
\smallskip\\
\textbf{ad (1):} Every instance $\vek{v}$ of $\CHECK$-$\varphi$ is a string of length
$N=\Theta(m\cdot m^3)=\Theta(m^4)$, and $f(\vek{v})$ is a string of length $N'=\Theta(m^4)$.
\smallskip\\
\textbf{ad (2):}
\begin{equation}\label{eqn:short-proof}
 \begin{array}[b]{cl}
   & \text{$\vek{v}$ is a ``yes''-instance of $\CHECK$-$\varphi$}
\\[1ex]
     \ \iff \
   & (v_1,\twodots,v_m) \ = \ (v'_{\varphi(1)},\twodots,v'_{\varphi(m)})
\\[1ex]
     \iff
   & (v_{\varphi^{-1}(1)},\twodots,v_{\varphi^{-1}(m)}) \ = \ (v'_1,\twodots,v'_m)
\\[1ex]
     \iff
   & \text{for all $i\in\set{1,\twodots,m}$, } \
     (w_{\varphi^{-1}(i),1},\twodots,w_{\varphi^{-1}(i),\mu}) \ = \ (w'_{i,1},\twodots,w'_{i,\mu}).
 \end{array}
\end{equation}
It is straightforward to see that (\ref{eqn:short-proof}) holds if, and only if, $f(\vek{v})$ is a ``yes''-instance of
\textsc{Short-(Multi)Set-Equality}. Furthermore, as the list of 0-1-strings in the second half of $f(\vek{v})$ is
\emph{sorted} in ascending order, $f(\vek{v})$ is a ``yes''-instance of
$\textsc{Short-}\CHECKSORT$ if, and only if, it is a ``yes''-instance of \textsc{Short-(Multi)Set-Equality}.
\smallskip\\
\textbf{ad (3):} In a first scan of the input tape, a deterministic Turing machine can compute the number $m$ and
store its binary representation on an internal memory tape.
\\
Now recall from Remark~\ref{rem:permut} that the permutation $\varphi=\varphi_m$ is chosen in such a way
that for every $i\in\set{1,\twodots,m}$, the binary representation of $\varphi(i)$ is exactly the
reverse binary representation of $i$ --- and for each particular $i$, this can be computed on the
internal memory tapes. Therefore, during a second scan of the input tape, the
machine can produce the string $f(\vek{v})$ on a second external memory tape (without performing any further
head reversals on the external memory tapes).
\smallskip\\
Altogether, the proof of Corollary~\ref{cor:tight-bounds-for-short-versions} is complete.
\qed
\end{proofcomment}
\mbox{}
\parno
%
\begin{proofcomment}{of Corollary~\ref{cor:DetVsRandVsNdet}}\mbox{}\\
(a) is an immediate consequence of Theorem~\ref{theo:set-equality} and
Theorem~\ref{thm:efficient-randomized/ndet-solutions}\;(\ref{thm:efficient-randomized/ndet-solutions:coRST}).
\\
The second inequality in (b) follows directly from Theorem~\ref{theo:set-equality} and
Theorem~\ref{thm:efficient-randomized/ndet-solutions}\;(\ref{thm:efficient-randomized/ndet-solutions:NST}).
\\
The first inequality in (b) holds because, due to Theorem~\ref{thm:efficient-randomized/ndet-solutions}\;(\ref{thm:efficient-randomized/ndet-solutions:coRST}),
the \emph{complement} of the $\MULTISETEQUALITY$ problem belongs to $\RST(2,O(\log N),1)$. Since the deterministic
$\ST(\cdots)$ classes are closed under taking complements, Theorem~\ref{theo:set-equality} implies that the complement
of the $\MULTISETEQUALITY$
does not belong to $\ST(O(r),O(s),O(1))$.
\qed
\end{proofcomment}%
\mbox{}
\parno
%
\begin{proofcomment}{of Corollary~\ref{cor:sort}}\mbox{}\\
Of course, the $\CHECKSORT$ problem can be solved for input $x_1\#\cdots x_m\#y_1\#\cdots y_m\#$ by
(1) sorting $x_1\#\cdots\#x_m$ in ascending order and writing
the sorted sequence, $x'_1\#\cdots \# x'_m$ onto the second external memory tape, and
(2) comparing $y_1\#\cdots \#y_m$ and the (sorted) sequence $x'_1\#\cdots \# x'_m$ in parallel.
\par
Therefore, if the sorting problem could be solved in
$\FRST(o(\log N),O(\frac{\sqrt[4]{N}}{\log N}),O(1))$, i.e., by an
$(o(\log N),O(\frac{\sqrt[4]{N}}{\log N}),O(1))$-bounded \emph{LasVegas}-RTM $T$,
then we could solve the $\CHECKSORT$ problem by an
$(o(\log N),O(\frac{\sqrt[4]{N}}{\log N}),O(1))$-bounded $(\frac{1}{2},0)$-RTM $T'$ which uses
$T$ as a subroutine such that $T'$ rejects whenever $T$ answers \emph{``I don't know''} and $T'$ accepts
whenever $T$ produces a sorted sequence that is equal to the sequence $y_1\#\cdots \#y_m$.
This, however, contradicts Theorem~\ref{theo:set-equality}.
\qed
\end{proofcomment}%

\clearpage
}
\end{document}